\documentclass[twoside,aas_macros,12pt]{article} 

\usepackage[colorlinks=true,citecolor=blue,linktoc=all]{hyperref}
\usepackage{graphicx,aas_macros,isotope,color}
\usepackage[T1]{fontenc}
\usepackage{cite}       
\usepackage{dsfont}               
\usepackage{mathrsfs}             
\usepackage{slashed}              
\usepackage{amsmath}
\usepackage{bbold}
\usepackage{amssymb}
\usepackage{amsbsy}
\usepackage{amsfonts}
\usepackage{bm}

\numberwithin{equation}{section}
\numberwithin{table}{section}
\numberwithin{figure}{section}

\usepackage{titlesec,multirow}
\usepackage{sectsty}
\titleformat{\section}{\normalfont\Large\bfseries}{\thesection}{1em}{}
\titleformat{\subsection}{\normalfont\large\bfseries}{\thesubsection}{1em}{}
\titleformat{\subsubsection}{\normalfont\normalsize\bfseries}{\thesubsubsection}{1em}{}
\newcommand{\be}{\begin{equation}}
\newcommand{\ee}{\end{equation}}
\newcommand{\bea}{\begin{eqnarray}}
\newcommand{\eea}{\end{eqnarray}}

\newcommand{\vecp}{{\bm p}}
\begin{document}
\title{{\bf Heavy Baryons in Compact Stars}}
\author{
 Armen Sedrakian$^{1,2}$, Jia Jie Li$^{3}$, Fridolin Weber$^{4,5}$\\
\\
$^1$Frankfurt Institute for Advanced Studies, Ruth-Moufang-Str. 1, \\60438 Frankfurt am Main, Germany\\
$^2$Institute of Theoretical Physics, University of Wroc\l{}aw, pl. M. Borna 9,\\ 50-204 Wroc\l{}aw, Poland\\
$^3$School of Physical Science and Technology, Southwest University, \\Chongqing 400700, China\\
$^4$Department of Physics, San Diego State University, San Diego,\\ CA 92182, USA\\
$^5$Center for Astrophysics and Space Sciences, \\University of California at San Diego, La Jolla, \\
 CA 92093, USA\\
}
\maketitle

\begin{abstract}
We review the physics of hyperons and $\Delta$-resonances in dense matter in compact stars. The covariant density functional approach to the equation of state and composition of dense nuclear matter in the mean-field Hartree and Hartree-Fock approximation is presented, with regimes covering cold $\beta$-equilibrated matter, hot and dense matter with and without neutrinos relevant for the description of supernovas and binary neutron star mergers, as well as dilute expanding nuclear matter in collision experiments. We discuss the static properties of compact stars with hyperons and $\Delta$-resonances in light of constraints placed in recent years by the multimessenger astrophysics of compact stars on the compact stars' masses, radii, and tidal deformabilities. The effects of kaon condensation and strong magnetic fields on the composition of hypernuclear stars are also discussed.  The properties of rapidly rotating compact hypernuclear stars are discussed and confronted with the observations of 2.5-2.8 solar mass compact objects in gravitational wave events. We further discuss the cooling of hypernuclear stars, the neutrino emission reactions, hyperonic pairing, and the mass hierarchy in the cooling curves that arises due to the onset of hyperons. The effects of hyperons and $\Delta$-resonances on the equation of state of hot nuclear matter in the dense regime, relevant for the transient astrophysical event and in the dilute regime relevant to the collider physics is discussed. The review closes with a discussion of universal relations among the integral parameters of hot and cold hypernuclear stars and their implications for the analysis of binary neutron star merger events. 
\end{abstract}

		
\newpage
\thispagestyle{empty}

{
  \hypersetup{pdfborder = {0 0 0},
    colorlinks,
    citecolor=red,
    linkcolor=blue
  }
  \tableofcontents
}

	
\newpage
\section{Introduction}
\label{sec:Intro}

Compact (or neutron) stars represent the endpoints of the evolution of ordinary massive stars.  They are natural astrophysical laboratories for particle and nuclear physics, under conditions that are very different from those in terrestrial laboratories.  For example, the densities in compact stars are a factor of 5-10 higher than in ordinary nuclei. In the 1960s and 1970s, it was suggested that they might contain non-nucleonic constituents of matter, for example, hyperons~\cite{Ambartsumyan1960SvA,Ambartsumyan1961AZh,Leung1971ApJ,Pandharipande1971NuPhA,Moszkowski1974PhRvD,Bethe1974NuPhA} or deconfined quark matter~\cite{Itoh1970PThPh,Collins1975PhRvL,Fritzsch_1973}. The presence of (hyper)nuclear matter in compact stars may exhibit some extraordinary properties, such as superfluidity and superconductivity, a trapped neutrino component in the early stages of
evolution, or superstrong magnetic fields, see~Refs.~\cite{Shapiro:1983du,Glendenning2012compact,weber1999pulsars,Sedrakian2007PrPNP,Oertel_RMP_2017}.

\begin{figure}[tb]
\begin{center}
\includegraphics[width=14cm]{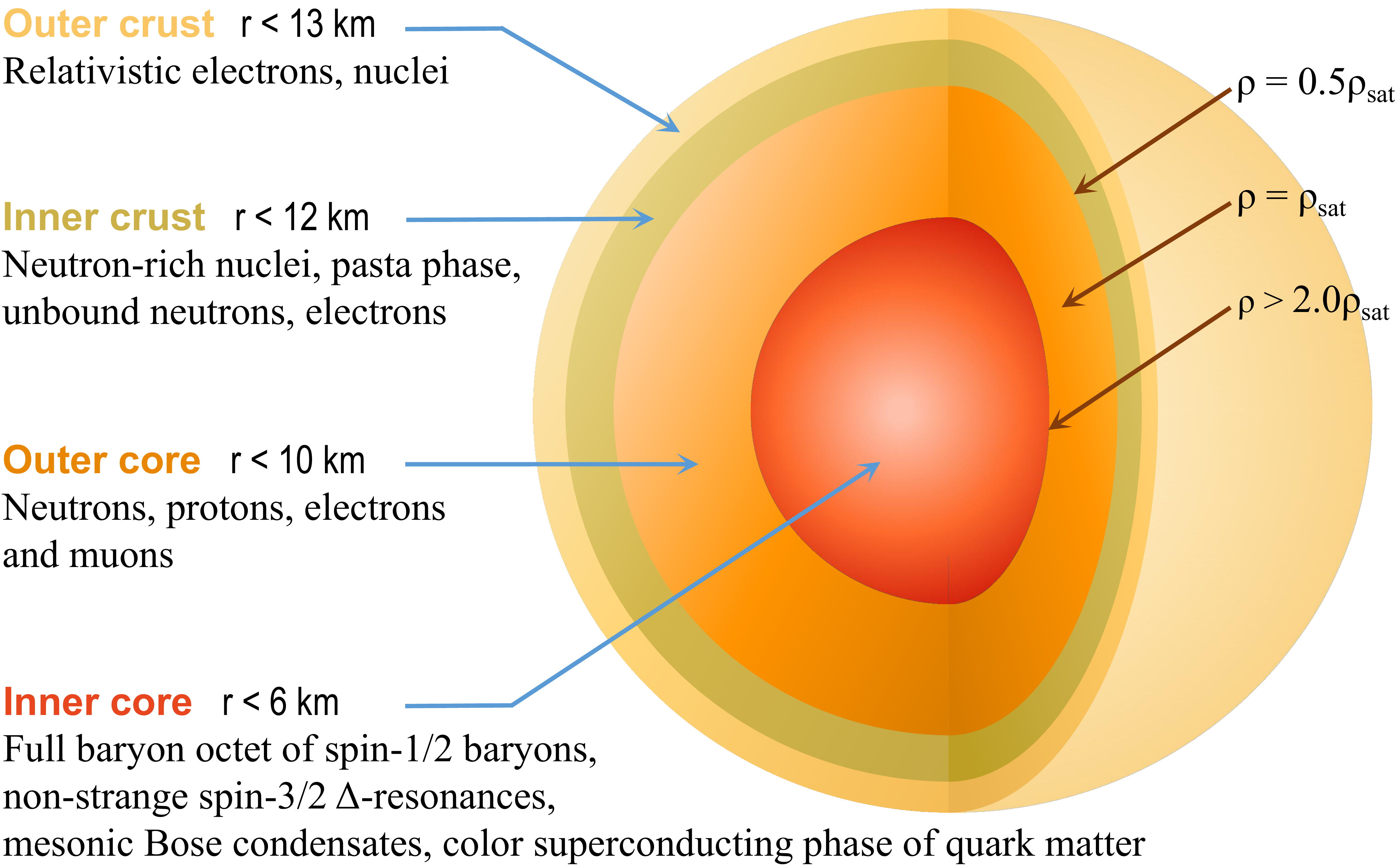}
\end{center}
\caption{Schematic illustration of the interior of a $M = 1.4\,M_{\odot}$ mass neutron star. The transition densities between different regions (in units of nuclear saturation density $\rho_{\rm sat}$) and the corresponding radial distance
$r$
from the stellar center are indicated. The particle content of each region is indicated as well. Note the possibility of mixed phases involving deconfined quark matter and hypernuclear matter, which are not shown. }
\label{fig:1.1}
\end{figure}

One of the main theoretical challenges in describing compact stars is the variety of possible phases at high densities.
A compact star has a mass (conservatively estimated) in the range $1.1\lesssim M/M_{\odot}\lesssim 2.3$,
where $M_{\odot}$ is the solar mass. Its central density can reach up to about 10 times the nuclear saturation density ($\rho_{\rm sat} \simeq 0.16$ fm$^{-3}$) and its radius lies within the range $11 \lesssim R\lesssim 14$~km. A compact star roughly consists of five main regions: the atmosphere, the outer and inner crust, and the outer and inner core, as schematically illustrated in Fig.~\ref{fig:1.1}. The outer and inner crusts are characterized by the presence of nuclear clusters immersed in electron gas and a sea of neutrons in the inner crust. The outer core consists of neutrons, protons, and leptons (mainly electrons with some admixture of muons).  The composition of the inner core which starts from about 2$\rho_{\rm sat}$
and extends up to densities $(5-10)\rho_{\rm sat}$  is not well understood: among the many possibilities are hyperonization, the phase transition to deconfined quark matter, the onset of meson condensation, etc.
For a textbook discussion of the phases of compact stars see, for 
example, Refs.~\cite{Shapiro:1983du,Glendenning2012compact,weber1999pulsars}.

The last decade has seen several groundbreaking observational advances that have reshaped our understanding of compact stars. These include the measurement of heavy pulsars in binary systems consisting of a neutron star and a white dwarf, the observation of gravitational waves in mergers of binary neutron stars (hereafter BNS), and the simultaneous measurement of the mass and radius of nearby X-ray-emitting neutron stars.  In recent years, these advances have provided crucial insights for the development of theoretical models of dense matter. Most importantly, they have allowed us to significantly constrain the theoretical parameter ranges of dense matter models.

The purpose of this review is to present recent advances in the study of the nuclear equation of state, the composition of superdense matter, and the physics of compact stars with hypernuclear cores. This is done in the framework of the covariant density functional (hereafter CDF) method for describing dense hypernuclear matter --- a method that provides a sufficiently flexible approach to accommodate astrophysical and laboratory information on hypernuclei. Lagrangian-based relativistic CDFs of nuclear systems (also known as nuclear relativistic mean-field models) work with effective degrees of freedom --- baryons and mesons. They provide a well-motivated and accessible way to determine the energy density of matter. 
The parameters of CDFs are determined from available nuclear data. The CDFs based on the relativistic mean-field model were applied to hypernuclear systems as early as in the 1980s and 1990s~\cite{Glendenning1985,Glendenning:1991ic,Glendenning:1991es,Huber1998,Huber:1994zz,Schaffner1996,Papazoglou:1997uw}.

The motivation to study hypernuclear stars and the interest in hypernuclear CDFs resurged after the observation of a two solar-mass pulsar in binary orbit with a white dwarf in 2010~\cite{Demorest2010Nature}. This radio-astronomical observation provided us with a significant lower bound on the maximum mass of a compact star for the first time since
the discovery of pulsars in 1967. 
The hyperons become energetically favorable once the Fermi energy of the neutrons exceeds their effective/in-medium rest masses.
The appearance of hyperons reduces the degeneracy pressure of cold hypernuclear matter so that its equation of state (hereafter EoS) becomes softer than that of nucleonic matter.
As a result, the maximum possible mass of a compact star with hyperons decreases to values below $2\,M_{\odot}$~\cite{Baldo_2000,SchulzePhysRevC}, which is in direct contradiction with the observations of two solar-mass pulsars. This contradiction is known as the ``hyperon puzzle''.
As discussed below, modern density functionals for hypernuclear matter not only solve the hyperon puzzle, but also explain the data from gravitational wave physics and X-ray observations of nearby pulsars~\cite{Bednarek:2011gd,Bonanno2012,Weissenborn2012a,Tsubakihara:2012ic,Jiang:2012hy,Massot_2012,Providencia:2012rx,Colucci2013,Dalen2014,Gusakov_MNRAS2014,Gomes:2014aka,Banik:2014qja,LopesPhysRevC2014,Maslov2015PhLB,Oertel2015,Drago:2015cea,Maslov:2015wba,Miyatsu_2015,Tolos2016,Oertel:2016xsn,Torres:2016ydl,Fortin2017,Lijj2018a,Lijj2018b,Gomes:2018eiv,Fortin:2020qin,Stone_FrASS_2022}.  The possibility of nucleation of $\Delta$-resonances was considered along with hyperonization, based on essentially the same arguments favoring heavy baryons over high-energy neutrons~\cite{Sawyer1972ApJ,Waldhauser1987,Waldhauser1988,Weber1989JPG,Choudhury1993,Schurhoff2010}. Again, after the discovery of massive pulsars, CDF methods were invoked to treat $\Delta$-resonance admixed (hyper)nuclear matter~\cite{Drago:2014oja,Caibj2015,ZhuPhysRevC2016,Kolomeitsev2017,Lijj2018b,Sahoo:2018xeu,Ribes_2019,Spinella2020:WSBook,Raduta2020,Thapa:2021kfo,Thapa:2020ohp,Dexheimer_2021_EPJA}.

Compact star properties will be discussed below exclusively within Einstein's theory of general relativity. However, there has been substantial work in recent years on the interpretation of the astrophysics of compact stars within alternative theories of gravity, see, for example, Refs.~\cite{Blazquez-Salcedo:2015ets,Motahar:2017blm,Astashenok:2014pua,Astashenok:2014gda,Astashenok:2014nua} and references therein, which we will not cover in this review.  Nevertheless, we will discuss universal (EoS-independent) relations between integral parameters of compact stars which are of great importance for testing the gravity theories in the strong-field regime.

This review is organized as follows. Section~\ref{sec:Constraints} discusses the astrophysical constraints on compact star properties that became available in recent years. The key ideas of the CDF theory are presented in Sec.~\ref{sec:Hyper_DFT}.  Section~\ref{sec:HNS_properties} is devoted to the properties of hypernuclear stars, including the EoS, composition, and global parameters. The recent results on rapidly rotating hypernuclear stars are discussed in Sec.~\ref{sec:Rapid_rotation}.  The finite temperature extension of the CDF theory is discussed 
in Sec.~\ref{sec:FiniteT}, which first considers hypernuclear matter at high densities 
with and without neutrinos (Subsec.~\ref{ssec:FT_high_densities}) and then examines the interplay between clustering and heavy-baryon degrees of freedom in the warm and low-density nuclear matter (Subsec.~\ref{ssec:Clusters}). The closely related problems of pairing among the hyperons and the cooling of hypernuclear stars are addressed in Secs.~\ref{sec:Pairing} and \ref{sec:Cooling}, respectively. The universal relations among the global parameters of static and rotating compact stars at zero and finite temperature are discussed in Sec.~\ref{sec:Universality}. 
Our conclusions are given in Sec.~\ref{sec:Conclusions}.


\section{Astrophysical Constraints on Neutron Stars}
\label{sec:Constraints}

Pulsar timing is one of the methods that provides information about the masses of pulsars, through measurements of Kepler parameters and observed spin properties of pulsars in binaries. These could be either neutron star--neutron star or a neutron star--white dwarf binaries. Neutron star--black hole binaries have also been observed in gravitational waves, but they did not place significant constraints on neutron star properties so far~\cite{Abbott_2021_BHNS}.  The Shapiro delay method for measuring pulsar's mass is based on the observation that its electromagnetic radiation experiences a time delay when passing through the gravitational field of the companion object (a neutron star or a white dwarf)~\cite{Shapiro1964PhRvL}. This method has been successfully applied to binary stellar systems involving a millisecond pulsar, where the pulsar's periodic pulse signal travels trajectories of different lengths in the space-time continuum depending on whether the pulsar passes in front of or behind its binary companion relative to a distant observer. In general relativity, the Shapiro time delay depends on the mass of the companion and the degree of inclination of the binary star system.  The first measurement of a massive pulsar was made for PSR J1614-2230, a 3.2 ms pulsar in an 8.7-day orbit with a massive white dwarf companion in a highly inclined orbit~\cite{Demorest2010Nature,Fonseca2016}.  Improved measurements from NANOGrav put the mass of this pulsar at $1.908(16)\,M_{\odot}$~\cite{Arzoumanian2018ApJS}. The second massive pulsar with high-precision mass measurements is J0348+0432, a 39 ms pulsar in a 2.46-hour orbit with a white dwarf. In this case, optical observation and modeling of the companion white dwarf were used in addition to pulsar timing measurements of the binary's Kepler parameters to determine the mass of this millisecond pulsar as $ 2.01\pm 0.04\,M_{\odot}$~\cite{Antoniadis2013Sci}.  The third and most massive neutron star with a high precision measurement of the mass is PSR J0740+6620~\cite{Cromartie2020NatAs}, a 2.89-ms pulsar in a 4.77-day orbit with a white dwarf. Timing analysis which measured the Shapiro delay puts the mass of this pulsar at $2.08\pm{0.07}\,M_{\odot}$ at $68.3\%$ credibility~\cite{Fonseca2021}. Thus, the timing observations of the three millisecond pulsars J1614-2230, J0348+0432, and J0740+6620 indicate that compact stars with masses $\simeq 2 M_{\odot}$ exist in nature.  On the other hand, general relativity predicts that stable compact stellar sequences terminate at some maximum mass for any EoS used.  The stable configurations are determined by the Bardeen--Thorne--Meltzer criterion \cite{Bardeen1966ApJ}, which states that a star is stable only as long as its mass increases with central density. However, the stability can also be inferred from the oscillation mode analysis of the stellar pulsations in the vicinity of the maximum mass of the star. From this, we can conclude that the maximum mass predicted by an EoS must be at least as large as those found in
 the timing observations. In other words, the millisecond pulsar observations quoted above set a lower limit on the maximum mass of a compact star.

 With the advent of gravitational wave astronomy and the first measurement by the LIGO and Virgo Collaboration of gravitational waves from a BNS merger in the event named GW170817~\cite{Abbott2017a}, it became possible to constrain the properties of compact stars by studying their tidal response. A fainter signal was later detected in the event GW190425, which is likely a BNS coalescence~\cite{Abbott2020ApJ}. These measurements, made by the second-generation LIGO and Virgo Collaboration ground-based gravitational observatories,
 employ the idea that a neutron star is deformed in the tidal gravitational field of its companion. At the lowest order, such deformations are described by the induced quadrupole moment of the star.  The gravitational wave signal emitted before the merger of stars in a binary contains direct information about the tidal properties of compact stars.  The tidal deformability $\lambda$ is defined as the coefficient relating the induced quadrupole moment $Q_{ij}$ to the perturbing tidal field ${\cal E}_{ij}$ acting on a star and perturbing its shape $Q_{ij} = -\lambda {\cal E}_{ij}$, where $i$ and $j$ denote the spatial coordinates~\cite{Flanagan:2007ix,Hinderer:2007mb}. It is related to the gravitational Love number $k_2$ via the relation
\begin{equation}
\lambda = \frac{2}{3} k_2 R^5,
\end{equation}
which exhibits its sensitivity to the radius of the star $R$.  
Both $k_2$ and $R$ depend on the EoS of matter~\cite{Baiotti:2019sew}.  Frequently, one uses the dimensionless tidal deformability defined as
\begin{equation}
\label{eq:Lambda}
\Lambda = \frac{\lambda}{M^5} = \frac{2}{3}k_2\,C^{-5},
\end{equation}
where $C= M/R$ is the compactness of the star. Often, one also defines an effective tidal deformability
as~\cite{Flanagan:2007ix,Hinderer:2007mb}
\begin{equation}
\tilde \Lambda = \frac{16}{13} \frac{(M_1 + 12M_2) M_1^4\Lambda_1 +
(M_2 + 12M_1)M_2^4\Lambda_2}{(M_1 + M_2)^5},
\end{equation}
involving the masses and tidal deformabilities of both stars. The analyses of the events GW170817 and GW190425 were performed for high- and low-spin priors. Below, we give some characteristic numbers only for low-spin priors suggested by Galactic observations. The total binary masses $2.73^{+0.04}_{-0.01}\, M_{\odot}$ and $3.4^{+0.3}_{-0.1}\, M_{\odot}$ were derived for these event, respectively.  The component masses are in the range of $1.16 -1.6 M_{\odot}$ for the GW170817 event and $1.46 -1.87 M_{\odot}$ for the GW190425 event.  The analysis of the GW170817 event resulted 
in an estimate $\tilde \Lambda \simeq 300^{+500}_{-190}$  ($90\%$ confidence)~\cite{Abbott_PRX_2019}. In the case of the event GW190425, an upper limit $\tilde \Lambda\le 600$ was derived for the mass range given above.

The surfaces of neutron stars emit $X$ rays due to their thermal heating by the currents associated with particle flows in the magnetosphere. The locations of the emitting hot spots reflect the structure and topology of the magnetosphere.  Due to the rotation of the star, the hot spots produce pulsed emission. The thermal $X$-ray radiation pulse profile of the millisecond pulsar J0030+0451 was used by the NICER team to constrain simultaneously its mass and radius~\cite{Miller2019ApJ,Riley2019ApJ}. Their procedure involves modeling the soft $X$-ray pulses produced by the rotation of hot spots on the surface of the star and fitting them to the NICER waveform data. It was assumed that the emitting atmosphere of the star consists of ionized hydrogen and that the magnetic field has no influence on the properties of the atmosphere. The two independent analyses concluded that $M = 1.44^{+0.15}_{-0.14} \, M_{\odot}$, $R = 13.02^{+1.24}_{^-1.06}$~km ~\cite{Miller2019ApJ}, and $M=1.34^{+0.15}_{-0.16} \, M_{\odot}$, $R= 12.71^{+1.14}_{-1.19}$~km~\cite{Riley2019ApJ} ($68.3\%$ credible interval).  The same collaboration also measured the radius of PSR J0740+6620 to be $R= 13.7^{+2.6}_{-1.5}$~km~\cite{Miller2021} and $R= 12.39^{+1.3}_{-0.98}$~km~\cite{Riley:2021pdl} ($68.3\%$ credible interval).  

In addition to the above constraints, the masses of neutron stars in binaries have been measured with high precision to be in the range of $1.2\lesssim M/M_{\odot}\lesssim 1.6$ with a significant concentration around  $1.4\, M_{\odot}$~\cite{Ozel:2016oaf,Lattimer:2019eez}. Furthermore, the moment of inertia of a neutron star is expected to be measured in the double pulsar system PSR J0737-3039~\cite{Lattimer_ApJ_2005}, where both stellar masses are already accurately determined by observations.

\section{Density Functionals for Hypernuclear Matter}
\label{sec:Hyper_DFT}

Density functional theory provides a flexible framework for the study of equilibrium thermodynamics of nuclear and neutron star matter. In nuclear physics, a class of density functionals can be obtained in the framework of the so-called ``relativistic mean-field models'' of nuclear matter. By construction, they possess the main feature of density functional theory: the potential energy of the zero temperature system is a function of (energy) density alone. The intrinsic energies in these theories are evaluated in the Hartree or Hartree-Fock theories of many-particle theory, with the first approximation being the simplest implementation of such a scheme. The Hartree-Fock theories include the pion contribution in the density functional explicitly, which could be advantageous for a detailed treatment of the tensor force. Once the density functional is constructed, the relativistic Lagrangian parameters are adjusted to reproduce the laboratory data in a range consistent with other constraints, such as those from observations of compact stars. The microscopic {\it ab initio} many-body calculations are treated as data, i.e., they constrain the allowed range of parameters that enter the functional.
(As an aside, density functionals have been derived directly from microscopic theories, but this is not discussed here.) An advantage of the approach taken here is the straightforward extension of the density functional from nuclear matter to hypernuclear matter, to matter containing the $\Delta$-resonance. The fast numerical implementations allow us to scan a large parameter space associated with the density functional, which can increase significantly when heavy baryons are included in the functional.

\subsection{Hartree CDFs with density-dependent couplings}
\label{ssec:DD_Hartree}

\begin{figure}[t]
\begin{center}
\includegraphics[width=13cm]{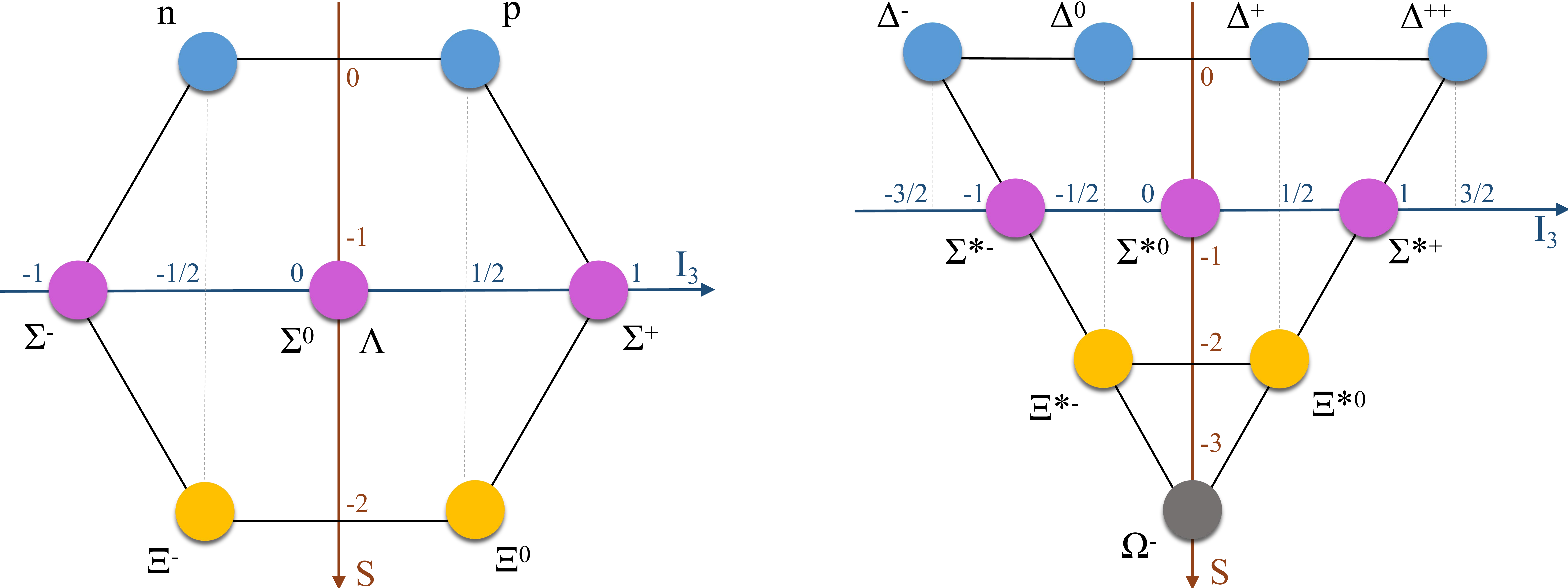}
\end{center}
\vspace{-0.8cm}
\caption{An illustration of the spin-1/2 octet of baryons (left) and the spin-3/2 decuplet of resonances (right), where the vertical axis is the strangeness and the horizontal axis is the third component of the isospin.}
\label{fig:1.2}
\end{figure}

In this section, we give a brief overview of the construction of the CDF of hypernuclear matter, starting from a relativistic Lagrangian and using the Hartree approximation. We will use a particular class of such functionals that assign a density dependence to the coupling constants describing the meson-baryon interaction, which incorporates modifications of the interaction due to changes in the density of the medium in which baryons and mesons 
are embedded~\cite{Typel1999,Lalazissis2005,Typelparticles2018}. The scheme discussed below will successively involve the $J=1/2$ baryon octet and the $J=3/2$ resonances; Figure~\ref{fig:1.2} illustrates the octet and decuplet ordering of the heavy baryons and their quantum numbers.

The Lagrangian of stellar matter with baryonic degrees of freedom can be written as
\begin{equation}
\label{eq:Lagrangian}
\mathscr{L} = \mathscr{L}_b + \mathscr{L}_m + \mathscr{L}_l + \mathscr{L}_{\rm em},
\end{equation}
where the baryonic Lagrangian is given by
\begin{eqnarray}
\label{eq:Lagrangian_B}
\mathscr{L}_b = \sum_b\bar\psi_b\big[\gamma^\mu \left(i\partial_\mu-g_{\omega b}\omega_\mu
- g_{\rho b}\bm{\tau}_b\cdot\bm{\varrho}_\mu\right)
- (m_b - g_{\sigma b}\sigma)\big]\psi_b ,
\end{eqnarray}
where the $b$-sum is over the $J^P = \frac{1}{2}^+$ baryon octet, $\psi_b$ are the Dirac fields of baryons with masses $m_b$, and $\sigma$, $\omega_\mu$, and $\bm{\varrho}_\mu$ are the mesonic fields which mediate the interaction among the baryon fields, $\bm{\tau}_b$ is the vector of Pauli matrices in the isospin space. Eq.~\eqref{eq:Lagrangian_B} contains the minimal set of mesons necessary for a quantitative description of nuclear phenomena. The isoscalar–scalar $\sigma$ meson mediates the medium-range attraction between baryon fields, the isoscalar–vector $\omega_\mu$ meson describes the short-range repulsion, and the isovector–vector $\bm{\varrho}_\mu$ meson accounts for the isospin dependence of baryon–baryon interactions. The baryon--meson vertex also takes the simplest possible form consistent with the Lorentz and isospin structures.  The coupling constants $g_{\sigma b}$, $g_{\omega b}$, and $g_{\rho b}$ are, in general, density-dependent.

The mesonic part of the Lagrangian is given by
\begin{eqnarray}
\label{eq:meson_H}
\mathscr{L}_m = \frac{1}{2}
\partial^\mu\sigma\partial_\mu\sigma-\frac{1}{2}m^2_\sigma \sigma^2 -
\frac{1}{4}\omega^{\mu\nu}\omega_{\mu\nu} + \frac{1}{2}m^2_\omega
                  \omega^\mu\omega_\mu - \frac{1}{4}{\bm\varrho}^{\mu\nu}\cdot
                  {\bm\varrho}_{\mu\nu}
               + \frac{1}{2}m^2_\rho {\bm\varrho}^\mu\cdot{\bm\varrho}_\mu,
\end{eqnarray}
where the boldface symbols denote vectors in the isospin space, 
$m_{\sigma}$, $m_{\omega}$, and $m_{\rho}$ are the meson masses,
and $\omega_{\mu\nu}$ and ${\bm\varrho}_{\mu\nu}$ represent the 
strength/field tensors of vector mesons
\begin{eqnarray}             
\omega_{\mu\nu} = 
\partial_{\mu}\omega_{\nu} - \partial_{\nu}\omega_{\mu},\qquad
\boldsymbol{\varrho}_{\mu\nu} = 
\partial_{\mu}\bm{\varrho}_{\nu} - \partial_{\nu}\bm{\varrho}_{\mu}.
\end{eqnarray}
The leptonic Lagrangian is given by
\begin{eqnarray}             
\mathscr{L}_l = \sum_{\lambda}\bar\psi_\lambda(i\gamma^\mu\partial_\mu -
      m_\lambda)\psi_\lambda,
\end{eqnarray}
where $\psi_\lambda$ are the lepton fields (in stellar matter the $\lambda$-summation includes electrons, muons, and at high temperatures the three flavors of neutrinos when neutrinos are trapped) and $m_\lambda$ are their masses. 

Finally, if the stellar matter is permeated by sizeable magnetic fields (which is the case, for example, in magnetars) then we need to include electromagnetism via its gauge-field Lagrangian
\begin{equation}
\label{eq:L_em}
\mathscr{L}_{\rm em} = - \frac{1}{4}F^{\mu\nu}F_{\mu\nu},
\end{equation}
where $F_{\mu\nu}$ is the electromagnetic field strength tensor, and replace the partial derivatives by the gauge-invariant expression $D_{\mu}= \partial_{\mu} + ie Q A_{\mu}$ where $A_\mu$ is the electromagnetic vector potential and $eQ$ is the charge of the particle with $e$ being the charge unit~\cite{Broderick2000,Broderick2002PhLB,Thapa:2020ohp,Dexheimer_2021_EPJA}.

The baryonic Lagrangian can be extended to include the non-strange $J=3/2$ members of the baryons decuplet which is the quartet of $\Delta$-resonances, see Fig.~\ref{fig:1.2}, by adding to the Lagrangian~\eqref{eq:Lagrangian_B} the term
\begin{eqnarray}
\mathscr{L}_d = \sum_{d}\bar\psi^{\nu}_{d}\bigg[\gamma^\mu \left(i\partial_\mu-g_{\omega d}\omega_\mu
  - g_{\rho d}{\bm\tau}_d\cdot{\bm\varrho}_\mu\right)
                   - (m_d- g_{\sigma d}\sigma)\bigg]\psi_{d\nu} ,
\end{eqnarray}
where the $d$-summation is over the resonances described by the Rarita--Schwinger 
fields $\psi_{d\nu}$. 

Furthermore, to describe the interactions between the strange particles the 
Lagrangian~\eqref{eq:Lagrangian} can be extended to include  two additional (hidden strangeness) mesons $\sigma^*$ and $\phi_{\mu}$ described by the Lagrangian
\begin{eqnarray}
\label{eq:meson_Hstr}
\mathscr{L}^{(s)}_{m} = \frac{1}{2}
\partial^\mu\sigma^*\partial_\mu\sigma^*
-\frac{1}{2}m^2_{\sigma^*} \sigma^{*2}
- \frac{1}{4}\phi^{\mu\nu}\phi_{\mu\nu} + \frac{1}{2}m_{\phi}^2\phi^{\mu}\phi_{\mu},
\end{eqnarray}
with field strength tensor 
$\phi_{\mu\nu} = \partial_{\mu}\phi_{\nu} - \partial_{\nu}\phi_{\mu}$.
Correspondingly, the baryonic Lagrangian~\eqref{eq:Lagrangian_B} 
is extended by adding a term 
\begin{eqnarray}
\label{eq:Lagrangian_S}
\mathscr{L}^{(s)}_b = \sum_b\bar\psi_b 
\big[- g_{\phi b}\gamma^\mu\phi_\mu + g_{\sigma^* b}\sigma^*\big]\psi_b . 
\end{eqnarray}
These two mesons do not couple to nucleons in the SU(6) spin-flavor model, i.e., $g_{\sigma^* N}= g_{\phi N}=0$, see however Eq.~\eqref{eq:Rphi} below.

The stationarity condition of the action-integral
\bea
S = \int \mathscr{L}(x)d^4x
\eea
with respect to the variations of the physical fields leads to the Euler-Lagrange equations, from which one can deduce the equations of motion for the meson fields, which obey inhomogeneous Klein-Gordon  or Proca equations with source terms. A similar procedure for the baryon fields gives the in-medium Dirac equation 
\begin{equation}
(-i\gamma^\mu\partial_\mu + m +\Sigma)\psi(x)=0,
\end{equation}
where $\Sigma$ is the baryon self-energy. To close the system of equations one needs an approximate form of the self-energy in terms of interactions and densities of baryons. Below we discuss the self-energy for the relativistic Hartree and Hartree-Fock approximations.

To describe infinite nuclear matter it is convenient to work in the four-momentum space. The most general Lorentz decomposition of the self-energy in a rotationally and time-reversal invariant system is given by
\begin{equation}\label{eq:general_self_energy}
\Sigma(k)=\Sigma_S(k)+\gamma_0\, \Sigma_0(k)+\bm{\gamma}\cdot\bm{\hat{k}}\, \Sigma_V(k),
\end{equation}
where $\bm{\hat{\bm{k}}}$ is the unit vector along $\bm{k}$, and the scalar 
component $\Sigma_S$, time component $\Sigma_0$ and spatial component $\Sigma_V$ 
of the vector potential are functions of the four-momentum $k$ of the baryon.
Using Eq.~\eqref{eq:general_self_energy}, the Dirac equation for baryons in 
infinite matter can be written as
\begin{align}
(\bm{\gamma}\cdot\bm{k}^\ast + m^\ast )u(k,s,\tau)= \gamma^0 E^\ast u(k,s,\tau),
\end{align}
with positive energy spinors
\begin{align}
u(k,s,\tau)= \bigg[\frac{E^\ast + m^\ast}{2E^\ast}\bigg]^{1/2}
\begin{pmatrix} 1 \\[0.4em] \frac{\bm{\sigma}\cdot\bm{k}^\ast}{E^\ast + m^\ast}\end{pmatrix}
\chi_s\chi_\tau,
\end{align}
where $\chi_s$ and $\chi_\tau$, denote the spin and isospin wave functions, 
respectively, $\bm{\sigma}$ is the vector of Pauli matrices, and the following effective expressions of the momentum, mass, 
and energy were introduced 
\begin{align}
\bm{k}^\ast = \bm{k}+\hat{\bm{k}}\,\Sigma_V,\qquad
m^\ast = m +\Sigma_S,\qquad
E^\ast = E-\Sigma_0,
\end{align}
which permit us writing the energy spectrum of baryons in a form similar to the non-interacting case, i.e., $E^{\ast2}=\bm{k}^{\ast2}+m^{\ast2}$. It is also useful to introduce
\bea
\hat{K}\equiv\frac{\bm{k}^\ast}{E^\ast}\equiv\cos\eta(k),\qquad \hat{M}\equiv\frac{m^\ast}{E^\ast}\equiv\sin\eta(k),
\eea
which we will use in the expressions of densities and self-energies below.

The negative solutions of the Dirac equation for baryons are often neglected in the so-called ``no-sea approximation'' which is motivated by the large gap separating the positive and negative baryon states. If, however, the negative states are included, divergent terms emerge which can be removed by a cumbersome renormalization procedure. If the vacuum polarization is taken into count, the parameter sets of the effective Lagrangians need to be readjusted, which leads to  new parameter sets predicting results that marginally differ from the case where the vacuum polarization is neglected. Thus, the effects caused by negative energy states and vacuum polarization are effectively absorbed in the definition of the coupling constants~\cite{Glendenning1988,Glendenning1989}. 

Let us now turn to the couplings in the nucleonic sector. These are given by their values at the saturation density $\rho_{\rm sat}$ and by a functional form describing their dependence on net baryon density $\rho_B=\rho_b+\rho_d$ given by
\begin{equation}
g_{iN}(\rho_B) = g_{iN}(\rho_{\rm sat})h_i(x),
\end{equation}
where $x = \rho_B/\rho_{\rm sat}$ and 
\begin{eqnarray}\label{eq:h_functions}
h_i(x) =a_i\frac{1+b_i(x+d_i)^2}{1+c_i(x+d_i)^2},~i=\sigma,\omega,\qquad
h_\rho(x) = e^{-a_\rho(x-1)}.
\end{eqnarray}
The density dependence of the couplings implicitly takes into account many-body correlations that modify the interactions in the medium. In the following, we will adopt the DDME2 parameterization~\cite{Lalazissis2005}, whose parameter values are listed in Table~\ref{tab:1}. 
\begin{table}[t]
\centering
\caption{The values of parameters of the DDME2 CDF.}
{
\begin{tabular}{ccccccc}
\hline \hline
Meson ($i$) & $m_i$ (MeV) & $g_{iN}$& $a_i$& $b_i$& $c_i$& $d_i$ \\
\hline
$\sigma$ & 550.1238 & 10.5396 & 1.3881 & 1.0943 & 1.7057 & 0.4421 \\
$\omega$ & 783      & 13.0189 & 1.3892 & 0.9240 & 1.4620 & 0.4775 \\
$\varrho$   & 763      & 3.6836  & 0.5647 &        &        &        \\
\hline\hline 
\end{tabular}
}
\label{tab:1}
\end{table}

The masses of $\omega$ and $\varrho$ mesons are taken to be their free values.
The five constraints $h_i(1)=1$, $h_i^{\prime\prime}(0)=0$ and $h^{\prime\prime}_{\sigma}(1)=h^{\prime\prime}_{\omega}(1)$ allow one to reduce the eight free parameters in the isoscalar channel to three. Three additional parameters in this channel are $g_{\sigma N}(\rho_{\rm sat}), g_{\omega N}(\rho_{\rm sat})$ and $m_\sigma$. 
With two additional parameters in the isovector channel, the parameterization has in total eight parameters (seven entering the definition of the couplings and the mass of the $\sigma$ meson) which are adjusted to reproduce the properties of symmetric and asymmetric nuclear matter, binding energies, charge radii, and neutron skins of finite nuclei.

The meson fields obey in general inhomogeneous Klein-Gordon equations for scalar mesons and Proca equations for vector mesons. As well known, the current conservation implies that the respective Proca equations can be reduced to Klein-Gordon equations. For static and homogeneous infinite nuclear matter, the expectation values of space-like components of vector fields vanish, i.e., only the zero components survive, due to translational invariance and rotational symmetry. In addition, only the third component of isovector fields needs to be considered becavuse of the rotational invariance around the third axis in the isospin space.
In the mean-field approximation, the meson fields are replaced by their respective expectation values (which are not distinguished below notationally from the fields)
\begin{equation}
\begin{aligned}
m_{\sigma}^2\sigma &= \sum_{b} g_{\sigma b}\rho_{b}^s + \sum_{d} g_{\sigma d}\rho_{d}^s,
\qquad m_{\sigma^{*}}^2 \sigma^{*} = \sum_{b} g_{\sigma^{*} b} \rho_{b}^s,\\
m_{\omega}^2\omega_{0} &= \sum_{b} g_{\omega b}\rho_{b} +\sum_{d} g_{\omega d}\rho_{d},
\qquad m_{\phi}^2\phi_{0}= \sum_{b} g_{\phi b}\rho_{b},\\
m_{\rho}^2\varrho_{03} &= \sum_{b} g_{\rho b}
\tau_{3b}\rho_{b} + \sum_{d} g_{\rho d} \tau_{3d}\rho_{d} ,
\end{aligned}
\end{equation}
where the scalar and baryon (vector) number densities are defined
for the baryon octet as
\begin{align}
\rho^s_{b} \equiv 
\langle\bar{\psi}_b \psi_b\rangle=\frac{\gamma_b}{2\pi^2}\int dk\, k^2\hat{M}_b f^+(E^k_{b}),\qquad 
\rho_{b}   \equiv                                                       
\langle\bar{\psi}_b \gamma^0 \psi_b\rangle = \frac{\gamma_b}{2\pi^2}\int dk\, k^2 f^-(E^k_{b}),
\end{align}
where $\gamma_b$ is the spin degeneracy factor,  and
\begin{equation}
f^{\pm}(E^k_{b}) = 
\frac{1}{1+\exp[(E^k_b - \mu^{*}_b)/T]} 
\pm \frac{1}{1+
\exp[(E^k_b + \mu^{*}_b)/T]}
\end{equation}
is the sum and difference of the Fermi distribution functions of particles and anti-particle which represents their respective occupation probabilities at temperature $T$, with $E^k_{b} = \sqrt{k^2+m^{*2}_{b}}$ the single-particle energies and $\mu^*_b$ the effective chemical potential that preserves the particle number at average. It reduces to a step function at zero temperature.
For the $\Delta$-resonances, these are defined as $\rho_{d}^s \equiv \langle\bar{\psi}_{d\nu} \psi^\nu_d\rangle$ and 
$\rho_{d}\equiv \langle\bar{\psi}_{d\nu} \gamma^0 \psi^\nu_d\rangle$,
respectively. 

A straightforward computation of the direct (Hartree) contribution to 
the components of Lorentz decomposition~\eqref{eq:general_self_energy} 
of the self-energy gives
\begin{align}\label{eq:Hartree_Baryon}
\Sigma_{S, b} = - g_{\sigma b}\sigma   - g_{\sigma^\ast b}\sigma^\ast, \qquad
\Sigma_{0, b} = + g_{\omega b}\omega_0 + g_{\phi b}\phi + g_{\rho b}\tau_{3b}\varrho_{03}
                + \Sigma_R,
\end{align}
where $\tau_{3b}$ is the third component of baryon isospin, and 
\begin{align}\label{eq:Hartree_Delta}
\Sigma_{S, d} = - g_{\sigma d}\sigma, \qquad
\Sigma_{0, d} = + g_{\omega d}\omega_0 + g_{\rho d}\tau_{3d}\varrho_{03} + \Sigma_R.
\end{align}
The density dependence of the couplings leads to the 
rearrangement term $\Sigma_R$ in the vector self-energy, which is given by
\begin{align}\label{eq:Rearrangement}
\Sigma_R = \sum_b\Big[
-\frac{\partial g_{\sigma b}}{\partial \rho_b} \sigma \rho^s_b
-\frac{\partial g_{\sigma^* b}}{\partial \rho_b} \sigma^* \rho^s_b
+\frac{\partial g_{\omega b}}{\partial \rho_b} \omega \rho_b
+\frac{\partial g_{\phi b}}{\partial \rho_b} \phi \rho_b
+\frac{\partial g_{\rho b}}{\partial \rho_b} \varrho_{03} \rho_b \tau_{3b} 
\Big] + \sum_d [b \rightarrow d].
\end{align}
In addition, the effective (Dirac) baryon masses in the same
approximation are given by
\begin{equation}
m_{b}^* = m_b + \Sigma_{S, b}, \qquad
m_{d}^* = m_d + \Sigma_{S, d}.
\end{equation}

Given the Lagrangian density~\eqref{eq:Lagrangian}, the energy stress tensor can be constructed
\begin{eqnarray}
T^{\mu\nu} = \frac{\partial \mathscr{L}}{\partial (\partial_\mu\varphi_i)}\partial^\nu \varphi_i -  g^{\mu\nu}\mathscr{L},
\end{eqnarray}
where $\varphi_i$ stands generically for a boson or fermion field. Then,
its diagonal elements define the energy density and pressure
\begin{eqnarray}
{\cal E} = \langle T^{00}\rangle, \qquad P = \frac{1}{3}\sum_i\langle T^{ii}\rangle,
\end{eqnarray}
where the brackets refer to statistical averaging. Explicitly one finds
\begin{align}
{\cal E} = & + \frac{1}{2}m_\sigma^2\sigma^2 +\frac{1}{2}m_{\sigma^*}^{2}\sigma^{*2} +
               \frac{1}{2}m_\omega^2\omega_0^2 + \frac{1}{2}m_\phi^2\phi_0^2 + 
               \frac{1}{2}m_\rho^2\varrho_{03}^2 \nonumber \\ 
           & + \frac{1}{2\pi^2}\sum_{b,d}(2J_{b,d}+1) 
               \int_0^{\infty} dk\,k^2 E^k_{b,d} f^+(E^k_{b,d})
             + \frac{1}{\pi^2} \sum_{\lambda}
               \int_0^{\infty}dk\,k^2 E^k_\lambda f^+(E^k_\lambda) \, ,
\end{align}
and 
\begin{align}
P = & - \frac{1}{2}m_\sigma^2\sigma^2 -\frac{1}{2}m_{\sigma^*}^{2}\sigma^{*2}
      + \frac{1}{2}m_\omega^2\omega_0^2 + \frac{1}{2}m_\phi^2\phi_0^2 
      + \frac{1}{2}m_\rho^2\varrho_{03}^2 +\rho_B \Sigma_R \nonumber \\
    & + \frac{1}{6\pi^2}\sum_{b,d}(2J_{b,d}+1)
        \int_0^{\infty}\!\!\! dk \frac{k^4}{E^k_{b,d}} f^+(E^k_{b,d})
      + \frac{1}{3\pi^2} \sum_{\lambda} 
        \int_0^{\infty}\!\!\! dk \frac{k^4}{E^k_\lambda}f^+(E^k_{\lambda})\, , 
\end{align}
where $2J_{b,d}+1$ is the baryon degeneracy factor with $J_{b}=1/2$ for 
baryon octet and $J_{d}=3/2$ for $\Delta$ quadruplet, 
$E^k_{b,d} = \sqrt{k^2+m^{*2}_{b,d}}$ and 
$E^k_{\lambda}=\sqrt{k^2+m_\lambda^2}$ are the single particle 
energies of baryons and leptons respectively. The lepton mass 
$m_\lambda$ can be taken equal to its free-space value. 

The effective chemical potentials $\mu_{b,d}^*$ are related to baryon
chemical potentials $\mu_{b,d}$ via 
$\mu_{b}^*$ = $\mu_{b}-\Sigma_{0,b}$ and $\mu_{d}^*$ = $\mu_{d}-\Sigma_{0,d}$.
The chemical potentials are then expressed as
\begin{eqnarray}
\mu_{b} = \mu_b^* + g_{\omega b}\omega_{0} + g_{\phi b}\phi_{0} 
+ g_{\rho b} \tau_{b3} \varrho_{03} + \Sigma_{R}, \qquad
\mu_{d} = \mu_d^* + g_{\omega d}\omega_{0} +  g_{\rho d}
\tau_{d3} \varrho_{03} + \Sigma_{R}.
\end{eqnarray}
The rearrangement self-energy $\Sigma_R$ 
guarantees the thermodynamic consistency of the theory, i.e.,
the fact that the thermodynamic relation
\begin{equation}\label{eq:Thermodynamic_relation}
P = \rho^2 \frac{\partial }{\partial \rho}\left(\frac{\cal E}{\rho}\right)
\end{equation}
is fulfilled. Note that the rearrangement term contributes to the pressure, 
but not to the energy of the system.

\subsection{Hyperonic and \protect{$\Delta$}-resonance couplings}
\label{sec:Couplings}
The lack of reliable information on the hyperon--nucleon and hyperon--hyperon interactions prevents a highly precise determination of the parameters that enter the Lagrangian \eqref{eq:Lagrangian_B}. The SU(3) flavor-symmetric model allows one to determine the magnitude of the couplings in the Lagrangian \eqref{eq:Lagrangian_B} based on symmetry arguments and the ``eightfold way'' principles of elementary particle classification in particle physics~\cite{Swart1963}. The SU(3) symmetry in flavor space is, of course, broken by the mass of strange quarks at densities and temperatures relevant to compact stars. Nevertheless, it provides some guidance in cases where little or no information is available.

In this model, the spin $1/2$ baryons and mesons are arranged in octets, which is the lowest nontrivial irreducible representation of the symmetry group.  The interaction part of the SU(3)-invariant Lagrangian, describing the coupling of baryons and mesons, is constructed using matrix representations for the $J^P = 1/2^+$ baryon octet $B$ and the $J^P=1^-$ meson octet ($M_8$), which is complemented by a meson singlet ($M_1$) that allows the description of physical mesons via a mixing mechanism. The Lagrangian contains linear combinations of the antisymmetric ($F$-type), symmetric ($D$-type), and singular ($S$-type) scalar contributions (using the standard notations) 
\begin{eqnarray}
\mathscr{L}_{\rm {SU(3)}} = 
-g_8\sqrt{2}[\alpha\text{Tr}([\bar{B},M_8]B) + (1-\alpha)\text{Tr}(\{\bar{B},M_8\}B)]
- \frac{g_1}{\sqrt{3}}\text{Tr}(\bar{B}B)\text{Tr}(M_1),
\end{eqnarray}
where  $g_8$ and $g_1$ denote the meson octet and singlet couplings, respectively,
and $\alpha=F/(F+D)$ with $0\le \alpha \le 1$.

The physical mesons $\omega$ and $\phi$ then appear as a mixture of the 
$\omega_0$ and $\omega_8$ members of the vector meson nonet: 
\begin{equation}
\label{eq:vector_mixing}
\left(\begin{array}{c}
            \omega_8\\
            \omega_0 
            \end{array}
          \right)  =
            \left(  \begin{array}{cc}
            \cos\theta_V & \sin\theta_V\\
            -\sin\theta_V & \cos\theta_V
            \end{array}\right) \left(  \begin{array}{c}
            \omega\\
            \phi 
            \end{array}
        \right) \, , 
\end{equation}
where $\theta_V$ is the vector mixing angle. Within this mixing scheme, the coupling of a baryon to the physical $\omega$ meson is given by 
\begin{equation}\label{eq:Bomega}
g_{B\omega} = \cos\theta_V g_1 + \sin\theta_V \frac{g_8}{\sqrt{3}}\delta_B, \qquad B \in \{N, \Xi,\Lambda,\Sigma\}, 
\end{equation}
where $\delta_N = 4\alpha_V-1$, $\delta_\Xi = 1+2\alpha_V$ and $\delta_\Sigma = -\delta_\Lambda =2( 1-\alpha_V)$. It is convenient to express the coupling of hyperons to mesons by using nucleonic couplings as normalization. Then,
\begin{eqnarray}
   \label{eq:Romega}
  R_{\omega B} = \frac{g_{\omega B}}{g_{\omega N}} =
  \frac{1-\frac{g_8}{g_1\sqrt{3}}\delta_B\tan \theta_V}{1-\frac{g_8}{g_1\sqrt{3}}(1-4\alpha_V)\tan \theta_V}.
\end{eqnarray}
The $\phi$ meson couplings can be obtained from those for $\omega$ meson by the substitution $\cos\theta_V \rightarrow -\sin\theta_V$ and $\sin\theta_V\rightarrow\cos\theta_V$.  In general, it is possible for the $\phi$
meson to couple to the nucleon with a coupling given by
\begin{align}\label{eq:Rphi}
R_{\phi N} = -
\frac{\tan \theta_V -\frac{g_8}{g_1\sqrt{3}}\delta_N}{1+\frac{g_8}{g_1\sqrt{3}}\delta_N\tan \theta_V}.
\end{align}
The coupling for the isovector ${\varrho}$ meson is given by  
\begin{eqnarray}\label{eq:Rrho}
g_{\rho N} = g_8, \quad  R_{\rho\Xi} = - (1-2\alpha_V), \quad 
 R_{\rho\Sigma} = 2  \alpha_V, \quad R_{\rho\Lambda} = 0.
\end{eqnarray}
Note that the ${\varrho}$ couplings vanish exponentially at high densities according to Eq. (\ref{eq:h_functions})
and their effect on the properties of dense matter (beyond the threshold for the onset of hyperons) is small.

The approximate equality of the masses of $\omega$ and $\bm \varrho$ mesons implies that the mixing is ideal, in which case the
 $\phi$ meson is a pure $\bar{s}s$ state and the mixing angle is given by the 
\textit{ideal mixing} value $\tan \theta^{*}_V = 1/\sqrt{2}.$
Since the nucleon does not couple to the pure strange meson $\phi$.
Eq.~\eqref{eq:Rphi} implies that this is the case when 
$g_1 = \sqrt{6}\,g_8$ and  $\alpha_V = 1$, the latter being the universality assumption for the (electric) $F/(F + D)$ ratio, i.e., only $F$-type coupling is non-zero. In this case, the couplings
of heavy baryons are related to those of the nucleon as in the additive quark model.
\begin{table}[t]
\begin{center}
\caption{The ratios of the couplings of hyperons in the SU(6) spin-flavor model.}
\begin{tabular}{cccccc}
\hline\hline
$Y\backslash R$ & $R_{\sigma Y}$ & $R_{\sigma^\ast Y}$ & $R_{\omega Y}$                      & $R_{\phi Y}$   & $R_{\rho Y}$               \\
\hline
$\Lambda$ & 2/3  &  $-\sqrt{2}/3$ &  2/3 &  $-\sqrt{2}/3$  & 0 \\
$\Sigma $ & 2/3  &  $-\sqrt{2}/3$ &  2/3 & $-\sqrt{2}/3$  & 2 \\
$\Xi$     & 1/3  & $-2\sqrt{2}/3$ &  1/3 & $-2\sqrt{2}/3$ & 1 \\
\hline
\hline
\end{tabular}
\end{center}
\label{tab:2}
\end{table}
The couplings in the case of the scalar mesons $\sigma$ and $\sigma^\ast$, are obtained from those of $\omega$ and $\phi$ mesons, respectively, with the replacements $\omega\to \sigma$, $\phi\to \sigma^\ast$, and changing the vector indices to scalar 
ones, i.e., $V \to S$. In the case of $\sigma$ meson, the couplings are given by
\begin{equation}
g_{B\sigma} = \cos\theta_S~g_1 +  \delta_{B}g_8 \sin\theta_S/\sqrt{3},\\
\end{equation}
where in the definitions of $\delta_B$ the scalar ratio $\alpha_S$ appears instead of
its vector counterpart.  It can be shown that the coupling scheme defined in this manner obeys the following relation~\cite{Colucci2013}
\begin{eqnarray}
\label{coupling_variation}
2(g_{N\sigma} + g_{\Xi\sigma}) = 3g_{\Lambda\sigma} +
g_{\Sigma\sigma},
  \end{eqnarray}
which is valid for arbitrary values of the four parameters $\alpha_S$, $g_1$, $g_8$ and $\theta_S$.
It is easy to verify that Eq.~\eqref{coupling_variation} is satisfied for the coupling constants in the SU(6) spin-flavor symmetric quark model. Table~\ref{tab:2} lists the couplings within this model.

The magnitudes of the couplings of hyperons to the scalar mesons can be determined from the fits to their potentials in nuclear matter and to hypernuclei within a particular model.
\begin{figure}[t]
\begin{center}
  \includegraphics[width=10.5cm]{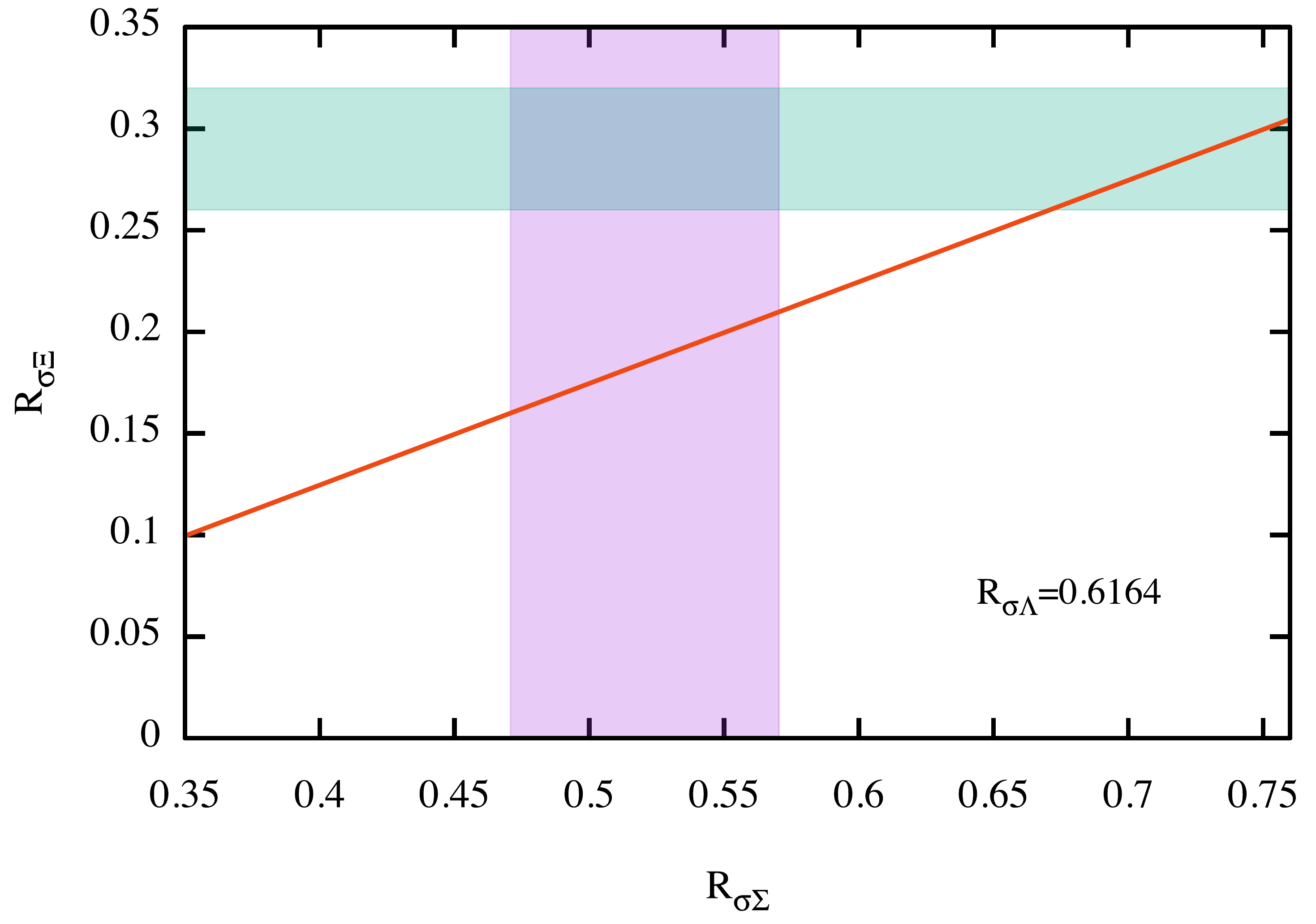} 
\end{center}
\caption{The ranges of coupling constants $R_{\sigma\Sigma}$ and $R_{\sigma\Xi}$ (shaded areas) compatible with experimental/theoretical information and relation \eqref{coupling_variation} (solid line) for DDME2 CDF and fixed $R_{\sigma\Lambda}= 0.6164$. The combined area of
$R_{\sigma\Sigma}$ and $R_{\sigma\Xi}$ has no overlap with the prediction based on the SU(3) flavor symmetry, indicating its breaking in the scalar-meson sector.}
\label{fig:1.3}
\end{figure}
For example, the binding energies of several single $\Lambda$-hypernuclei were used to determine the value of the $g_{\Lambda\sigma}$ coupling ~\cite{Dalen2014,Fortin2017}.  Similarly, the coupling of $\Lambda$ to the $\sigma^*$ meson is obtained from the fits to the binding of double $\Lambda$-hypernuclei \cite{Fortin2017}. In addition, the density functionals were adapted to treat hypernuclei with multiple strange cores~\cite{Khan:2015bxa,Margueron:2017eeq,Guven:2018sgo}.

Figure~\ref{fig:1.3} shows the ranges of the (dimensionless) couplings $R_{\sigma\Sigma}$ and $R_{\sigma\Xi}$, which cover the potential depths
  $U_{\Sigma}(\rho_{\rm sat}) = [-10: +30]$~MeV and $U_{\Xi}(\rho_{\rm sat}) = [-24:0]$~MeV
  at nuclear saturation density~\cite{Gala2016,Khaustov2000}.  
  The value $U_{\Xi}(\rho_{\rm sat}) = -24$~MeV
 was given in \cite{Friedman2021} and is much lower than the value obtained from the
 Lattice 2019 results~\cite{Inoue2019AIPC,Sasaki:2019qnh}. A fixed value of $R_{\sigma\Lambda}= 0.6164$~\cite{Dalen2014}
 was used in Fig.~\ref{fig:1.3}. It can be seen that, in contrast to the SU(6) model, the values of 
 $R_{\sigma\Xi}$ predicted by the relation \eqref{coupling_variation} do not intersect the overlapping 
 range of $R_{\sigma\Sigma}$ and $R_{\sigma\Xi}$ couplings, which
 indicates the breaking of the corresponding symmetry.
 The results for hyperonic compact stars reported in the following sections were obtained with parameters $R_{\sigma\Lambda} = 0.6106$, $R_{\sigma\Sigma} = 0.4426$, and $R_{\sigma\Xi} = 0.3024$~\cite{Lijj2018b} in the context of the DDME2-CDF extension to the hypernuclear sector, unless stated otherwise.
     
 Most studies of $\Delta$s in dense matter have been performed employing the Hartree CDF approach~\cite{Waldhauser1987,Waldhauser1988,Weber1989JPG,Choudhury1993,Caibj2015,Lavagno2010,Malfatti:2020,Lijj2018b,Drago:2014oja,Kolomeitsev2017}, the relativistic Hartree-Fock approach~\cite{Weber1989NPA,Lijj2018a,ZhuPhysRevC2016} and the quark-meson coupling model~\cite{Motta2020delta}. Recent work also includes models of $\Delta$-admixed nucleonic matter which  account for parity doubling and chiral restoration~\cite{Marczenko_2019_Uni,Marczenko_2022_ApJ,Marczenko_2022_PRD}. 
 The interactions of the $\Delta$-resonance within matter are not well known.
 Some information about the $\Delta$ potential in the isospin-symmetric nuclear matter is available from the analysis of the scattering of electrons and pions from nuclei with $\Delta$ excitation~\cite{Koch:1997ei,Connell1990PhRvC,Wehrberger1989NuPhA,Horikawa1980NuPhA,Nakamura2010PhRvC} and photoabsorption~\cite{Alberico1994PhLB,Riek2009}. The extracted value of the  potential is~\cite{Drago:2014oja} (note that the potentials are negative)
\begin{equation}
-30~\textrm{MeV}+V_{N} (\rho_{\rm sat})\le V_\Delta(\rho_{\rm sat})\le V_N(\rho_{\rm sat}),
\end{equation}
where $V_{N} (\rho_{\rm sat})$ is the nucleon isoscalar potential at the saturation density in symmetric nuclear matter.  The $\Delta$-resonance production in heavy-ion collisions is another channel of information, where however collective dynamics of nuclear matter comes into play~\cite{Cozma2016PhL,Cozma2021,Ono2019PhRvC,Xu2019PrPNP}. Numerical simulations provide hints towards the values of the potential in the range~\cite{Kolomeitsev2017}
\begin{equation}
V_{N} (\rho_{\rm sat}) \le V_\Delta(\rho_{\rm sat}) 
                       \le 2/3V_N(\rho_{\rm sat}).
\end{equation}
The couplings of $\Delta$-resonances to isovector mesons are not known. In the following, as in the case of hyperons,  we will use the ratios $R_{m\Delta } = g_{m\Delta}/g_{mN}$ (where $m$ stands for a meson) to quantify the  $\Delta$-resonance couplings. In recent works, these parameters have been varied in the range~\cite{Lijj2018b,Li2019PhRvC}
\begin{equation}
R_{\rho\Delta} = 1, \quad  0.8 \le R_{\omega\Delta} \le 1.6, 
\quad            
R_{\sigma\Delta} = R_{\omega\Delta} \pm 0.2,
\end{equation}
to explore the consequences of the inclusion of the $\Delta$-resonance in the CDF of (hyper)nuclear matter. It has been also observed that in a certain domain of coupling the $\Delta$-nucleon attraction may lead to spinodal instabilities as well as to an onset of direct Urca neutrino emission process involving nucleons, that is forbidden when $\Delta$-resonances are neglected~\cite{Raduta_PLB_2021}. The results pertaining the $\Delta$-resonance
shown in the following section were obtained for the couplings~\cite{Lijj2018b}
\bea
R_{\Delta\omega} = 1.10,\quad  R_{\Delta\rho} = 1.00 \quad R_{\Delta\sigma} = 1.10;
\, 1.16;\, 1.23;
\eea
where the three values of $R_{\Delta\sigma} $ correspond to the $x_{\Delta N}\equiv V_{\Delta}/V_N = 1,\, 4/3,\, 5/3$ (in the given order).  An exception are the results shown in Subsec.~\ref{ssec:FT_high_densities}, where the values $R_{\Delta\omega} = R_{\Delta\rho} = R_{\Delta\sigma} = 1$ have been used.

\subsection{Hartree-Fock CDFs with density-dependent couplings}
\label{ssec:RHF}
Next, we proceed to the construction of CDFs which are based on the Hartree-Fock approximation for the baryon self-energies. The hyperonization problem in the framework of the Hartree-Fock theories were studied in the quark-meson coupling models~\cite{Massot_2012,Whittenbury_2014,Miyatsu_2015} and CDF models~\cite{Long2012,Lijj2018a}. As mentioned earlier, the Fock diagram for such 
theories allows the pion contribution to be explicitly included in the CDF. Thus, the long-range contribution of the tensor force to the observables can be tracked explicitly. 
We first rearrange the first two terms of the Lagrangian~(\ref{eq:Lagrangian}) by writing it as a sum of free baryonic ($b$) and mesonic ($m$) Lagrangians plus the interaction part
\bea\label{eq:Lagrangian_density_Fock}
\mathscr{L}=\mathscr{L}_b+\mathscr{L}_{m}+\mathscr{L}_{\text{int}}.
\eea
The first term is the Dirac Lagrangian of free relativistic point-like baryons 
with rest masses $m_b$. The second term $\mathscr{L}_{m}$
consists of contributions of meson included in the Lagrangians~\eqref{eq:meson_H} 
and~\eqref{eq:meson_Hstr} plus the contribution from the $\bm{\pi}$-meson isotriplet,
\bea
\mathscr{L}^{(\pi)}_m = 
\frac{1}{2}\partial^\mu\bm{\pi}\cdot\partial_\mu\bm{\pi}-\frac{1}{2}m_\pi^2\bm{\pi}^2.
\eea
The meson--baryon interaction is described by the third term of 
Eq.~\eqref{eq:Lagrangian_density_Fock}, which now have a more complex  
form,
\bea\label{eq:interaction_Lagrangian}
\mathscr{L}_{\rm{int}} &= & \sum_b \bar{\psi}_b
\Big(-g_{\sigma b}\sigma-g_{\sigma^\ast b}\sigma^\ast
  -g_{\omega b}\gamma^\mu\omega_\mu-g_{\phi b}\gamma^\mu\phi_\mu
  -g_{\rho b}\gamma^\mu\bm{\varrho}_\mu\cdot\bm{\tau}_b \nonumber \\
  &+& \frac{f_{\omega b}}{2m_b}
  \sigma^{\mu\nu}\partial_\mu\omega^\nu
     + \frac{f_{\phi b}}{2m_b}\sigma^{\mu\nu}\partial_\mu\phi^\nu
     + \frac{f_{\varrho b}}{2m_b}\sigma^{\mu\nu}\partial_\mu\bm{\varrho}^\nu\cdot\bm{\tau}_b 
      -\frac{f_{\pi b}}{m_\pi}\gamma_5\gamma^\mu\partial_\mu\bm{\pi}\cdot\bm{\tau}_b \Big)\psi_b,
\eea
where $\sigma^{\mu\nu} =\frac{i}{2}[\gamma^\mu, \gamma^\nu]$, $\bm{\tau}_b$ is the vector of isospin Pauli matrices, with $\tau_{3b}$ being its third component. 

In the Hartree-Fock approximation, the self-energy $\Sigma(k)$ is a sum of the direct and exchange terms. The direct (Hartree) contribution to the self-energy is given by (the presence of $\Delta$-resonances is ignored for the sake of simplicity)
\begin{subequations}\label{eq:Hartree_terms}
\bea
\Sigma_{S,b}^H & =& -\frac{1}{m^2_\sigma}g_{\sigma b}
                   \sum_{b'} g_{\sigma b'} \rho^s_{b'}
                   -\frac{1}{m^2_{\sigma^\ast}}g_{\sigma^\ast b}
                   \sum_{b'} g_{\sigma^\ast b'} \rho^s_{b'}, \\
\Sigma_{0,b}^H & =& +\frac{1}{m^2_\omega}g_{\omega b}
                   \sum_{b'} g_{\omega b'} \rho_{b'}
                   +\frac{1}{m^2_\phi}g_{\phi b}
                   \sum_{b'} g_{\phi b'} \rho_{b'}
                   +\frac{1}{m^2_\rho}g_{\rho b}
                   \sum_{b'} g_{\rho b'}\tau_{3b}\rho_{b'},
\eea
\end{subequations}
which are, in principle, identical to the expressions~\eqref{eq:Hartree_Baryon}, but we reformulate them in terms of the scalar and vector densities. 

To write down the exchange (Fock) terms we further ignore the retardation effects,
i.e., we drop the energy transfer between interacting baryons, which gives at most 
a few percent contribution to the self-energy~\cite{Serot:1984}. 
Then, the Fock terms are given by
%
\begin{subequations}
\label{eq:Fock_terms}
\bea
\label{eq:Fock_terms_a}
\Sigma^F_{S, b}(k,\tau) &= &
\frac{1}{(4\pi)^2k}\sum_{\varphi,b^\prime}\tau^2_\varphi
\int dk^\prime k^\prime \big[\hat{M}_{b^\prime}(k^\prime)B_\varphi(k,k^\prime)
+\frac{1}{2}\hat{K}_{b^\prime}(k^\prime)D_\varphi(k,k^\prime)\big] f^+(E^{k'}_{b'}), \\
\label{eq:Fock_terms_b}
\Sigma^F_{0, b}(k,\tau) &= &
\frac{1}{(4\pi)^2k}\sum_{\varphi,b^\prime}\tau^2_\varphi
\int dk^\prime k^\prime  A_\varphi(k,k^\prime) f^+(E^{k'}_{b'}) ,\\
\label{eq:Fock_terms_c}
\Sigma^F_{V, b}(k,\tau) &= &
\frac{1}{(4\pi)^2k}\sum_{\varphi,b^\prime}\tau^2_\varphi
\int dk^\prime k^\prime \big[\hat{K}_{b^\prime}(k^\prime)C_\varphi(k,k^\prime)
+\frac{1}{2}\hat{M}_{b^\prime}(k^\prime)D_\varphi(k,k^\prime)\big] f^+(E^{k'}_{b'}),
\eea
\end{subequations}
%
where the symbol $\varphi$ refers to the individual mesons, and $\tau^2_\varphi$ is 
the isospin factor at the meson--baryon vertex. The explicit expression for the 
functions $A_{\varphi}$, $B_{\varphi}$, $C_{\varphi}$ and $D_{\varphi}$ in 
Eqs.~\eqref{eq:Fock_terms_a}-\eqref{eq:Fock_terms_c} are given in 
Table~\ref{tab:Functions_in_self_energy}, where the following functions are used:
\begin{subequations}
\bea
\Theta_\varphi(k,k^\prime) & \equiv & \ln \frac{m^2_\varphi+(k+k^\prime)^2}{m^2_\varphi+(k-k^\prime)^2},
\label{eq:Functions_a}\\
[0.em]
\Phi_\varphi(k,k^\prime)   & \equiv & \frac{1}{4 kk^\prime}(k^2+k^{\prime2}+m^2_\varphi)\Theta_\varphi(k,k^\prime)-1,\\
[0.6em]
\Pi_\varphi(k,k^\prime)   & \equiv & (k^2+k^{\prime2}-m^2_\varphi/2)\Phi_\varphi(k,k^\prime)- kk^\prime\Theta_\varphi(k,k^\prime),\\
[0.6em]
\Lambda_\varphi(k,k^\prime)& \equiv & (k^2+k^{\prime2})\Phi_\varphi(k,k^\prime)- kk^\prime\Theta_\varphi(k,k^\prime),\\
[0.6em]
\Omega_\varphi(k,k^\prime)    & \equiv & k\Theta_\varphi(k,k^\prime)-2k^\prime\Phi_\varphi(k,k^\prime).
\label{eq:Functions_e}
\eea
\end{subequations}

\begin{table*}[tb]
\centering
\caption{Functions $A_\varphi$, $B_\varphi$, $C_\varphi$ and $D_\varphi$ in 
  Eqs.~\eqref{eq:Fock_terms_a}-\eqref{eq:Fock_terms_c} for non-strange mesons.
  For the strange meson $\sigma^\ast$ these functions coincide with those
  for $\sigma$ and for $\phi$ they coincide with those for $\omega$.  The index $i$ indicates
  the meson-baryon interaction channels $S$ (scalar), $V$ (vector), $T$ (tensor), $VT$ (vector-tensor) and $PV$ (pseudo-vector) }
\setlength{\tabcolsep}{19pt}
\label{tab:Functions_in_self_energy}
\vspace{0.2cm}
\centering
\begin{tabular}{ccccc}
\hline\hline
$\varphi_i$   &$A_\varphi$                    & $ B_\varphi$                    & $C_\varphi$                     & $D_\varphi$\\
\hline
$\sigma_S$   & $ g^2_{\sigma b}\Theta_\sigma$ & $  g^2_{\sigma b}\Theta_\sigma$ & $-2g^2_{\sigma b}\Phi_\sigma$ & -\\
[0.6em]
$\delta_S$   & $ g^2_{\delta b}\Theta_\delta$ & $  g^2_{\delta b}\Theta_\delta$ & $-2g^2_{\delta b}\Phi_\delta$ & -\\
[0.6em]
$\omega_V$   & $2g^2_{\omega b}\Theta_\omega$ & $-4g^2_{\omega b}\Theta_\omega$ & $-4g^2_{\omega b}\Phi_\omega$ & -\\
[0.6em]
$\omega_T$   & $-\left(\frac{f_{\omega b}}{2M}\right)^2m^2_\omega\Theta_\omega$&$-3\left(\frac{f_{\omega b}}{2M}\right)^2m^2_\omega\Theta_\omega$      &
              $4\left(\frac{f_{\omega b}}{2M}\right)^2m^2_\omega\Lambda_\omega$& -\\
[0.6em]
  $\omega_{VT}$&    -     &    -    &    -                           &$
                                                                       12\left(
                                                                       \frac{f_{\omega b} g_{\omega b}}{2M}
                                                                       \right)\Omega_\omega$ \\
[0.6em]
$\varrho_V  $   & $ 2g^2_{\rho b}\Theta_\rho$    & $-4g^2_{\rho b}\Theta_\rho $    & $-4g^2_{\rho b}\Phi_\rho$   & -\\
[0.6em]
$\varrho_T$     & $-\left(\frac{f_{\rho b}}{2M}\right)^2m^2_\rho\Theta_\rho$&$-3\left(\frac{f_{\rho b}}{2M}\right)^2m^2_\rho\Theta_\rho$ &
              $4\left(\frac{f_{\rho b}}{2M}\right)^2m^2_\rho\Lambda_\rho$& -\\
[0.6em]
$\varrho_{VT}$  &    -     &    -    &    -                           &$12\left(\frac{f_{\rho b} g_{\rho b}}{2M}\right)\Omega_\rho$      \\
[0.6em]
$\pi_{PV}$   & $-f^2_{\pi b}\Theta_\pi$       & $-f^2_{\pi b}\Theta_\pi$        & $2\frac{f^2_{\pi b}}{m^2_\pi}\Pi_\pi$  & -\\
\hline\hline
\end{tabular}
\end{table*}
As in the case of the Hartree approximation, the density dependence of the meson-baryon couplings
requires taking into account the contribution of the rearrangement term $\Sigma_R$ to the self-energy component $\Sigma_0(k)$,
\begin{align}\label{eq:rearrangement}
\Sigma_R=
\sum_\varphi\frac{\partial g_{\varphi b}}{\partial\rho_b}
\sum_b\frac{1}{\pi^2}\int dk\, k^2 
\Big[\hat{M}_b(k)\Sigma^\varphi_{S,b}(k)
+\Sigma^\varphi_{0,b}(k)+\hat{K}_b(k)\Sigma^\varphi_{V,b}(k) \Big]
f^+(E^{k}_{b}).
\end{align}

Once the Fock self-energies are determined, the computation of the energy density 
and the pressure of hypernuclear matter can be carried out in full analogy to the
Hartree case. We note that the energy density can be expressed through the
self-energies which include the contributions from the Fock terms as
\begin{align}
{\cal E}_b = \frac{\gamma_b}{2\pi^2}\int dk\, k^2[T_b(k)+\frac{1}{2}V_b(k)]
f^+(E^{k'}_{b}),
\end{align}
where the kinetic and potential energies are given as follows, 
\begin{align}
T_b(k) = \hat{K}_b k_b + \hat{M}_b m_b, \qquad 
V_b(k) = \hat{M}_b\Sigma_{S,b}(k) + \hat{K}_b\Sigma_{V,b}(k) - \Sigma_{0,b}(k).
\end{align}
The pressure $P$ can be easily obtained from the thermodynamic relation~\eqref{eq:Thermodynamic_relation}. This completes the presentation of the CDF formalism in the Hartree-Fock approximation. 

The Hartree-Fock studies of hyperonic stars of Ref.~\cite{Lijj2018a}, which used the PKO1-3 and PKA1 parameterizations~\cite{Long2006PhLB,Long_2008,Long2010a} for the nucleonic sector, compared the Hartree and Hartree-Fock theories and their predictions for the properties of hypernuclear stars. This provides insight into the role played by the tensor force, pion exchange, and the spatial component of the vector self-energy which are explicit in the Hartree-Fock treatment. Firstly, it was found that the Fock terms generically make the hyperonic EoS softer. Secondly, the meson--hyperon tensor couplings provide additional attraction among hyperons in the Fock terms.   This softening of the EoS leads to maximum masses of hyperonic compact stars below the two-solar-mass limit, if the parameterization is based on hyperon couplings within the SU(6) spin-flavor quark model and the allowed range of the hyperon potentials. This, however, is no longer the case if the quark model symmetries are based on the SU(3) flavor symmetry~\cite{Lijj2018a}, in which case models of compact stars with masses above two solar masses are obtained, see Sec.~\ref{ssec:SU(3)}.

\subsection{Characteristics of nuclear matter close to saturation}
\label{ssec:Characteristics}

As is well known, the EoS of nuclear matter can be parameterized via an expansion in the vicinity of saturation density and isospin-symmetrical limit via a double-expansion in the Taylor series:
\begin{eqnarray}
\label{eq:Taylor_expansion}
  E(\chi, \delta) & \simeq &E(\chi) + E_{\rm sym}(\chi)  \delta^2 + O(\delta^4) \nonumber\\
  &\simeq &
  E_{\text{sat}} + \frac{1}{2!}K_{\text{sat}}\chi^2
  + \frac{1}{3!}Q_{\text{sat}}\chi^3 + \left(J_{\text{sym}} + L_{\text{sym}}\chi +\frac{1}{2!}K_{\text{sym}}\chi^2 + \frac{1}{3!}Q_{\text{sym}}\chi^3\right) \delta^2
 + {\mathcal O}(\chi^4),\nonumber\\
\end{eqnarray}
where $\chi=(\rho-\rho_{\text{sat}})/3\rho_{\text{sat}}$ and $\delta = (\rho_{n}-\rho_{p})/\rho$, with $\rho_n$ and $\rho_p$ being the neutron and proton densities. The coefficients of the expansion of the symmetric nuclear matter   are known as  the {\it binding
  energy} $E_{\rm sat}$, {\it incompressibility} $K_{\text{sat}}$, the {\it skewness} $Q_{\text{sat}}$;  the coefficients of density expansion of 
 $E_{\rm sym}(\chi)$ are the {\it symmetry energy} $J_{\text{sym}}$, the {\it slope and curvature parameters} $L_{\text{sym}}$
 and $K_{\text{sym}}$, and the {\it skewness of symmetry energy}  $Q_{\rm sym}$.
   These  are defined explicitly as 
\bea\label{eq:isoscalar_coeff}
K_{\text{sat}} =\left.9 \rho_{\text{sat}}^2 \frac{d^2 E(\chi)}{d \rho^2}\right|_{\rho=\rho_{\text{sat}}},\quad
Q_{\mathrm{sat}}=\left.27 \rho_{\text{sat}}^3 \frac{d^3 E(\chi)}{d \rho^3}\right|_{\rho=\rho_{\text{sat}}},
\eea
and
\bea\label{eq:isovector_coeff}
L_{\mathrm{sym}}  =\left.3 \rho_{\text{sat}} \frac{d E_{\mathrm{sym}}(\chi) }{d \rho}\right|_{\rho=\rho_{\text{sat}}}, \quad 
K_{\mathrm{sym}} =\left.9 \rho_{\text{sat}}^2 \frac{d^2 E_{\mathrm{sym}} (\chi)}{d \rho^2}\right|_{\rho=\rho_{\text{sat}}}, \quad  
Q_{\mathrm{sym}}=\left.27 \rho_{\text{sat}}^3 \frac{d^3 E_{\mathrm{sym}}(\chi)}{d \rho^3}\right|_{\rho=\rho_{\text{sat}}}.
\eea

Expansion \eqref{eq:Taylor_expansion} is rooted in the infinite matter limit of the Bethe-Weizs\"acker formula and, therefore, the low-order coefficients such as the binding energy per nucleon $E_{\text{sat}}\simeq -16.0\pm 1$ MeV and nuclear symmetry energy at the saturation density $J_{\text{sym}}\simeq 32.0\pm 2$~MeV as well as the value of the saturation density $\rho_{\rm sat}= 0.15-0.16$ fm$^{-3}$ are well constrained by experimental data on nuclei~\cite{Dutra2012,Dutra2014}.
The higher order coefficients of the density expansion both in the iso-scalar \eqref{eq:isoscalar_coeff} and iso-vector \eqref{eq:isovector_coeff} channels have been constrained via statistical Bayesian models, meta-modeling, and multimessenger astrophysics of compact stars~\cite{Li2019PhRvC,Zhang_2018,Zhangnb2019,Xie_2019,Choi_2021,Li:2021thg}. Generally, the validity of the expansion \eqref{eq:Taylor_expansion} with respect to density is limited by  condition $\chi < 1$, which translates into $\rho < 4\rho_{\text{sat}}$. The radius of convergence of this expansion depends on the parameterization of a particular CDF, but should be limited to the density range below a few $\rho_{\text{sat}}$.

The experimental data on giant dipole resonances allows one to deduce the value $K_{\rm sat}$. It has been suggested, that the density at which these oscillations probe the properties of matter is the so-called ``crossing density'' $\rho\approx 0.11$ fm$^{-3}$ at which the results for many models coincide. Currently, despite the progress that has been made over the years, still a broad range of value $200 \le K_{\rm sat}\le 300$~MeV is considered, see Refs.~\cite{Choi_2021,Piekarewicz_2004,Colo_2004,Khan_2012,Wangyj:2018,Grams_2022}.  Heavy-ion data provide constraints on the nuclear matter at intermediate densities that are relevant to compact stars, in particular, the KaoS experiment \cite{Miskowiec_1994} and flow data \cite{Danielewicz_2002} indicate a soft EoS with low values of $K_{\rm sat}$, albeit in a model dependent setting~\cite{Sagert_2012}. The skewness parameter is largely unknown and various authors suggest ranges within the range $-800 \le Q_{\rm sat} \le 400$~MeV~\cite{Tews2017,Xie_2019,Choi_2021}.  We emphasize that the information above applies to nucleonic matter and may not be valid for heavy-baryon matter, in particular,  the one found compact stars.
\begin{figure}[t]
\begin{center}
\includegraphics[width=10cm]{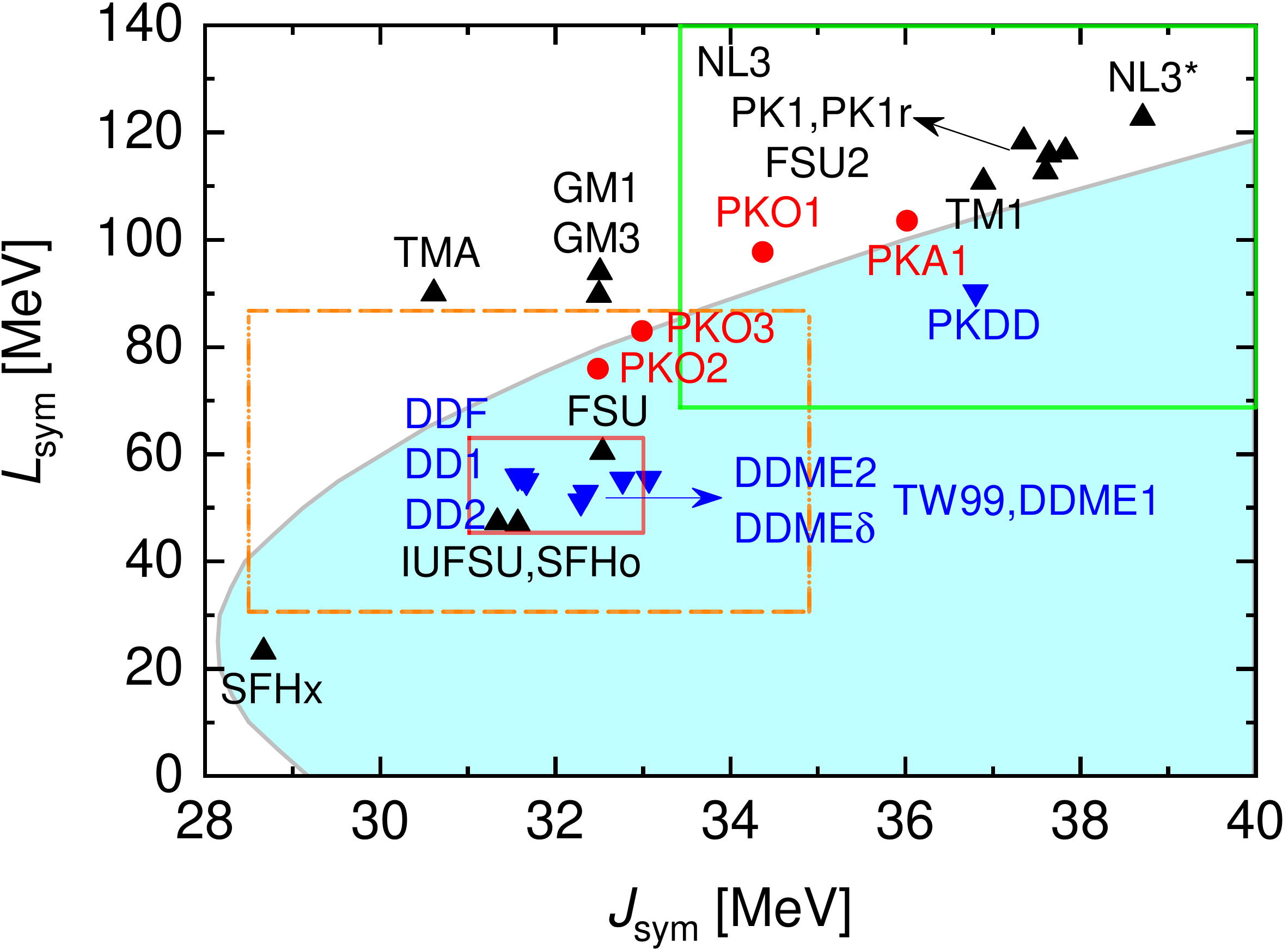} 
\end{center}
\caption{Predictions of different CDFs in the plane spanned by the symmetry energy $J_{\rm sym}$ and its slope $L_{\rm sym}$~\cite{Lijj2018a}. The orange rectangle shows the bounds on the most probable values of the symmetry energy $J_{\rm sym} = 31.7\pm 3.2$~MeV and the slope $L_{\rm sym} = 58.7\pm 28.1$~MeV obtained from the combined analysis of astrophysical constraints and terrestrial experiments~\cite{Oertel_RMP_2017}, the green and red rectangles are the ranges deduced from the analysis of the PREX-II experiment in
 Refs.~\cite{Reed_2021,Reinhard_2021}, respectively. The shaded region is the one allowed by the unitary gas bounds~\cite{Tews2017}. }
\label{fig:Esym-Lsym}
\end{figure}

The parameters in the isovector channel are constrained via studies of isospin asymmetrical systems by means already pointed out for the scalar channel. The symmetry energy can be constrained via the heavy-ion collisions~\cite{Tsang_2012,Wangyj:2020} and isobaric analog states in nuclei~\cite{Danielewicz_2009,Danielewicz_2014}. By combining the ranges suggested by various measurements, it is possible to isolate  the most probable range of  $J_{\rm sym}$ and $L_{\rm sym}$. Examples range are $30.2<J_{\rm sym}<33.7\,$MeV and $35<L_{\rm sym}<70\,$MeV~\cite{Lattimer_2013} and $J_{\rm sym} \simeq (31.7 \pm 3.2)$~MeV and $L_{\rm sym}\simeq (58.7 \pm 28.1)$~MeV~\cite{Oertel_RMP_2017}.
Most notable are the recent determinations of the symmetry energy and its slope from the PREX~\cite{Reed_2021,Reinhard_2021,Adhikari_2021} and CREX~\cite{CREX:2022kgg,Reed_2022,Mondal_2022} experiments involving parity-violating electron scattering off nuclei and their analysis. 

Figure~\ref{fig:Esym-Lsym} shows the results for various density functionals and experimental constraints. It also includes the recent analysis of the results of the Lead Radius Experiment Collaboration (PREX-II) \cite{Reed_2021,Reinhard_2021}, which reported the most precise measurement yet of the neutron skin thickness of the lead nucleus $R^{208}_{\rm skin}= 0.283 \pm 0.071$~fm (mean and $1\sigma$ standard deviation) in a parity-violating electron scattering experiment~\cite{Adhikari_2021}.  The authors of Ref.~\cite{Reed_2021} found $J_{\rm sym} = 38.1 \pm 4.7$~MeV and $L_{\rm sym} = 106 \pm 37$~MeV from a family of meson-exchange CDFs. Reference~\cite{Reinhard_2021} expanded the base (and the functional form) of employed density functionals to include non-relativistic Skyrme density functionals, CDFs with density-dependent meson-exchange couplings, and relativistic point coupling density functionals.  They found $J_{\rm sym} = 32 \pm 1$ MeV and $L_{\rm sym} = 54 \pm 8$ MeV. These values include the additional requirement on density functionals to be consistent with the experimental limits on the dipole polarizability of $^{208}$Pb, which prefer density functionals predicting a small value of $L_{\rm sym}$. The large central value and the range of $L_{\rm sym}$ found in the first analysis is in potential tension with the estimates of Refs.~\cite{Lattimer_2013,Danielewicz_2014,Oertel_RMP_2017,BaldoBurgio_2016}, whereas those obtained from the second analysis are within the range inferred previously. Note that the difference between the quoted two values of $L_{\rm sym}$ is $1.37\sigma$ which translates to about $83\%$ significance. Since $L_{\rm sym}$ is highly correlated with the radius of a compact star and its tidal deformability,  the rather large value found in the first analysis is in potential tension with the value of the deformability deduced from the analysis of GW170817 event if one assumes a purely nucleonic core composition~\cite{Reed_2021}.

While the CDFs provide full access to the microscopic information on matter, e.g., effective masses and chemical
potentials, an expansion of the type~\eqref{eq:Taylor_expansion} provides access to only a limited set of physical parameters. However, the uncertainties in the nuclear matter properties are easily characterized in terms of uncertainties in the characteristics entering the expansion \eqref{eq:Taylor_expansion}, therefore it is important to establish a one-to-one correspondence between the two descriptions.
\begin{table*}[tb]
 \caption{
The coefficients of the expansion \eqref{eq:Taylor_expansion} for the Hartree and Hartree-Fock CDFs. Also shown are the
Dirac mass $m^\ast_N$ in units of nucleon vacuum mass as well as the nonrelativistic Landau effective mass $m^\ast_{L}$. }
\setlength{\tabcolsep}{8.0pt}
\label{tab:NMP}
\vspace{0.2cm}
\begin{tabular}{cccccccccc}
\hline\hline
Method  &Parameter &$\rho_{\rm sat}$ & $E_{\rm sat}$ & $K_{\rm sat}$ & $Q_{\rm sat}$ & $J_{\rm sym}$ & $L_{\rm sym}$& $m^\ast_N$ & $m^\ast_{L}$  \\
 & set  & (fm$^{-3})$ & (MeV) &(MeV) &(MeV) &(MeV) &(MeV) &   \\ \hline
Hartree     & DD-ME2  & 0.152 & $-16.14$ &  251.15 &  479   &  32.31 &  51.27 & 0.57 & 0.65     \\
            & GM1     & 0.153 & $-16.33$ &  300.22 & $-222$ &  32.51 &  93.96 & 0.70 & 0.77    \\
\\
Hartree-Fock& PKA1  & 0.160 & $-15.83$ &  229.96 & 950  &  36.02 & 103.50 & 0.55 & 0.68      \\
            & PKO1  & 0.152 & $-16.00$ &  250.28 & 355  &  34.37 &  97.71 & 0.59 & 0.75      \\
            & PKO3  & 0.153 & $-16.04$ &  262.44 & 622  &  32.99 &  82.99 & 0.59 & 0.74      \\
\hline\hline
\end{tabular}
\end{table*}
Five of the macroscopic characteristics in  the expansion \eqref{eq:Taylor_expansion} which appear in the leading order  terms, namely $E_{\rm sat}$, $K_{\rm sat}$, $Q_{\rm sat}$ in the isoscalar channel and $J_{\rm sym}$, $L_{\rm sym} $ in the isovector channel, 
together with the preassigned values of the saturation density $\rho_{\rm sat}$ and the nucleon Dirac mass $m^\ast_N$ uniquely determine the seven adjustable parameters of the CDFs of the DDME2 type. This allows one to generate new parameterizations which reproduce desired values of the characteristics~\cite{Li2019PhRvC}, especially those that are associated with high-density and large-isospin behavior. As finite nuclei do not probe these regimes, the loss of accuracy of the new CDFs in reproducing the binding energies, charge radii, and neutron skins of finite nuclei is marginal. 

Having such a tool at hand, one can now vary the individual properties within their acceptable ranges and extract the EoS of dense matter and thus the properties of compact stars. This allows one to explore the correlation(s) between specific properties of nuclear matter and/or compact stars~\cite{Li2019PhRvC,Li:2021thg,Cai:2020hkk,Li:2020ass}. Detailed investigations of the EoS were carried out where the higher-order coefficients of the expansion, specifically, $Q_{\text{sat}}$ and $L_{\text{sym}}$ were varied since their values are weakly constrained by the conventional fitting protocol of CDFs~\cite{Margueron2018a,Margueron2018b,Margueron2019,Zhang_2018,Li2019PhRvC}.  It is interesting that the one-to-one mapping described above allows one to {\it predict} the higher-order terms in the expansion \eqref{eq:Taylor_expansion}~\cite{Li2019PhRvC}, which are highly model dependent~\cite{Dutra2012,Dutra2014}. Table~\ref{tab:NMP} shows the low-order characteristics of nuclear matter for five Hartree and Hartree-Fock parameterizations.


\subsection{Alternative approaches}
\label{ssec:Micro}
The microscopic methods use the experimental information from scattering experiments and binding energies of bound states with few nucleons ($N$) as well as those containing additional hyperons ($Y$). Once the $NN$, $NY$, and $YY$ interactions are fitted to the measured phase shifts and the binding energies of systems with few baryons, a potential is obtained on which a many-baryon theory can be built. While $NN$ potentials are based on comprehensive data on the scattering and bound states, this is not the case for potentials involving  single or multiple hyperons. 

The microscopic approaches are mainly based on lattice studies~\cite{Nemura:2007cj,Inoue2019AIPC,Sasaki:2019qnh}, many-particle theories employing the Br\"uckner scattering matrix~\cite{Baldo_2000,Yamamoto:2014jga,Bombaci:2016xzl,Haidenbauer2017EPJA}, Monte Carlo simulations~\cite{Lonardoni:2012rn,Lonardoni:2013rm,Lonardoni:2014bwa,Gandolfi:2015rvc} and variational methods~\cite{Togashi:2016fky,Shahrbaf:2019wex,Shahrbaf:2020uau}, see also the reviews~\cite{Chatterjee:2015pua,Vidana:2018bdi,Blaschke:2018mqw,Providencia2019,Burgio2021PrPNP}.  A key shortcoming of the potential-based approaches is the softness of the EoS derived from the most advanced $YN$ and $YY$ potentials, e.g., the Nijmegen potentials~\cite{Rijken1989}. Consequently, the maximum masses of hyperonic stars are below the $2M_{\odot}$ lower bound on the maximum mass of a compact star. This is clearly seen in the Br\"uckner calculations of cold hypernuclear stars, which yield maximum masses around $1.4M_{\odot}$~\cite{Baldo_2000}.  The tension between the observations of two-solar mass pulsars and the low maximum masses of hypernuclear stars obtain in various theories, notably microscopic ones based on potentials, led to the "hyperon puzzle", mentioned in Sec.~\ref{sec:Intro}. The hyperon puzzle can be addressed in microscopic theories by invoking three-body or, more generally, multi-body forces involving hyperons. Some recent studies have found them to be sufficient to produce heavy enough compact stars. For example, Ref.~\cite{Yamamoto:2014jga} introduces multi-pomeron exchange potential for the universal many-body repulsion in the baryonic matter in the spirit of the soft-core hyperon-nucleon Nijmegen potentials.  Reference~\cite{Haidenbauer2017EPJA} derived a $NN\Lambda$ force based on chiral effective field theory to sufficiently high order and implemented them in the Br\"uckner calculations of hypernuclear matter. Density-dependent $\Lambda$-nucleus optical potentials have been constructed which include a three-body repulsive $NN\Lambda$ force, which is repulsive at $\rho_{\rm sat}$ and higher densities thus allowing for massive compact stars~\cite{Friedman:2022bpw}.  Finally, we recall that the hyperon puzzle has also been extensively studied in the CDF approaches and has been solved by using parameterizations with strong vector repulsion, see Sec.~\ref{sec:Intro}. As we will discuss below, there is also a universal limit $2.4-2.5 M_{\odot}$ on the maximum mass of a hyperon-rich matter star in these theories, see Secs ~\ref{ssec:SU(6)} and ~\ref{ssec:SU(3)}.

\section{Static Hypernuclear Stars}
\label{sec:HNS_properties}

\subsection{EoS and composition}
\label{ssec:EoS_comp}
Given a CDF, one computes the EoS of the stellar matter by implementing the additional conditions of weak equilibrium and electric change neutrality that prevail in compact star matter. The thermodynamic conditions in the first moments after the birth of a neutron star are different from those of cold neutron stars and will be discussed later in Sec.~\ref{sec:FiniteT}. The strangeness-changing weak equilibrium conditions are given by
\begin{subequations}
\begin{eqnarray}
\label{eq:c1}
&&  \mu_{\Lambda}=\mu_{\Sigma^0}=\mu_{\Xi^0}=\mu_{\Delta^0}=\mu_n=\mu_b,\\
\label{eq:c2}
&&  \mu_{\Sigma^-}=\mu_{\Xi^-}=\mu_{\Delta^-}=\mu_b-\mu_Q,\\
\label{eq:c3}
&&  \mu_{\Sigma^+}=\mu_{\Delta^+}=\mu_b+\mu_Q,\\
\label{eq:c4}
&&  \mu_{\Delta^{++}}=\mu_b+2\mu_Q,
\end{eqnarray}
\end{subequations}
where the quantities $\mu_b$ and $\mu_Q=\mu_p-\mu_n$ are the baryon and charge chemical potentials, and $\mu_i$ with $i \in \{\Lambda, \Sigma^{0,\pm}, \Xi^{0,\pm}, \Delta^{0,\pm,++}\}$ are the chemical potentials of hyperons and $\Delta$-resonances.  The baryonic charge is given by
\begin{eqnarray}
\rho_p+\rho_{\Sigma^+}+2\rho_{\Delta^{++}}
+\rho_{\Delta^{+}}-(\rho_{\Sigma^-}+\rho_{\Xi^-}+\rho_{\Delta^-})=\rho_Q,
\label{eq:nQcharge}
\end{eqnarray}
where $\rho_i$ denotes the number density of baryon $i$.
The condition of electric charge neutrality is then satisfied by equating $\rho_Q$ of \eqref{eq:nQcharge} to the total lepton charge
\begin{equation}
\label{eq:charge_neutraility}
\rho_Q = \rho_e + \rho_\mu,
\end{equation}
where $\rho_{e}$ and $\rho_{\mu}$ are the number densities of electrons and muons,
respectively.
\begin{figure}[t]
\begin{center}
\includegraphics[width=10.5cm]{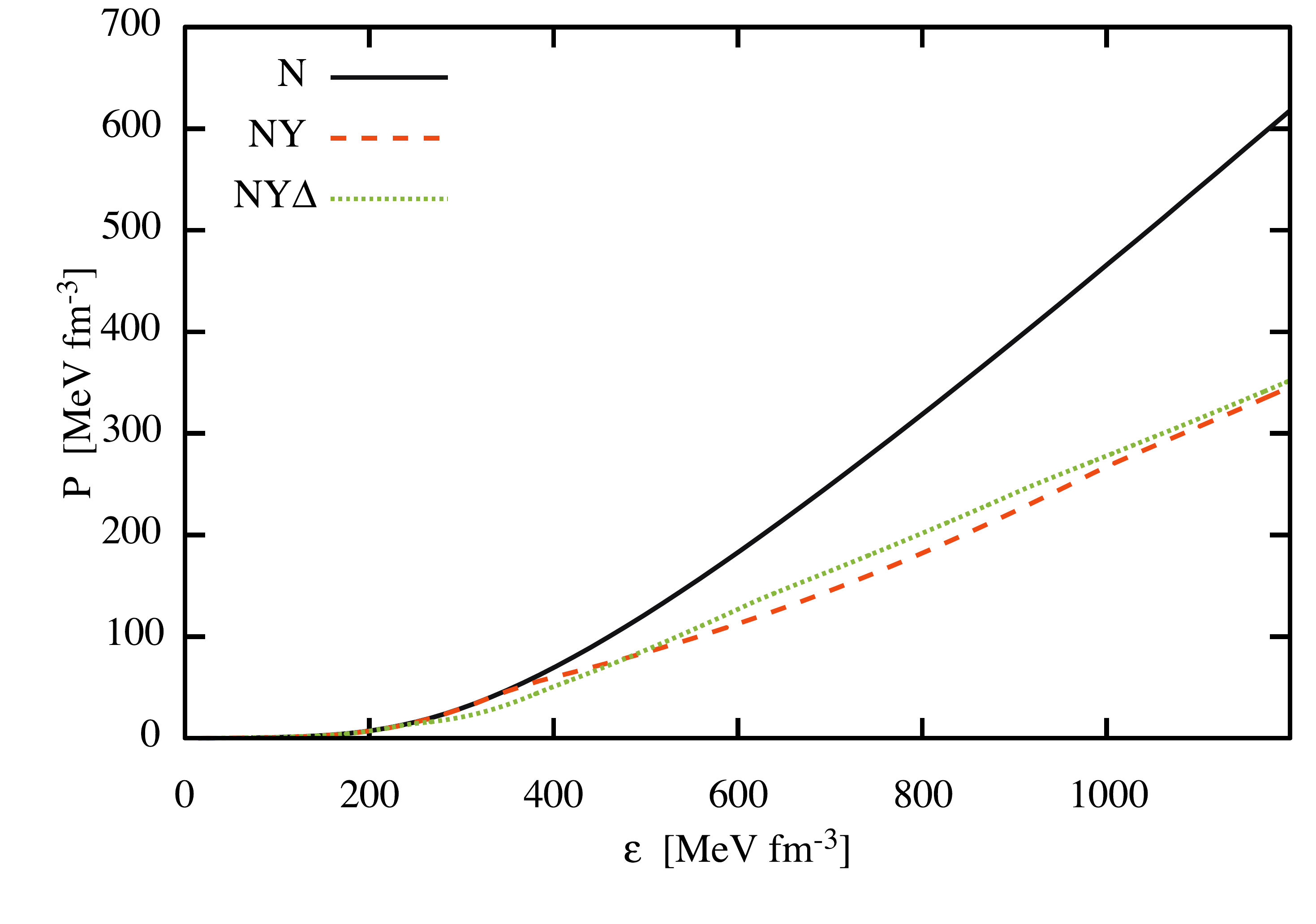}
\end{center}
\vspace{-0.5cm}
\caption{
  The EoS of nucleonic ($N$), hyperonic ($NY$) and $\Delta$-admixed hyperonic ($NY\Delta$) matter at zero temperature and in $\beta$-equilibrium.  The $\Delta$-potential is fixed by $V_\Delta(\rho_{\rm sat}) = V_N(\rho_{\rm sat})$~\cite{Sedrakian2020}.}
\label{fig:1.4}
\end{figure}
\begin{figure}[!]
\begin{center}
\includegraphics[width=10.5cm]{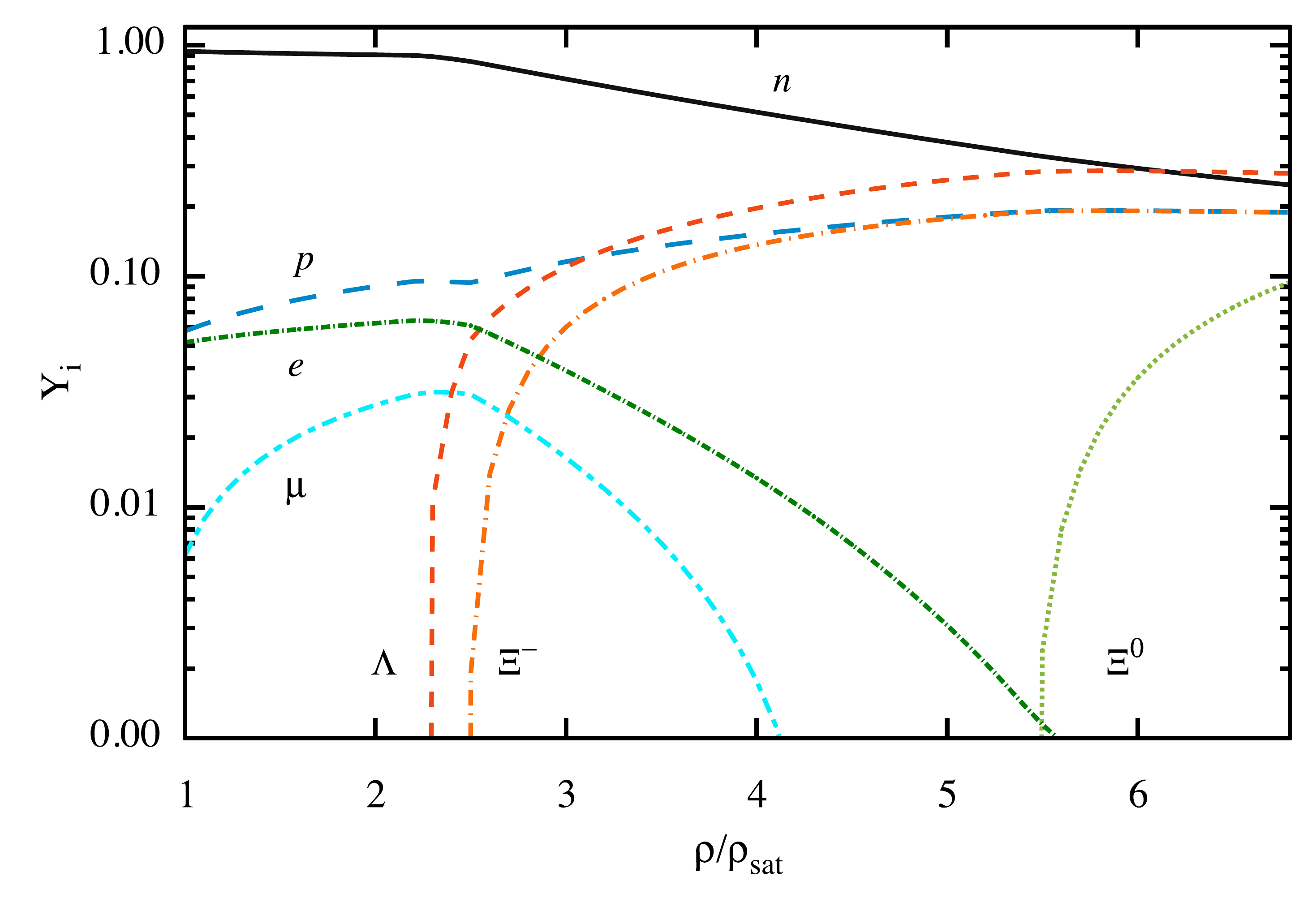}
\vspace{-0.5cm}
 \end{center}
 \caption{Particle fractions $Y_i = \rho_i/\rho$ as a function of
 density  where $\rho_i$ is the number density of particle species $i$ and $\rho$ is the total number density, in zero-temperature $\beta$-equilibrium hypernuclear matter 
  according to the DDME2 model~\cite{Sedrakian2020}. 
  }
\label{fig:1.5}
\end{figure}
Figure~\ref{fig:1.4} shows the EoS for nucleonic ($N$), hyperonic ($NY$) and $\Delta$-admixed hyperonic ($NY\Delta$) matter at zero temperature and in (weak) $\beta$-equilibrium derived on the basis of Hartree CDF with density-dependent couplings~\cite{Sedrakian2020}. The appearance of hyperons and $\Delta$-resonances can be seen by the change in the slope of the pressure above the saturation density and by the softening of the EoS at their onset. In the case where $\Delta$'s are included in the composition, the EoS softens at low and stiffens at high densities compared to the purely hyperonic case.  To compute stellar configurations the EoS needs to be supplemented by a model of the low-density matter. High-precision calculations would required to add a crustal EoS which is based on the same CDF as used for the core region. We do not use such a ``unified'' EoS, see~Ref.~\cite{Fortin_PRC2016}, rather we smoothly interpolate the high-density EoS to the EoS of inner crust from Ref.~\cite{Baym1971b} at the density of $\rho_{\rm sat}/2$. The EoS of inner crust is then matched to the EoS of outer crust of Ref.~\cite{Baym1971a} at the neutron drip density $4\times 10^{11}$~g~cm$^{-3}$.  
\begin{figure}[t]
\begin{center}
\includegraphics[width=10.5cm]{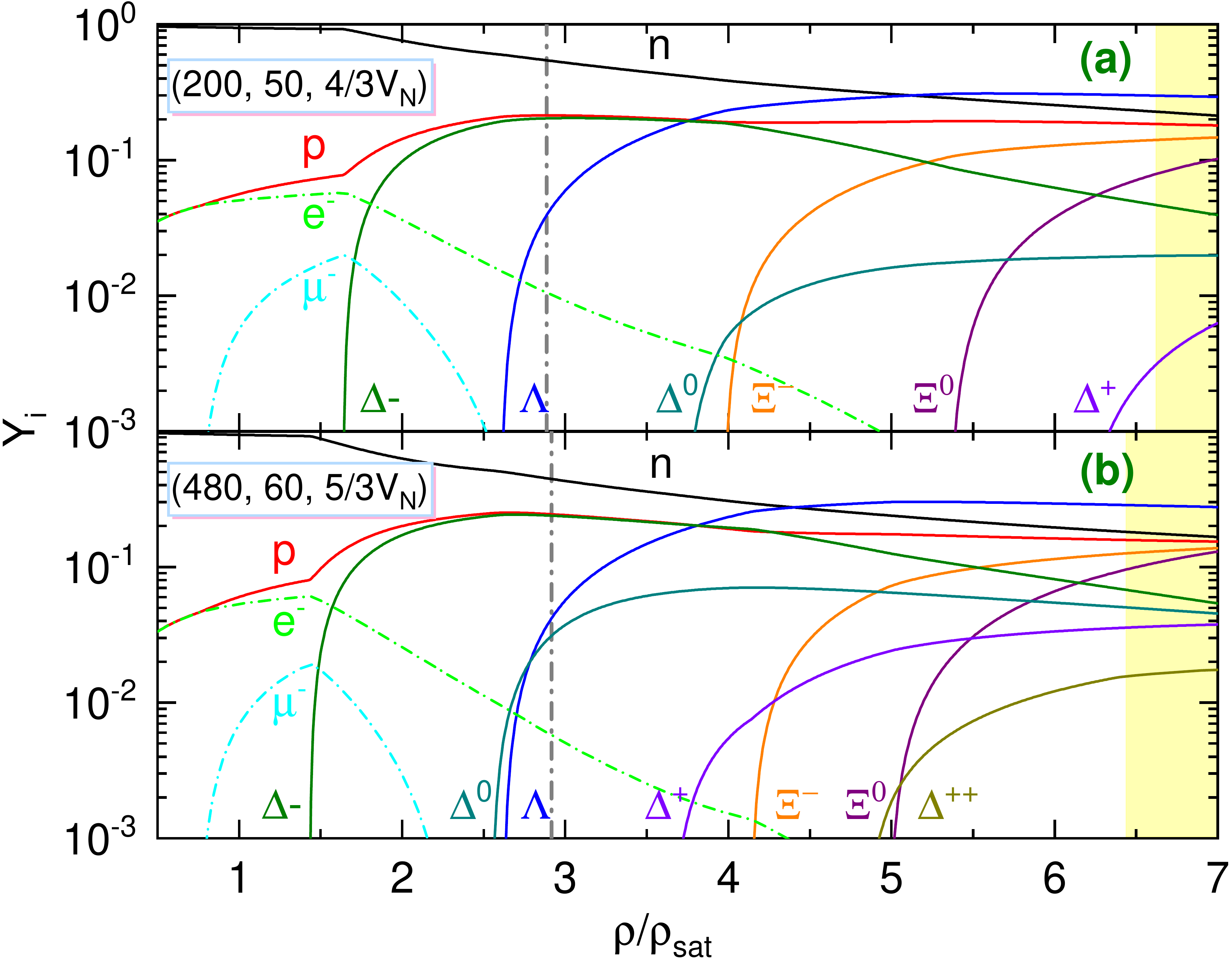} 
\end{center}
\vspace{-0.5cm}
\caption{
Particle fractions in zero-temperature  $NY\Delta$ stellar matter in the parameter space
($Q_{\text{sat}}$, $L_{\text{sym}}$, $V_\Delta$) (in MeV) and for $x_{\Delta N}= 4/3$ (panel a) and
$x_{\Delta N}= 5/3$ (panel b). The corresponding EoS  supports a
$1.4 M_\odot$ neutron star with a radius of about 12.5 km~\cite{Li2019PhRvC}. 
The thick vertical lines indicate the central density of the respective canonical 
$1.4 M_\odot$ neutron star, and the yellow shadings show densities greater than 
those reached in the maximum mass configurations.
}
\label{fig:1.6_b}
\end{figure}

Figure~\ref{fig:1.5} shows the particle abundances of hyperonic matter for the EoS shown in Fig.~\ref{fig:1.4}.  The first hyperon to appear in nucleonic matter as the density is increased is the $\Lambda$ hyperon, which is followed by the $\Xi^-$ hyperon.  The $\Sigma^-$ hyperons do not appear because they are disfavored by their repulsive potential at nuclear saturation density~\cite{Gomes:2014aka,LopesPhysRevC2014,Maslov:2015wba,Miyatsu_2015,BartPhysRevLett,DOVER1984171}. A similar arrangement of the hyperon thresholds has been found in other hypernuclear CDFs~\cite{Weissenborn2012a,Fortin2017,Lijj2018a,Lijj2018b,Weissenborn2012b}. We also note that this picture is in strong contrast to the case of a free hyperonic gas, where the $\Sigma^-$ is the first hyperon to nucleate~\cite{Ambartsumyan1960SvA}.

The modification of the particle abundances when $\Delta$-resonances are included in the composition are shown in Fig.~\ref{fig:1.6_b}~\cite{Lijj2018b}.  One observes that for a sufficiently large by absolute value $\Delta$ potential $\vert V_\Delta\vert >\vert V_N\vert$, the $\Delta$ threshold density is much lower than the one for the $\Lambda$ hyperon.  The larger $\vert V_\Delta\vert $ is, the lower $\Delta$'s threshold density is. The first $\Delta$-resonance to appear is the $\Delta^-$ because of its negative charge, as it compensates for the negatively charged (energetic) leptons. It shifts the threshold for the $\Xi^-$ to higher densities and effectively eliminates the $\Sigma^-$ (if present at all). The $\Lambda$ hyperon abundance is only weakly affected by the $\Delta$'s.  For $\vert V_\Delta\vert \geq \vert V_N\vert $ -- the case shown in Fig.~\ref{fig:1.6_b} -- the remaining $\Delta^{0,+,++}$ resonances also nucleate.  Due to the electric charge neutrality between the baryons and leptons, the appearance of negatively charged $\Delta^-$ and $\Xi^-$ depletes the electron-muon population. For a sufficiently large ($\vert V_\Delta \vert\geq \vert V_N\vert $) potential $\Delta$'s appear already in $1.4\,M_{\odot}$ compact stars.

\begin{figure}[t]
 \begin{center}
\includegraphics[width=13.cm,keepaspectratio]{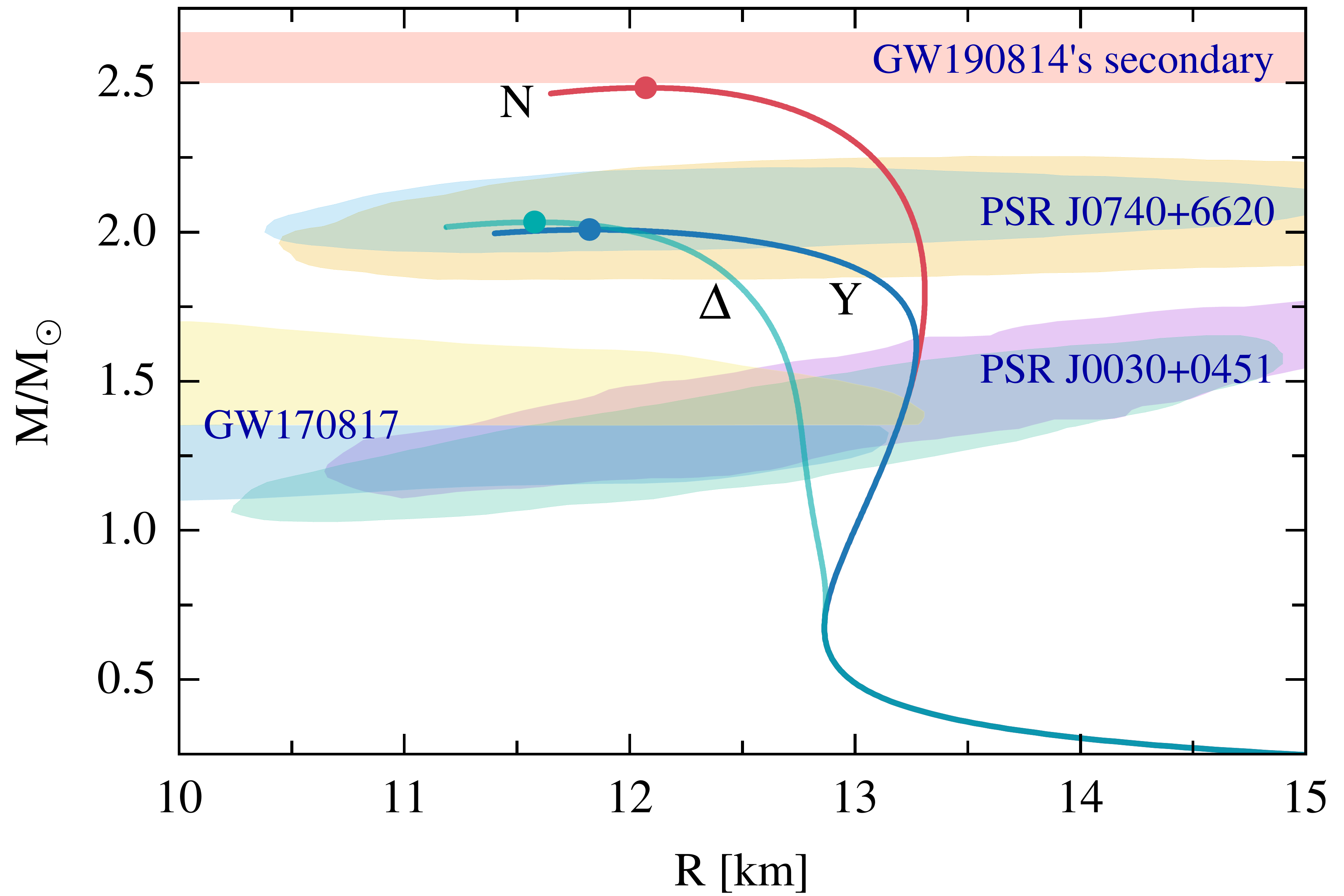}
\end{center}
\vspace{-0.5cm}
\caption{MR relations for cold  EoS  with $N$, $NY$, and $NY\Delta$ 
  compositions for the DDME2 nucleonic models and
  $x_{\Delta N} =4/3$~\cite{Sedrakian2020}. The ellipses show 90\% CI regions for pulsars PSR J0030+0451~\cite{Miller2019ApJ,Riley2019ApJ}, PSR
  J0740+6620~\cite{Miller2021,Riley:2021pdl} and the gravitational-wave event GW170817~\cite{Abbott_PRX_2019} as labeled. The mass range extracted from the GW190814 event is shown as well~\cite{Abbott:2020khf}. The maximum-mass star
  of each stellar sequence is marked by a solid dot.}
\label{fig:1.6}
\end{figure}

\subsection{Global properties of static stars}
\label{ssec:TOV}
The static properties of compact stars are obtained in spherical symmetry (assuming no rotation and no significant magnetic fields) from the integration of the Tolman-Oppenheimer-Volkoff (TOV) equations~\cite{Tolman1939,Oppenheimer1939}, which are the solution of Einstein's equation for a spherically symmetric mass distribution.  It is often useful to compare theoretical predictions with observations on diagrams containing only observable quantities, i.e., combinations of mass, radius, the moment of inertia, spin frequency, etc. As an example, we show in Fig.~\ref{fig:1.6} the mass-radius (MR) relations for the EoS plotted in Fig.~\ref{fig:1.4} for purely nucleonic, hyperonic, and hyperon-$\Delta$ admixed EoS. 

The following observational constraints are included: (a) the 90\% CI regions of the NICER experiment on the mass {\it and} radius of PSR J0030+0451~\cite{Miller2019ApJ,Riley2019ApJ} and J0740+6620~\cite{Miller2021,Riley:2021pdl};  (b) the 90\% CI ellipses derived from the analysis of the gravitational-wave event GW170817~\cite{Abbott_PRX_2019};  (c) the mass range derived for the light companion of the binary observed in the GW190814 event~\cite{Abbott:2020khf}, detailed in Sec.~\ref{sec:Rapid_rotation}. The softening of the EoS when heavy baryons are allowed is reflected in the significant reduction of the maximum mass of compact stars when hyperons ($NY$) and hyperons and $\Delta$-resonances ($NY\Delta$) are present.  The additional softening of the EoS at intermediate densities in the case when $\Delta$-resonances are allowed shifts the radii of stellar configurations to smaller values. Since, in this case, the stiffness of the EoS at high densities is comparable to (and even asymptotically exceeds) that of pure hyperonic matter, the values of the maximum masses ($M_{\text{max}} \gtrsim 2.0\,M_\odot$) of stars made of pure hyperonic matter and $\Delta$-admixed hypernuclear matter are comparable with each other, as can be seen in Fig.~\ref{fig:1.6}.

\begin{figure}[t]
\begin{center}
\includegraphics[width=12.cm,keepaspectratio]{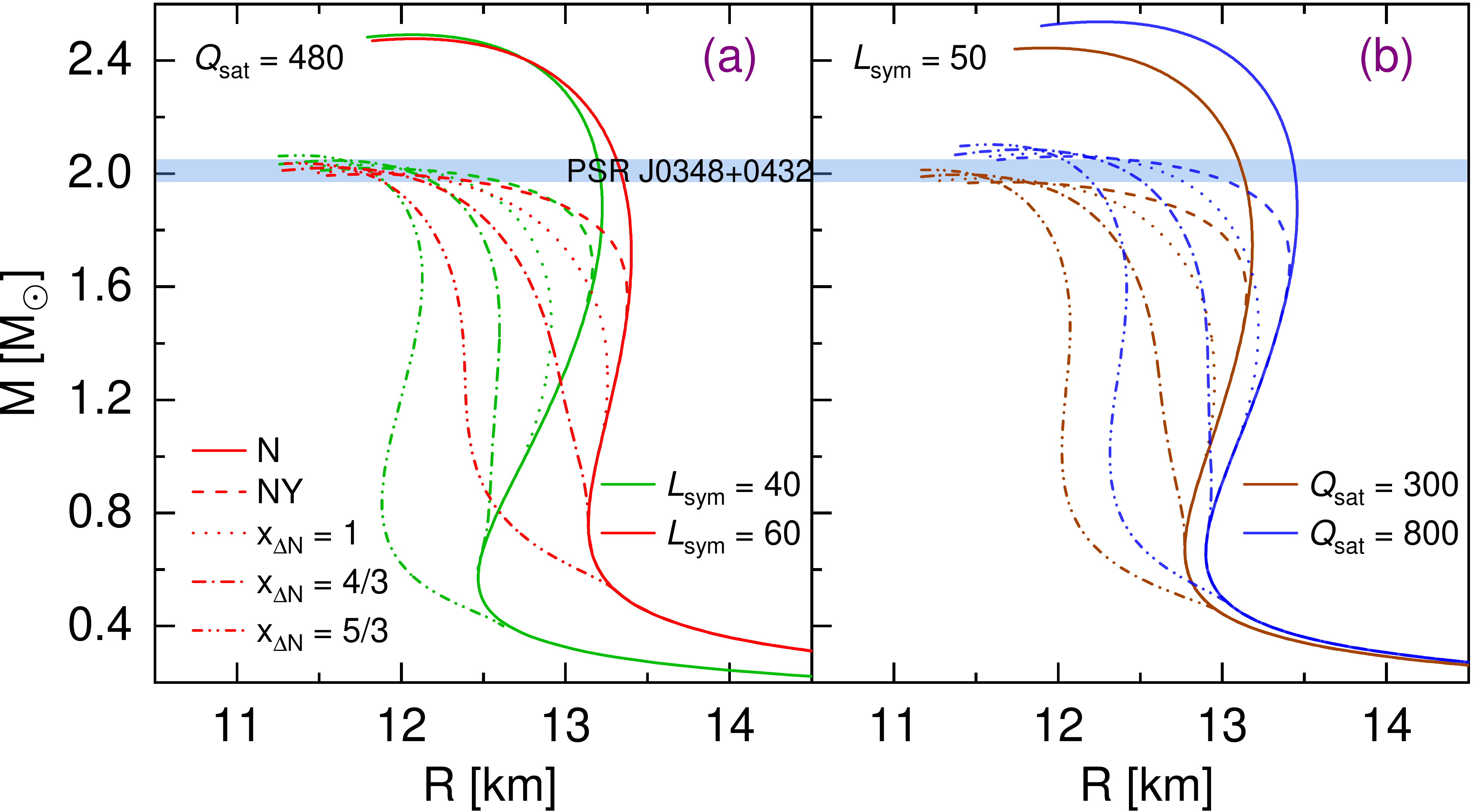}
\end{center}
\vspace{-0.5cm}
\caption{MR relations for a set of EoS with varying
  $L_{\text{sym}}$ (a) and $Q_{\text{sat}}$ (b) and assuming
  purely nucleonic ($N$), hyperonic ($NY$), and hyperon-$\Delta$
  admixed ($NY\Delta$) compositions of zero-temperature stellar matter~\cite{Lijj2019}. Three
  values of the $\Delta$-potential have been used:
  $x_{\Delta N}  =1$, 4/3, and 5/3. }
\label{fig:1.7}
\end{figure}
\begin{figure}[!]
\begin{center}
\includegraphics[width=9cm,
keepaspectratio]{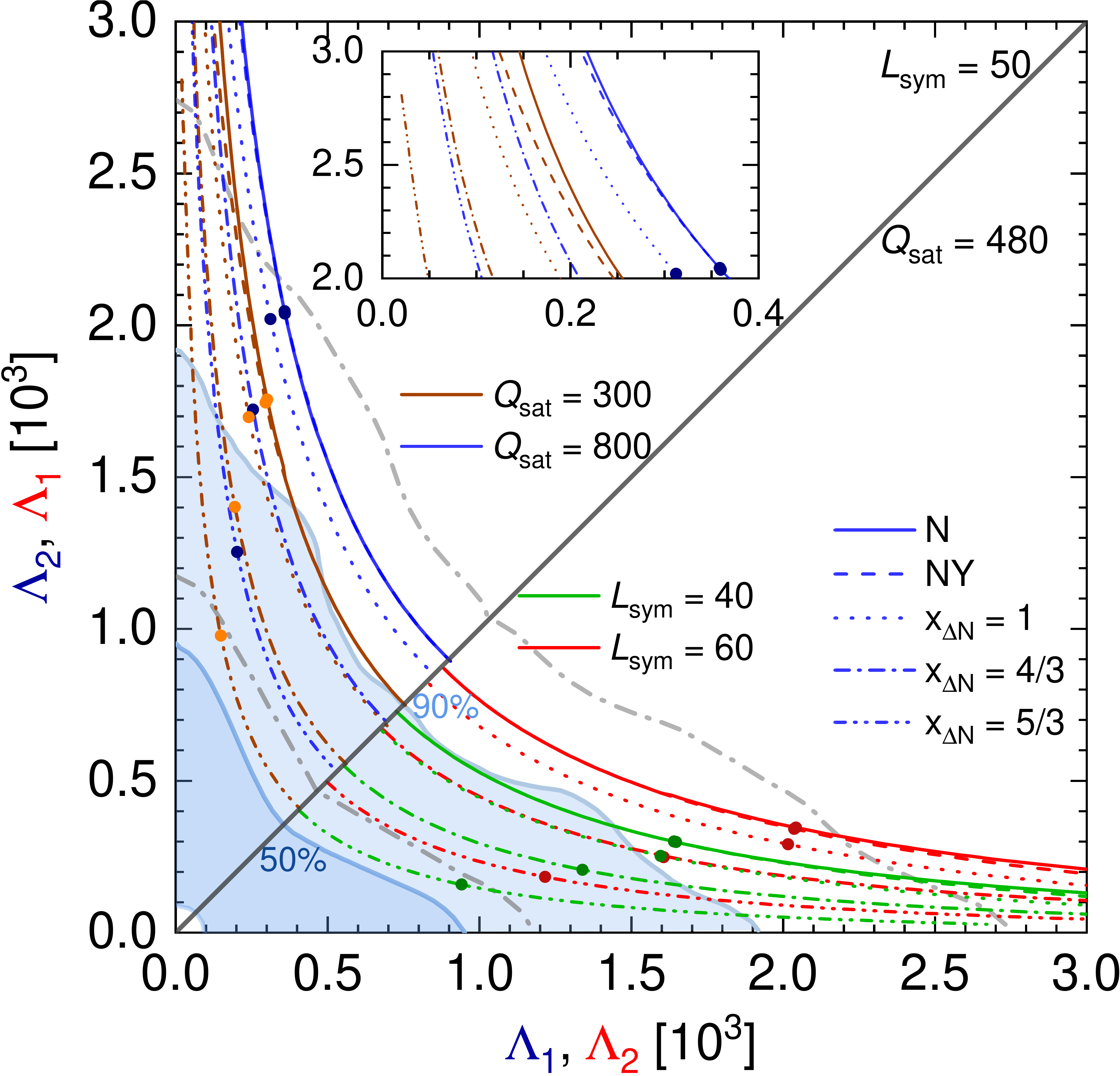}
\end{center}
\caption{Dimensionless tidal deformabilities extracted observationally
  from the GW170817 event (shaded areas) are compared with the
  predictions of EoS with various compositions and varying values of 
  $L_{\text{sym}}$, $Q_{\text{sat}}$, and $x_{N\Delta}$~\cite{Lijj2019} for zero-temperature stellar
  models with MR relation shown in Fig.~\ref{fig:1.7}.
  The light and heavy shadings show the 50\% and 90\% credibility
  regions, respectively \cite{Abbott_PRX_2019}. The results of the earlier
  analysis by the LIGO and Virgo
  Collaboration\cite{Abbott2017a} are shown by the gray dash-dotted curves.
  The dots correspond to the predictions for a BNS with a mass ratio of $q = 0.73$. }
\label{fig:1.8}
\end{figure}
To illustrate the effects of variations of the nucleonic properties on the mass and radius of a compact star, we show in Fig.~\ref{fig:1.7} the MR relations for the three cases of $N$-, $NY$-, and $NY\Delta$-matter, where the isoscalar $Q_{\text{sat}}$ and isovector $L_{\text{sym}}$ characteristics are varied while the remaining parameters in Eq.~\eqref{eq:Taylor_expansion} are kept fixed at their values corresponding to DDME2 parameterization (see Table~\ref{tab:NMP}). Note that the variation of $Q_{\text{sat}}$ is achieved by varying the three density-dependent parameters entering Eq.~\eqref{eq:h_functions}, therefore, these variations do not affect the meson--hyperon and meson-$\Delta$ couplings at the nuclear saturation density. It is seen that smaller values of $L_{\text{sym}}$ and/or $Q_{\text{sat}}$ imply a smaller stellar radius for any given mass. At the same time, a smaller value of $Q_{\text{sat}}$ predicts a smaller maximum mass, since the values given by Eq.~\eqref{eq:Taylor_expansion} are smaller for asymptotically high densities. The effects of varying the $\Delta$-meson couplings are shown by using the values for the ratio of $\Delta$-potential scaled by the nucleonic potential: $x_{\Delta N} = V_\Delta(\rho_{\text{sat}}) /V_N(\rho_{\text{sat}})=1, 4/3, 5/3$. It can be seen that the larger the value of $x_{\Delta N}$, the smaller the radius of the predicted stellar configuration, as expected from the discussion of Fig.~\ref{fig:1.6}. Moreover,
 the differences between the MR curves for $N$, $NY$- and $NY\Delta$-matter remain qualitatively intact, because the changes in the values of $L_{\text{sym}}$ and/or $Q_{\text{sat}}$ characteristics affect  the properties of nucleonic matter itself.  The differences between the two compositions (i.e., $NY$ vs. $NY\Delta$) in Fig.~\ref{fig:1.7} (e.g., in the radius of a canonical neutron star) arise from the factors already discussed in connection with Fig.~\ref{fig:1.6} and we do not repeat them here.  We conclude by noting that the parameter space used in Fig.~\ref{fig:1.7} implies stellar maximum masses larger than $2\,M_{\odot}$ and radii in the range of $12\lesssim R\lesssim 14$~km, independent of composition.

\subsection{Tidal deformabilities}
\label{ssec:TD}
Since the first observation of gravitational waves from a BNS merger -- the GW170817 event -- the tidal deformability of compact stars has become observationally accessible~\cite{Abbott2017a,Abbott_PRX_2019} and can thus be confronted with the theoretical predictions, as already mentioned in Sec.~\ref{sec:Constraints}.
Figure~\ref{fig:1.8} shows the tidal deformabilities $\Lambda_1$ and $\Lambda_2$ given by Eq.~\eqref{eq:Lambda}
for stars containing  $N$-, $NY$-, and $NY\Delta$-matter in their cores. Three different values of the $\Delta$-resonance potential in nuclear matter expressed as $x_{\Delta N} = V_\Delta/V_N$ are used. The diagonal line corresponds to the case $M_1 = M_2 = 1.362 M_\odot$. The light and dark shaded regions represent the 90\% and 50\% confidence limits extracted from the analysis of the GW170817 event~\cite{Abbott_PRX_2019}. 
We show the tidal deformabilities for selected values of  $Q_{\text{sat}}$ and $L_{\text{sym}}$ characteristics. It is seen that the  
 data favor low values of $Q_{\text{sat}} \lesssim 400$~MeV and $L_{\text{sym}} \lesssim 50$~MeV for purely nucleonic compact stars. Otherwise their tidal deformabilities are incompatible with the observation.
The inclusion of hyperons and $\Delta$'s in the CDF reduces the tidal deformability, as can be clearly seen from the inset in Fig.~\ref{fig:1.8}. The most significant reduction results from the inclusion of $\Delta$'s with a large attractive potential corresponding to large $x_{\Delta N}$ values. Therefore, the presence of $\Delta$-resonances enlarges the range of allowed 
$L_{\text{sym}}$ and $Q_{\text{sat}}$ and mitigates the potential tension between the theoretical predictions for purely nucleonic stars and observation of GW170817 event. Purely hyperonic models are compatible with the $90\%$ confidence limits for our selected values of $Q_{\text{sat}}$ and $L_{\text{sym}}$.  We note that the softening of the EoS at intermediate densities caused by the appearance of $\Delta$'s is similar in effect to a first-order phase transition to quark matter, which in turn gives rise to more compact configurations, see e.g. 
Refs.~\cite{Bonanno2012,Zdunik:2012dj,Alford:2017qgh,Alvarez-Castillo:2018pve,Otto2020EPJST,Fukushima:2020cmk,Kojo2020,Tan_PRL_2022,Li2023}. In conclusion, we find that the information obtained from MR and $\Lambda_1-\Lambda_2$ relations are complementary to each other. Indeed,
as demonstrated by our examples, different EoS models predicting different MR relationships could have $\Lambda_1-\Lambda_2$ relationships that are quite close. The opposite is also true in general: models which predict similar MR relations may have quite different tidal deformabilities, as exemplified by the hybrid star models see, e.g., Ref.~\cite{Li2023}.

\subsection{Kaon condensation}
\label{ssec:Kaons}
The possibility of condensation of mesons (pions, kaons, ${\bm \varrho}$  mesons) has been considered since Migdal's seminal paper on pion condensation~\cite{Migdal_1972}. Initial work focused on pion condensation and its implications for neutron star physics~\cite{Sawyer1973,Brown1976,Haensel1982} (for a recent review, see Ref.~\cite{Mannarelli2019}).  While the condensation of pions was due to the $P$-wave interaction and occurred at finite momentum (in contrast to the ordinary Bose-Einstein condensation), later it was discovered that $S$-wave interaction
driven condensation of (anti)kaon ($\bar K$) condensates was favored in compact stars for a range of potential values~\cite{Kaplan1988,Nelson1987,Brown1994,Lee1994,Knorren1995,Pons2000PRC,Pons2001ApJ,YueShen2008PRC}. The formation of kaon condensates
in hypernuclear matter was studied in a large number of works, see, for example,  Refs.~\cite{Schaffner1996,Prakash_1997,Glendenning1999,Banik2001,Menezes2005PRC,Malik2021,Thapa:2021kfo}. Whether or not (anti)kaons
appear in compact star matter depends sensitively on the $K^-$ optical potential in nuclear matter.
The presence of hyperons generally shifts the threshold of kaon condensation to higher densities, see, e.g., Ref.~\cite{Char2014}.  Also, for sufficiently attractive $K^-$ optical potential a first-order phase transition has been observed with the possibility of a formation of mixed phase~\cite{Schaffner1996,Pons2000PRC,Pons2001ApJ}. This feature is not observed in more recent models, which have been tuned to predict sufficiently stiff EoS that allows for stable two-solar mass compact stars to exist~\cite{Malik2021,Thapa:2021kfo}.
Condensation of charged ${\bm \varrho}$ mesons  has also been considered as a possible phenomenon that could occur in compact star matter~\cite{Voskresensky_1997,Aharony_2008,Kolomeitsev:2017gli}.

It is now well established that the appearance of these condensates softens the EoS of dense matter (albeit to varying degrees), leading to lower maximum masses of stars with meson condensates, than without them.
 The onset of (anti)kaon condensation may also affect the transport properties and hence thermal features
of compact stars such as the bulk viscosity~\cite{Chatterjee2007PhysRevD,Chatterjee_2008} and cooling~\cite{Kubis2003}.

Next, let us sketch the inclusion of kaon condensation in the models discussed above (see also Ref.~\cite{Thapa:2021kfo}).
The kaonic Lagrangian is given by 
\begin{equation} \label{eq:kaon_1} 
\begin{aligned}
  \mathscr{L}_{(\bar{K})}  & = 
  D^{(\bar{K})*}_\mu \bar{K} D^\mu_{(\bar{K})} K - m^{*^2}_K \bar{K} K , 
\end{aligned} \end{equation}
where $K$ is the kaonic field and $m^*_K$ denotes the (anti)kaon mass. The covariant derivative in Eq.\eqref{eq:kaon_1} reads
\begin{equation}\label{eqn:kaon_2}
D_{\mu (\bar{K})} = \partial_\mu + ig_{\omega \bar{K}} \omega_\mu + ig_{\rho \bar{K}} \boldsymbol{\tau}_{\bar{K}} \cdot \boldsymbol{\varrho}_{\mu} + ig_{\phi \bar{K}} \phi_\mu ,
\end{equation}
where the isovector nature of the $\bm\varrho$-meson field requires  
the presence of  $\boldsymbol{\tau}_{\bar{K}}$.
The Dirac effective (anti)kaon mass is given by
\begin{equation} \label{eqn:kaon_4}
m_{K}^*  = m_K - g_{\sigma K}\sigma - g_{\sigma^* K}\sigma^*.
\end{equation}
The changes ($\delta$) in the expectation values of the meson fields due to their coupling
to kaons are given by 
\begin{equation}\label{eqn:kaon_5}
\begin{aligned}
\delta\sigma & = + \sum_{\bar{K}} \frac{1}{m_{\sigma}^2} g_{\sigma K}\rho_{\bar{K}}^s,
\quad
\delta\sigma^* = + \sum_{\bar{K}} \frac{1}{m_{\sigma^*}^2} g_{\sigma^* K}\rho_{\bar{K}}^s, \\
\quad
\delta \omega_{0} & = - \sum_{\bar{K}} \frac{1}{m_{\omega}^2} g_{\omega K}\rho_{\bar{K}}, 
\quad
\delta \phi_{0} = - \sum_{\bar{K}} \frac{1}{m_{\phi}^2} g_{\phi K}\rho_{\bar{K}}, \\
\delta \rho_{03} & = \sum_{\bar{K}} \frac{1}{m_{\rho}^2} g_{\rho K} \boldsymbol{\tau}_{\bar{K}3}\rho_{\bar{K}} ,
\end{aligned}
\end{equation}
where in the case of $S$-wave (anti)kaons, the number density is given as
\begin{equation} \label{eqn:kaon_7}
\begin{aligned}
\rho_{K^-, \bar{K}^0} = 2 \left( \omega_{\bar{K}} + g_{\omega K} \omega_0 
 + g_{\phi K} \phi_0 \pm \frac{1}{2} g_{\rho K} \varrho_{03} \right) 
 = 2 m^*_K \bar{K} K.
\end{aligned}
\end{equation}
Here, $\omega_{\bar{K}}$ represents the in-medium energies of (anti)kaons  given by
\begin{equation} \label{eqn:kaon_8}
    \omega_{K^{-} , \bar{K}^0} = m^*_K - g_{\omega K} - g_{\phi K} \phi_0 \mp \frac{1}{2} g_{\rho K} \rho_{03} ,
  \end{equation}
upon assigning isospin projections of $\mp 1/2$ to $K^-$ (upper sign) and $\bar{K}^0$ (lower sign).  In the case of (anti)kaons, their equilibrium is governed by 
strangeness-changing processes such as
\bea
n\to p + \bar{K}^-, \qquad l^- \to  K^-+\nu_l, \qquad l=e,\mu ,
\eea
and their inverse reactions. These imply that the chemical potentials of nucleons and leptons are given by
\begin{equation} \label{eqn:kaon_12}
\begin{aligned}
    \mu_n - \mu_p = \mu_e =\mu_{\mu} =\omega_{\bar{K}^-}, \qquad \omega_{\bar{K}^0} = 0.
\end{aligned}
\end{equation}

The contribution of (anti)kaons to the  total energy density is then given by
\begin{equation} \label{eqn:kaon_14}
{\cal E}_{\bar{K}} = m^*_K (\rho_{K^-} + \rho_{\bar{K}^0}) ,
\end{equation}
where $\rho_{K^{-}}$ and $\rho_{\bar K^{0}}$  are their respective densities. The pressure of the condensate vanishes at zero temperature since the (anti)kaons then occupy states with zero momentum. In charge neutral stellar matter the neutrality condition \eqref{eq:charge_neutraility} must now include the negatively charged kaon condensate.
The final step is to fix the meson-(anti)kaon couplings (see Ref.~\cite{Char2014}). For vector mesons one finds
\begin{equation}\label{eq:kaon_vector_meson}
g_{\omega K} = \frac{1}{3} g_{\omega N}, \quad g_{\rho K} = g_{\rho N},
\end{equation}
and for hidden strangeness mesons 
$
g_{\sigma^* K} = 2.65$ and $g_{\phi K} = 4.27.
$
The scalar meson-(anti)kaon coupling can be extracted from the value of the real part of the $K^-$ optical potential at nuclear saturation density~\cite{Mannarelli2019,Waas1997,Friedman1999}. It is noted that the anti-kaon potential in the nuclear matter is attractive, but the kaon potential is repulsive~\cite{LI_1997,Pal2000}.
To study kaon condensation in hypernuclear matter with $\Delta$-resonances, the authors of Ref.~\cite{Thapa:2021kfo} chose for the $K^-$ optical potential values in the interval  $-150\leq U_{\bar{K}}\leq -120$ MeV, which led to a range of values for the coupling constant $0.4311\le g_{\sigma K}\le 1.2175$.
\begin{figure}[!]
\begin{center}
\includegraphics[width=10.5cm,keepaspectratio]{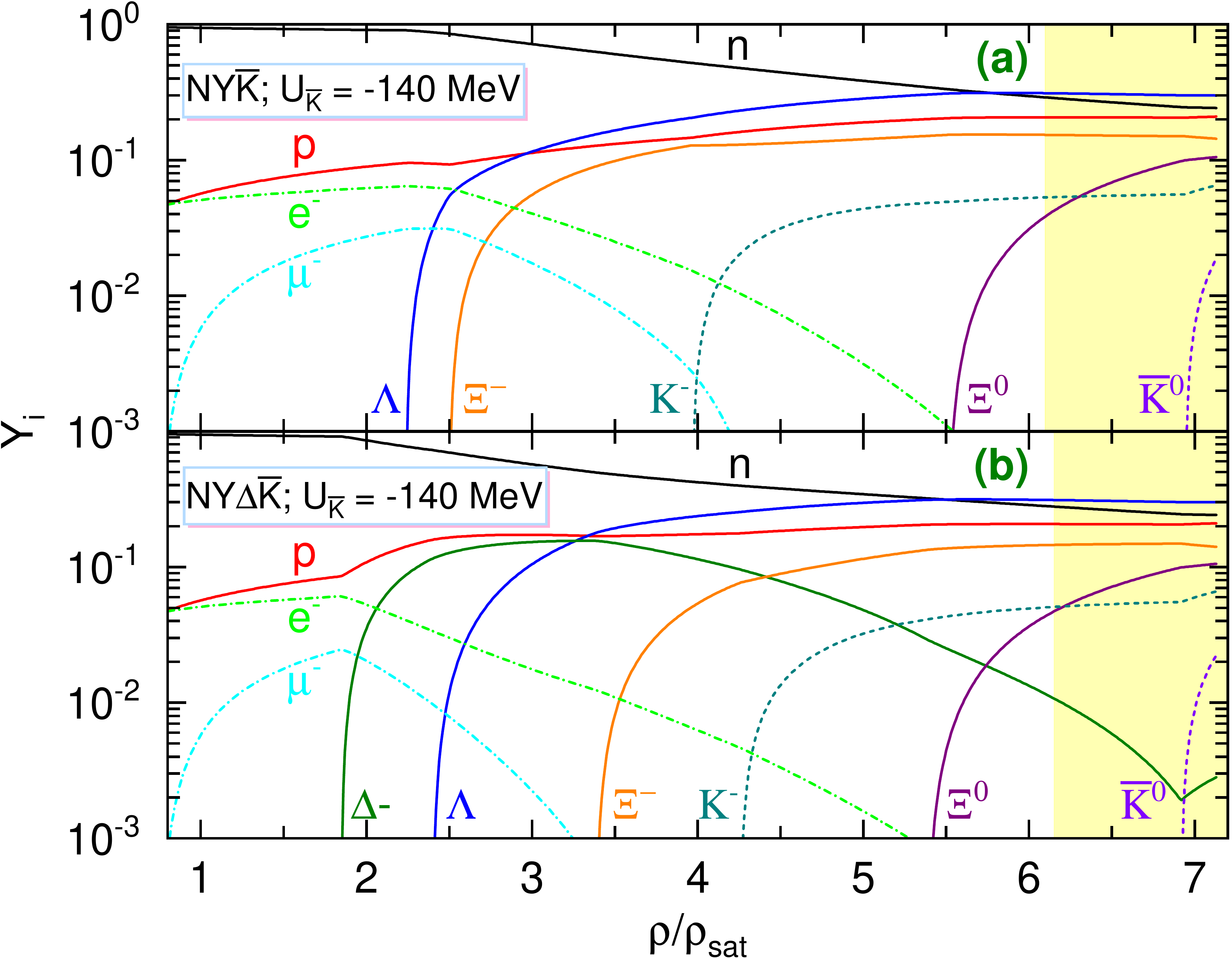}
\end{center}
\vspace{-0.5cm}
\caption{Particle fractions $Y_i$ as a function of  $\rho/\rho_{\rm sat}$
  in zero-temperature hypernuclear matter containing a kaon condensate,
  without (upper panel) and with (lower panel) $\Delta$-resonances for
  $U_{\bar{K}}= -140$ MeV,  $x_{\Delta N}=1$.  and the DDME2 parameterization
  of the nucleonic sector~\cite{Thapa:2021kfo}. The yellow shadings show densities greater
  than those reached in the maximum mass configurations.
  The underlying CDF model is based on the DDME2 parameterization.  }
\label{fig:kaon_1}
\end{figure}
The effects of kaon condensation on the global parameters (mass, radius, etc.) of compact stars 
are within the current uncertainties in the values of these parameters, so the bulk properties of compact stars cannot be used to either rule out or confirm the presence of kaon condensates in compact stars. However, the presence of such a condensate may affect the dynamical properties of compact stars. It is therefore interesting to explore the effect a condensate has on the composition of a star. This is done
in Fig.~\ref{fig:kaon_1} with and without the inclusion of the $\Delta$-resonance. At low densities, before the onset of strange particles, electric charge neutrality is maintained among the protons, electrons, muons, and the $\Delta^{-}$ resonance, if present. At somewhat higher density ($\ge 2.5\rho_{\rm sat}$) $\Lambda$ and $\Xi^{-}$ appear in the matter (recall that because of the repulsive $\Sigma$-potential at saturation, the $\Sigma$ thresholds are shifted to very high densities). The next particle species to appear is the $K^-$ which is favored by its negative charge and the possibility to replace 
negatively charged energetic electrons. The $\bar K^0$ appear at much higher densities. The presence of the $\Delta^-$ shifts the onset of $K^-$ to higher densities as they effectively take over the role of negatively
charged low-energy-cost constituents at low to intermediate densities.

To summarize, the recent work shows that kaon condensation may become phenomenologically relevant to the physics of compact star models that are consistent with two-solar mass constraint on the star's maximum mass. (Early work on kaon condensation predicted maximum masses of kaon-condensed stars that are significantly below this limit.) This opens the field for exploration of the effects of kaon condensation at high densities on evolution of compact stars, for example, their cooling. Kaon condensate may lead to intense neutrino emission from  the core of a compact star that is comparable to direct Urca process~\cite{Thorsson1995PhysRevD,Kubis2003}. Rapid cooling via kaon condensation thus becomes an option for massive stars with central densities exceeding several times the nuclear saturation density~\cite{Page1990ApJ,Tatsumi1998}. Also, kaon condensation is known to contribute to the transport properties of a  compact star's core, for example, to the bulk viscosity~\cite{Chatterjee2007PhysRevD,Chatterjee_2008}, which is important for the damping of $r$-mode instability.

\subsection{Strong magnetic fields}
\label{ssec:Bfields}
The features seen from anomalous X-ray pulsars and soft gamma repeaters can be explained in terms of the concept of a magnetar --- a highly magnetized neutron star with a surface magnetic field in the range of $10^{14}$--$10^{15}$ G~\cite{Harding2006,Turolla2015}. Magnetars are good candidates for repeating fast radio bursts (FRBs) as their periodicities can be generated by the rotation or precession of the neutron stars radiating the burst~\cite{Margalit2020,Beniamini2020,Beloborodov2020,Levin2020,Zanazzi2020}. In this context, the studies of dense matter, and in particular hypernuclear $\Delta$-admixed matter, in strong fields is of interest. An upper limit on the magnitude of the magnetic field can be set by considering the gravitational equilibrium of a compact star. A very general idea about the magnitude of the limiting field can be obtained from the virial theorem, which sets it at about $B \leq 10^{18}$--$10^{19}$~G. The influence of the magnetic field on the mass, radius, and moment of inertia of a compact star has been studied for some time~\cite{Chakrabarty1997,Broderick2000,Chen2007,Rabhi2008}, but has been found to be insignificant. However, magnetic fields can be important in dynamic settings such as supernova explosions and BNS mergers. The magnetic deformations could be significant enough to make solitary neutron stars important sources of gravitational waves~\cite{Suvorov2016MNRAS,Mastrano2012MNRAS}. The solutions of general relativistic equations with magnetic field for stationary stars have been initially obtained in Refs.~\cite{Bonazzola1993,Bocquet1995,Cardall2001} assuming various forms of the poloidal and toroidal field configurations. Later on, combined toroidal and poloidal fields were considered and were found to be more stable~\cite{Ciolfi2009,Ciolfi2010,Ciolfi2013}.  The magnitude and the structure of the interior fields in magnetars are not known, but some ``universal'' field profiles have been generated from studies of equilibria in general relativity coupled
 with the Maxwell equations. These profiles permit one to assess the local quantities, for example, the EoS, from the given values of the field at the boundaries (the surface or the center of the star). 

 The magnetic field effects could be twofold depending on whether they affect Fermi surface phenomena (which are typically of the order of MeV or less) or the Fermi sphere's volume, in which case one needs 100 MeV to GeV scale fields. In the first case, the changes  in the quasiparticle spectrum of baryons lead to the suppression of the superfluidity of protons via Landau's diamagnetism~\cite{Sinha2013,Sinha2015} and superfluidity of neutrons via Pauli paramagnetism~\cite{Stein2016,Sedrakian2017JPhCS}. Such modifications can alter the neutrino emissivity of compact stars through the modifications of the neutrino production reaction rates. Fields of this magnitude also introduce anisotropy in transport coefficients which in turn affect the magnetic, rotational, and thermal evolutions of magnetars~\cite{Pons2019,Sedrakian2016}. Larger fields, up to GeV magnitude, affect the stellar equilibrium
 via the modifications of the pressure.

 Recent work on hypernuclear matter with or without an admixture of $\Delta$-resonances
 within the CDF approach has incorporated strong magnetic fields~\cite{Sinha2013,Thapa:2020ohp,Rather_2022,Marquez_2022,Dexheimer_2021_EPJA} along the lines of the earlier work in Refs.~\cite{Chakrabarty1997,Broderick2000,Chen2007,Rabhi2008}. Let us outline the modification introduced by the magnetic fields to the CDFs discussed above. Firstly, under the assumption of minimal coupling of electromagnetic fields, the covariant derivative becomes $D^{\mu}= \partial^{\mu} + ieQ A^{\mu}$, where $A^{\mu}$ is the electromagnetic vector potential, $eQ$ is the charge of the particle.  A further step is to account for the way the strong magnetic fields modify the phase space sampling.  As well known, the orbits of the charged particles in the direction perpendicular to the field become Landau quantized, and therefore, the integration in the orthogonal to the $B$-field plane must be replaced by a summation over the Landau levels.  The spectrum of charged fermions (both baryons and leptons) is modified to
\bea
E = \sqrt{p^2+m^2+ 2\nu e\vert Q\vert B}.
\eea
The scalar and baryonic densities of charged particles labeled by the index $c$ are given at $T=0$ by
\begin{subequations}\label{eqn.11}
\begin{eqnarray}
\rho^{s}_c & =& \frac{e|Q|B}{2 \pi^2} m_{c}^{*} \sum_{\nu=0}^{\nu_{\rm max}}(2-\delta_{\nu,0}) \ln \left( \frac{p_{{F}_c} + E_{F_c}}{\sqrt{m^{*^2}_c + 2\nu e|Q|B}} \right), \\
\rho_c & = &\frac{e|Q|B}{2 \pi^2} \sum_{\nu=0}^{\nu_{\rm max}}(2-\delta_{\nu,0})
p_{{F}_{c}},
\end{eqnarray}
\end{subequations}
where $p_{{F}_{c}}$,  $E_{F_c}$, and $m^{*}_c$ are, respectively, the Fermi
momentum of the $\nu^{\text{th}}$-Landau level, Fermi energy, and effective mass of a baryon carrying an electric charge $c$. The summation over Landau levels is limited by the maximal value 
\begin{equation}\label{eqn.15}
\begin{aligned}
\nu_{\rm max} = \text{Int} \left( \frac{p_{{F}_{c}}}{2e|Q|B} \right).
\end{aligned}
\end{equation}
A similar expression can be obtained for spin-3/2 $\Delta$-resonances.  Finally the energy-density of the field itself, i.e., the contribution from the term \eqref{eq:L_em} should be taken into account.

Given these modifications, an EoS for strongly magnetized matter can be obtained. 
It turns out that the modifications to the global parameters of the compact stars due to the magnetization of matter are small, at the level of a few percent even for very strong magnetic fields. However, the composition of matter and the effective masses of baryons undergo visible oscillations as the density changes. These are akin to the de Haas-van Alphen oscillations known from condensed matter systems (e.g., oscillations in the magnetic susceptibility of electronic systems).  These oscillations are generated by the discontinuous occupation of the Landau levels. Because the charged and neutral baryons are coupled by the baryon number and charge conservation, the oscillations are coupled as well and affect the fractions of neutral particles.  Comparing the oscillations in the strange and non-strange sectors, one finds that the hyperon fractions are more affected by the magnetic fields than the non-strange baryon fractions because the latter have lower Fermi momenta and energies, so the same magnetic field has a larger effect. Because the Dirac masses of baryons undergo oscillations as well, one would expect that some thermodynamic and kinetic quantities such as the specific heat, baryon mean-free path, thermal conductivity, etc.  show analogous oscillations as well.

Observationally, magnetic fields expected in magnetars are in the range of  $10^{15}-10^{17}$~G and can induces sufficient deformation (ellipticity of the order of $10^{-6}$) in order to produce continuous gravitational waves from rotating solitary neutron stars~\cite{Suvorov2016MNRAS,Mastrano2012MNRAS}. The suppression of proton superconductivity occurs for fields of the order of $10^{14}-10^{15}$~G via Landau's diamagnetic effect.
 Larger magnetic fields of the order of $10^{16}-10^{17}$~G are needed to suppress the neutron $S$- and $P$-wave pairings in the crust and core of a compact star via Pauli paramagnetic effect. These suppression mechanisms may be also effective for the  hyperonic pairing. Macroscopically, the suppression of pairing will affect the rotational crust-core coupling and precession in magnetars~\cite{Sedrakian2016,Wasserman2022ApJ,Gao2023MNRAS} as well as their cooling in the 
neutrino-dominance era~\cite{Sinha2013,Sinha2015}.

\section{Rapidly Rotating Hypernuclear Stars}
\label{sec:Rapid_rotation}

\subsection{SU(6) parameterisation and the GW190814 event}
\label{ssec:SU(6)}
A general feature of Einstein's gravity is the existence of a maximum mass of static compact stars under the assumption that the pressure comes from ``ordinary'' baryonic matter. The exact value of this maximum mass for non-rotating objects is currently unknown. Rotating compact stars host masses about $20\%$ larger than their static (non-rotating) counterparts because the centrifugal force provides additional support against the gravitational pull toward the center of the star. There are several publicly available codes for computing stellar configurations of rapidly rotating compact stars (see Refs.~\cite{RNS} and~\cite{LoreneCode}). They are based on an iterative method for solving the Einstein equations~\cite{Nozawa1998,Cook1994} in axial symmetry and use a tabulated EoS as input.
The solution method is iterative: it starts with a ``guess configuration'' (density profile), integrates the equations of the stellar structure, and uses this result as input for a new iteration.
This procedure is repeated until convergence to the desired accuracy is achieved at every point on the spatial grid. Rotating compact stars with hyperonic cores have also been considered in various contexts~\cite{Rosenfield2006,Wen2011,Sekiguchi2012,Zhang2013,Marques_PRC_2017,Lenka2018,Tuzh:2022,Liang:2022} and more recently in relation to the $2.5 M_{\odot}$ 
mass compact object observed in gravitational waves~\cite{Sedrakian2020,Dexheimer:2020,Lijj:2020,Fu:2022eeb}.
  
\begin{figure}[t]
\begin{center}
\includegraphics[width=13.cm,keepaspectratio]{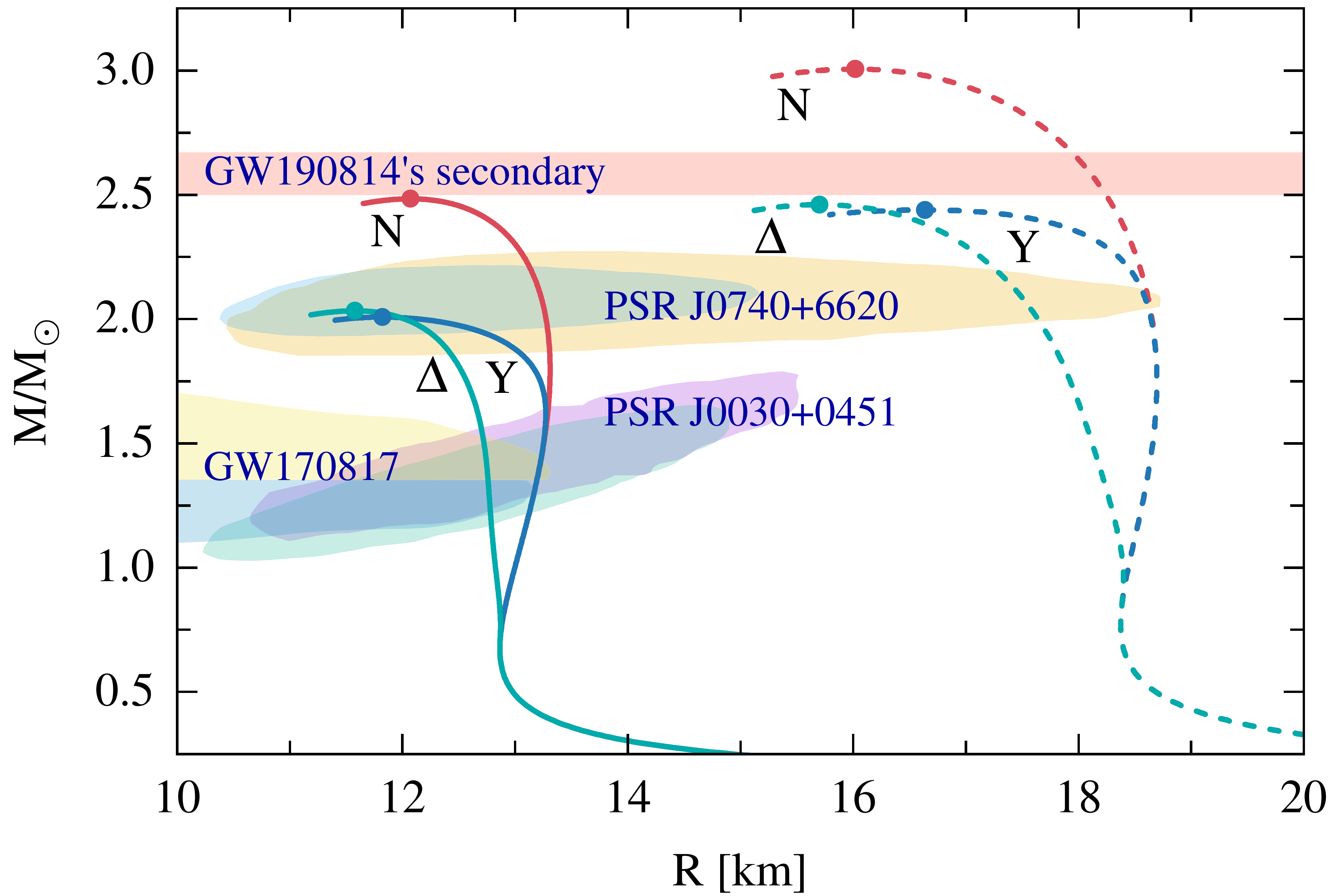} 
\end{center}
\vspace{-0.5cm}
\caption{MR relations for nonrotating (solid lines) and
  maximally rotating (dashed lines) nucleonic ($N$), hypernuclear
  ($Y$) and $\Delta$-admixed-hypernuclear ($\Delta$) zero-temperature
  stars~\cite{Sedrakian2020}. The parameters of the CDF, the experimental ellipses and the mass range for GW190814 are the same as in Fig.~\ref{fig:1.6}. 
}
\label{fig:1.13}
\end{figure}

Renewed interest in the study of rapidly rotating stellar models, as discussed in the previous sections, arose after the LIGO and Virgo Collaboration  measured
gravitational waves originating from the binary coalescence of a $22.2-24.3 M_{\odot}$ black hole with a compact object in the mass range $2.50-2.67M_{\odot}$ (the GW190814 event)~\cite{Abbott:2020khf}.  Later similarly asymmetric event GW200210~\cite{LVKC:2021} was observed with component masses of $24.1^{+7.5}_{-4.6}\,M_{\odot}$ and $2.83^{+0.47}_{-0.42}\,M_{\odot}$. The light members of these binaries have masses that lie in the so-called ``mass gap'' of $2.5\lesssim M/M_{\odot}\lesssim 5$, where neither a neutron star nor a black hole has ever been observed and whose existence is not obvious from the point of view of stellar evolution scenarios. A natural question that arises in this context is the possible compact star nature of the light companions of these binaries. In the following, we discuss the conjecture of very massive compact star rotating at a frequency close to the Keplerian limit~\cite{Sedrakian2020,Lijj:2020}, based on an EoS containing hyperons and $\Delta$-resonances. This issue has also been discussed in Refs.~\cite{Dexheimer:2020,Tan_PRD_2022} from a somewhat different perspective using alternative models.

Figure~\ref{fig:1.13} shows the MR relations of compact stars using 
nucleonic ($N$), hyperonic ($NY$), and hyperonic admixed with $\Delta$-resonances
($NY\Delta$) EoS and the  DDME2 parameterization of the 
CDFs as described in Sec.~\ref{sec:Hyper_DFT}.  The nucleonic models cover the mass range $2.48\le M/M_{\odot}\le 3$ in the spin frequency range $0\le \Omega \le \Omega_K$, where $\Omega_K$ is the Kepler frequency.
It is seen that, the nucleonic models account for the mass of a compact star in GW190814 even without rotation.  In the case of $NY$ and $NY\Delta$ compositions, the maximum masses (but not the radii) are quite similar in the static case, and this property extends to the case of rapidly rotating stars.  As noted, the softening of the EoS in the case of $NY$- and $NY\Delta$-compositions implies a lower maximum mass compared to the nucleonic case.  The maximum masses for these compositions are close to $2.0M_{\odot}$, so the corresponding EoS are inconsistent with the interpretation of the light companion in GW190814 as a compact star.  Their maximally rotating (Keplerian) counterparts, on the other hand, have maximum masses  $\le 2.4M_{\odot}$, suggesting that the maximum rotation is insufficient to increase the masses to the required value of $2.5M_{\odot}$. Thus, there is considerable tension between hyperonization (with or without the admixture of $\Delta$-resonances) and the interpretation of the light companion of GW190814 as a compact star. 

The results shown so far were obtained with a particular density functional, and the next question is whether the changes in the CDF can change this conclusion.
\begin{figure}[t]
\centering
\includegraphics[width=12cm]{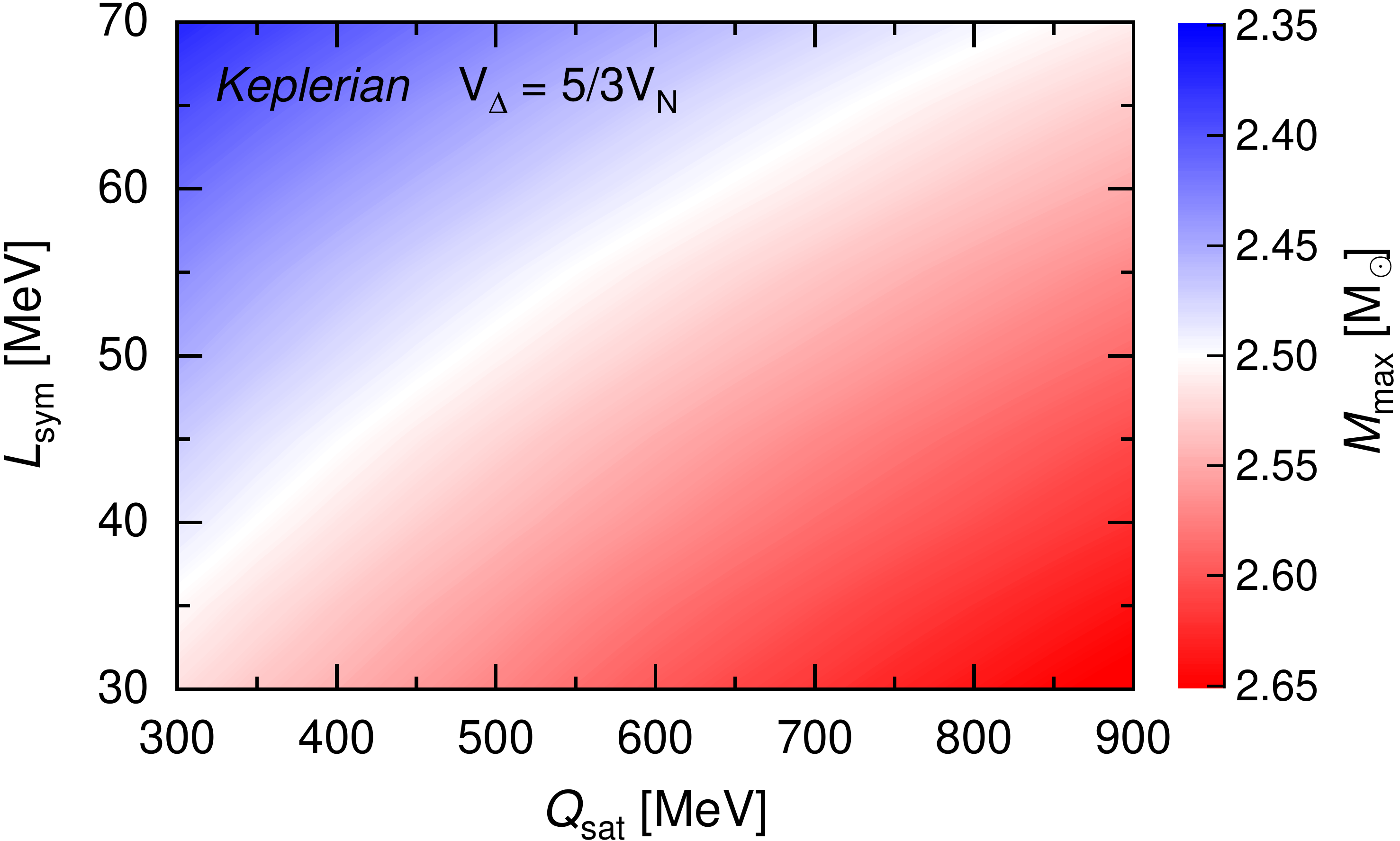}
\caption{The maximum masses of Keplerian sequences
(color-coded column on the right) for the parameter space
spanned by $Q_{\rm sat}$ and $L_{\rm sym}$~\cite{Lijj:2020}. The $\Delta$-resonance potential is fixed at $x_{N\Delta} =5/3$.
The large-$Q_{\rm sat}$ and small-$L_{\rm sym}$ range corresponds to compact stars
with masses exceeding $2.5\,M_{\odot}$.}
\label{fig:1.14}
\end{figure}
In~Ref.~\cite{Lijj:2020} a study of rapidly rotating stars was carried out for the case where the modifications of the EoS of $NY\Delta$ matter were formulated in the language of the characteristics appearing in equation \eqref{eq:Taylor_expansion}, in particular the values of the parameters $Q_{\rm sat}$ and $L_{\rm sym}$. The value of $x_{N\Delta}=5/3$ was assumed, although, as discussed above, the occurrence of $\Delta$-resonances only slightly increases the maximum mass of a configuration.  Figure~\ref{fig:1.14} shows the dependence of the maximum masses of the Keplerian models on the parameters $Q_{\rm sat}$ and $L_{\rm sym}$ for $x_{N\Delta} =5/3$. Large masses compatible with the compact star in GW190814 arise in the region of large $Q_{\rm sat}$, implying larger pressure at large energy densities. Large masses are also favored for smaller values of the parameter $L_{\rm sym}$ (implying smaller radii of the stars and thus more compact objects). One can compare the required values of $Q_{\rm sat}$ with other existing functionals (to avoid, in a first approximation, a complete calculation with a given functional).  A comparison shows that the range of $Q_{\rm sat}$ values compatible with a compact star in GW190814 has no overlap with values predicted by large samples of nonrelativistic and relativistic density functionals~\cite{Dutra2012,Dutra2014}. Exceptions are those values of $Q_{\rm sat}$ predicted by  the DDME2  parameterization~\cite{Lalazissis2005} and those found in some of the recently proposed functionals~\cite{Taninah:2019cku,Fattoyev:2020}.
Thus, the compact star interpretation of the secondary object in GW190814 is in strong tension with the idea of hyperonization of dense matter (with and without $\Delta$-resonances). To argue otherwise requires several extreme assumptions, such as maximally fast rotation and $Q_{\rm sat}$ and $L_{\rm sym}$ values outside the range covered by most density functionals.

\subsection{SU(3) parameterization, massive stars, and vanishing hyperon population}
\label{ssec:SU(3)}
We have seen so far that the hyperonization in the nucleonic matter makes it difficult to achieve high masses of compact stars, with an immediate conclusion that objects with masses of the order $2.5M_{\odot}$ are necessarily black holes~\cite{Sedrakian2020,Lijj:2020}. 
 However, the softening due to hyperonization can be mitigated if additional repulsion is introduced in the vector-meson sector.
A systematic way to implement this program is to break the SU(6) symmetry down to SU(3) using the more general expressions for the couplings given by Eqs. \eqref{eq:Romega}-\eqref{eq:Rrho}.
\begin{figure}[tb]
\centering
 \includegraphics[width=13cm,keepaspectratio]{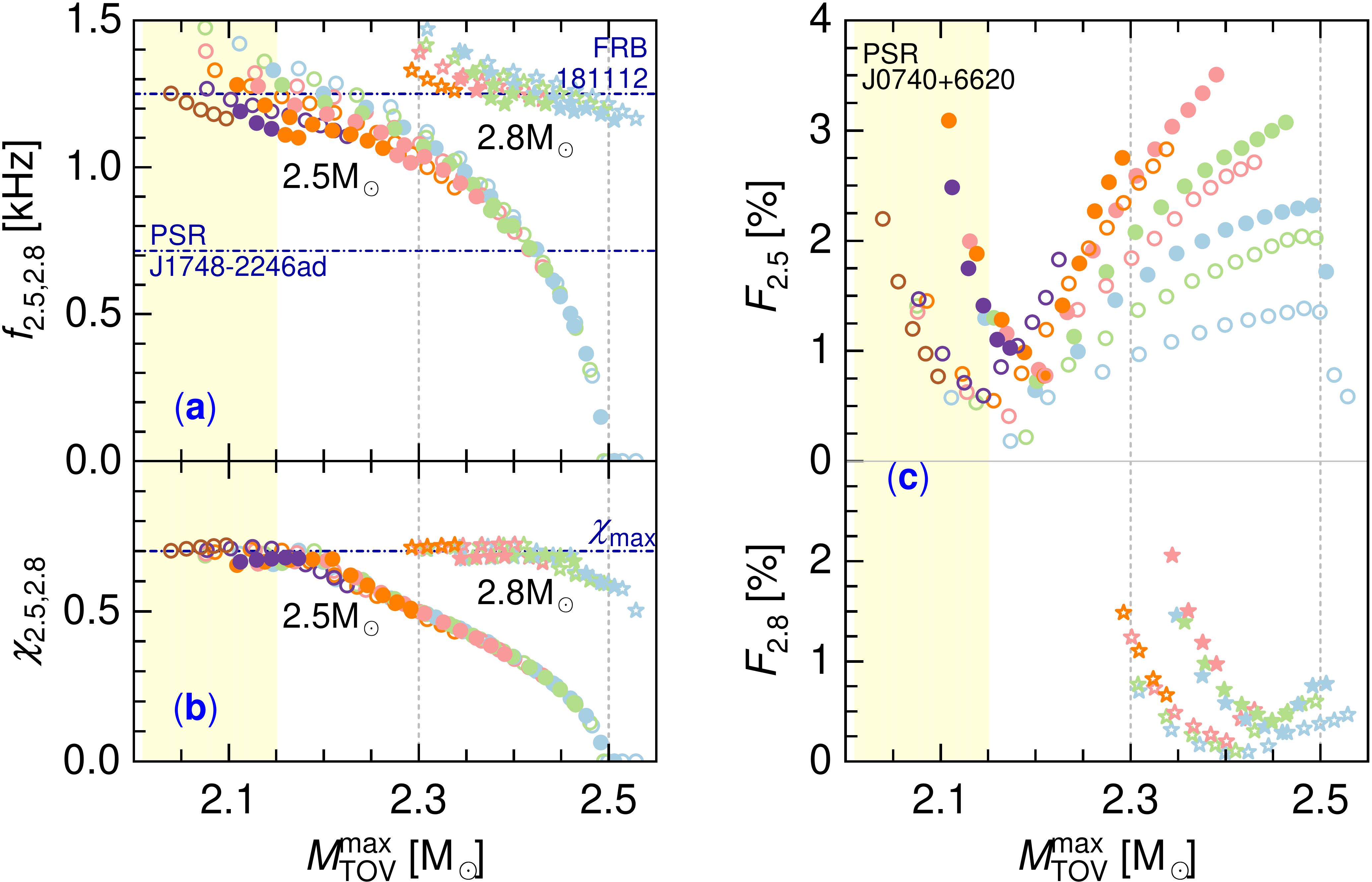}
 \caption{(a) The minimum rotational frequencies $f_{2.5, 2.8}$, (b) dimensionless spin parameters $\chi_{2.5, 2.8}$ ($\chi \equiv J/M^2$, where $J$ is the angular momentum of the star), and (c) strangeness fractions $F_{2.5, 2.8}$ of the models with masses
   $M = 2.5$ and $2.8\,M_{\odot}$ as a function of the value of the maximum mass of
   the non-rotating star
  $M^{\rm max}_{\rm TOV}$~\cite{Fu:2022eeb}. In panel (a) the lower horizontal line
  corresponds to the frequency of PSR J1748-2246ad~\cite{Hessels:2006}
  and the upper one to that of FRB 181112~\cite{Yamasaki:2020}. In
  panel (b), the horizontal line denotes the upper bound on the spin
  parameter $\chi_{\rm max} = 0.7$ deduced in Refs.~\cite{Cook:1994,Haensel:1995}.
   The open circles correspond to models with $L_{\rm sym} = 40$ MeV, while the filled ones --- to 
   $L_{\rm sym} = 100$ MeV. The different colors correspond to different $z$ values drawn from the interval $0<z< z_{SU(6)}=1/\sqrt{6}$. The same color symbols represent models with the
  same fixed $z$ but varying $Q_{\rm sat}$ value in the interval [-600, 1000] with a step size of 100 MeV.  The yellow shading indicates the mass range of PSR J0740+6620.
  }
\label{fig:1.14a}
\end{figure}

The parameters available in the SU(3) model are $\alpha_V$ (with SU(6) value $\alpha_V = 1$), the mixing angle ~$\theta_V $, see Eq.~\eqref{eq:vector_mixing}, and the parameter $z$, which is the ratio of the meson octet and singlet couplings~\cite{Dover:1985,Rijken:1998,Weissenborn2012b,Lijj2018a}.  The exploration of the SU(3) model and its comparison to SU(6) can be reduced to variation of the last parameter $z$. The argumentation is as follows: one first assumes the ideal mixing value in which case $\theta_V = \rm tan^{-1} (1/\sqrt{2})$. This assumption is substantiated by the observation that the mixing between nonstrange and strange quark wave functions in the $\omega$  and $\phi$ mesons is ideal.  Experiments on meson masses suggest that $\theta \approx 40^\circ$~\cite{ParticleDataGroup:2020} which is very close to the ideal mixing angle $\theta \approx 35.3^\circ$. The effects of the remaining parameters were explored by fixing one of them and varying the other, see Ref.~\cite{Lijj2018a}. This study shows that they have the same effect on the observables, so it would be reasonable to vary only one of them to see the effects of broken SU(6) symmetry. Reference~\cite{Fu:2022eeb} kept $\alpha_V=1$ fixed at its SU(6) value and varied $z$ parameter to obtain a series of EoS of hyperonic matter and the structure of non-rotating and rotating hypernuclear stars.

It turns out that non-rotating compact stars within the SU(3) symmetric model can have maximum
masses between $2.3$ and $2.5\,M_{\odot}$ for a range of values of the $z$ parameter and a set of stiff high-density nucleonic EoS, which can be parameterized in terms of the $Q_{\rm sat}$ coefficient in Eq.~\eqref{eq:Taylor_expansion}. However, for such objects the hyperon fractions are reduced to several percent, i.e., one is dealing essentially with nucleonic stars.  The global parameters of these stars (mass, radius, etc.) are consistent with those  of stars based on purely nucleonic EoS models~\cite{Lijj:2020,Fattoyev:2020,Zhangnb:2020,Tsokaros:2020,Huangkx:2020,Biswas:2021,Tews:2021} (we exclude hybrid star models requiring deconfinement~\cite{Dexheimer:2020,Tanhuang:2020,Bombaci:2020,Roupas:2020,Christian:2020,Demircik:2020,Malfatti:2020,Jumin:2021}).  Therefore, we are led to the conclusion that stars with a high hyperonic fraction (10-20\% ) have maximal masses that are significantly smaller~\cite{Sedrakian2020,Lijj:2020,Dexheimer:2020,Tuzh:2022,Liang:2022}.
Rotation will increase the masses of stars, as is well known, by up to about 20$\%$. In the Keplerian
limit, the models with small $z$ but large $Q_{\rm sat}$ parameter values (i.e., nearly nucleonic models) achieve masses up to $3.0\,M_{\odot}$. One may now ask for minimum frequencies which can
accommodate stars that are compatible with the mass range $2.5-2.8\,M_{\odot}$, which would then
imply the involvement of compact stars in the gravitational wave events GW190814 and GW200210. Figure~\ref{fig:1.14a}
shows the minimum frequencies $f_{2.5}$ and $f_{2.8}$   needed to support compact stars with masses of $2.5$ and $2.8\,M_{\odot}$. 

One may conclude that the most extreme models featuring large-$Q_{\rm sat}$ and small-$z$ values can account for a stellar origin of smaller mass members of the binaries in the events GW190814 and GW200210.  However, the resulting objects are nearly nucleonic stars, since the amount of hyperons characterized by the strangeness fractions $\lesssim 3\%$ is extremely small. 


\section{Finite Temperatures}
\label{sec:FiniteT} 

\subsection{Warm hypernuclear matter at high-densities}
\label{ssec:FT_high_densities}

Several astrophysical processes result in the formation of transient warm, neutrino-rich compact objects. The description of these objects requires an understanding of the properties of nuclear and hypernuclear matter at non-zero temperatures which, in general, range from a few tens to around 100 MeV. Let us give a few examples. A well-known and long-studied process is the core collapse of massive stars which leads to type-II supernovae. The whole process of collapse, bounce, and explosion is sensitive to input EoS, the composition of matter, and neutrino interactions with baryons. In addition, these processes may take place within strong magnetic fields, therefore electromagnetic effects, such as anisotropy of the transport in strong fields, may play a role as well.  Type-II supernovae leave behind a compact object -- a proto-neutron star -- which is hot and neutrino rich. Its subsequent thermal, magnetic, and rotational evolution depends sensitively on the finite temperature EoS and the composition of matter. Furthermore, the process of nucleosynthesis in the ejecta, which is gravitationally detached from proto-neutron star, also depends on the neutron-to-proton ratio, alpha-particle abundances, and other factors which are determined by the properties of baryonic matter~\cite{Prakash_1997,Pons_ApJ_1999,Janka_PhysRep_2007,Mezzacappa2015,Connor2018ApJ,Malfatti_PRC_2019,Burrows2020MNRAS,Dexheimer_2022}. Type-II supernovae may leave behind a black hole if the masses of the progenitor stars, which undergo core collapse, are large enough. The exact threshold value is not known accurately but is estimated to be in the range of a few tens of solar masses~\cite{Sumiyoshi_2007,Fischer_2009,OConnor_2011,Schneider_2020}.  In this case, the role of baryonic matter is limited to the dynamics of the formation of black holes, the relevant time scales of formation, and the features of the neutrino signal that may be detected from such an event.  Warm baryonic matter plays a role also in  BNS mergers~\cite{Shibata_11,Faber2012:lrr,Rosswog:2015nja,Rosswog:2017sdn,Baiotti:2019sew}.\footnote{Note that there are also many parallels to the physics of heavy ion collisions, as emphasized, for example, in Ref.~\cite{Most_2022_arxiv_jan}.} The event rate for these processes is higher than for type-II supernovae and the prospects of detecting ``loud'' multimessenger merger events like GW170817 will be good with the increased sensitivity of gravitational wave detectors in the upcoming decade. Interestingly, such processes can simultaneously test properties of warm and cold baryonic matter in a single event: the pre-merger gravitational signal, as the one detected from GW170817, carries information about the properties of cold baryonic matter; the post-merger transient hypermassive compact star is hot and probes the properties of warm baryonic matter. Once gravitational waves are detected from this phase of a BNS merger, the properties of matter can be constrained by modeling the oscillations and dynamics of the merger remnants.

The distinctive aspect of hot baryonic matter is the presence of a neutrino component which is trapped in the matter for temperatures above the trapping temperature  $T_{\rm tr}\simeq 5$~MeV~\cite{Alford2018b}.  The matter does not reach beta-equilibrium and, therefore, instead of imposing the beta-equilibrium condition, one is led to fix the lepton number in each lepton family. Since the tau-leptons are too heavy to be relevant, only the lepton numbers of electrons and muons need to be fixed. Furthermore, since the processes governing Type II supernova collapses and explosions and BNS mergers proceed, in a first approximation,  adiabatically, the  
matter is locally characterized by the density and entropy, whereas the temperature exhibits gradients.  The role of neutrinos in the hot stages of evolution is twofold: they affect the EoS and composition of matter, as we will discuss below, and they affect the dynamics of explosions and mergers by transporting energy and momentum across the system.
\begin{figure}[tb] 
\includegraphics[width=0.9\linewidth]{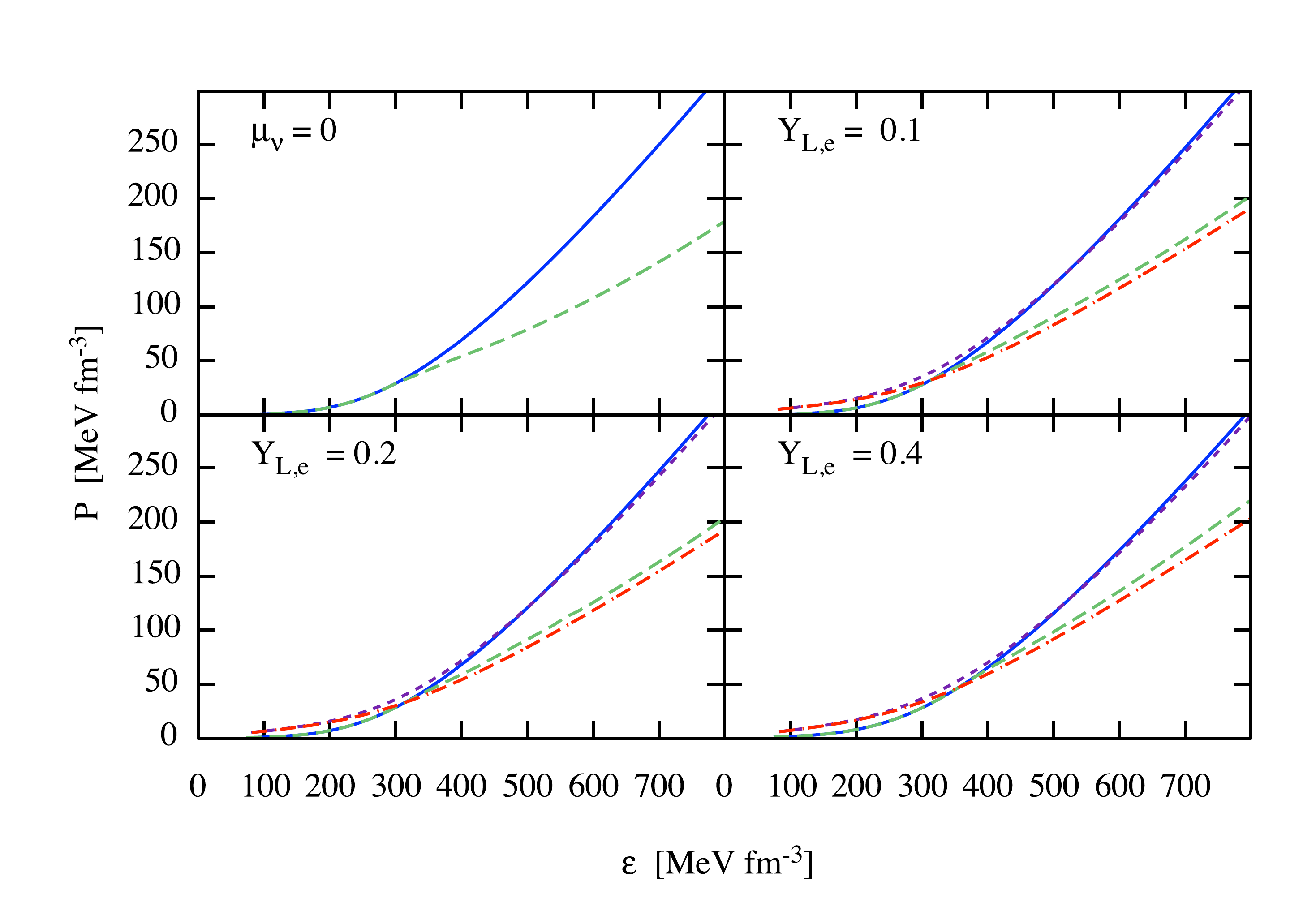}
\caption{Pressure as a function of the energy density for
  finite-temperature hypernuclear matter~\cite{Sedrakian2021Univ}. The upper-left panel
  corresponds to neutrino-free matter in $\beta$-equilibrium
  without (solid) and with (dashed) hyperons at
  $T=0.1$~MeV. The remaining three panels show the EoS of
  neutrino-trapped matter for $Y_{L,e}= 0.1, \, 0.2, \, 0.4$. 
  Two values of temperature are used:
  $T=5$~MeV (solid lines--without hyperons, 
  long-dashed--with hyperons); 50~MeV (short-dashed--without
  hyperons, double-dash-dotted--with hyperons). The muon fractions are
  adjusted to the conditions of supernova and BNS merger cases as follows:
  $Y_{L,\mu} = Y_{L,e} = 0.1$ (upper-right panel) and $Y_{L,\mu} = 0$
  for $Y_{L,e}=0.2$ and 0.4 (bottom panels).  The~case $Y_{L,e} = 0.1$ is
  characteristic of a merger remnant, whereas
  $Y_{L,e} = 0.2,\, 0.4$ correspond to supernovae.}
\label{fig:EoS_Uni} 
\end{figure}
Initial work on the hot hyperonic matter was motivated by the type-II supernova and proto-neutron star modeling~\cite{Prakash_1997,Pons_ApJ_1999,Bednarek2004IJMPD,Rios2005PhRvC,Yasutake2009PhRvD,Ohnishi2010NuPhA,2012PhRvC..85e5806O,Nakazato2012,Peres2013,Colucci2013,Masuda2016EPJA,Marques_PRC_2017,Fortin_PASA_2018,Lenka_JPG_2019,Roark_MNRAS_2019,Malfatti_PRC_2019,Kochankovski2022MNRAS,Dexheimer_EPJC_2022}, for reviews see~\cite{Oertel_RMP_2017,Raduta2022EPJA}. More recent work included also the conditions that are relevant for BNS mergers~\cite{Raduta2020,Sedrakian2021Univ,Sedrakian2022EPJA}.  Let us next specify the thermodynamic conditions that are prevailing in the two regimes that arise depending on the ratio of the neutrino mean free path to the size of the system.  The neutrino-free regime arises when the neutrino mean-free path is much larger than the size of the system.  The neutrino-trapped regime arises in the opposite case when the neutrino mean free path is much smaller than the size of the system. When trapped, neutrinos are in thermal equilibrium with respect to the matter: in equilibrium, their distribution is given by the Fermi distribution function at temperature $T$, which is perturbed when the dynamics of the matter is considered. Then, the distribution functions are the solution of kinetic equations for neutrinos coupled to matter. We will restrict the discussion to the equilibrium neutrino component at some temperature, which then provides the background equilibrium upon which time-dependent perturbations are applied. For static computations we will adopt typical values of lepton fractions that are required in numerical simulations; more general results involving grids in the lepton fractions are provided by three-dimensional tables for the matter at finite temperatures~\cite{Dexheimer2022Parti,Typel2022}.  If the lepton number is conserved in each family, i.e., if the neutrino oscillations are neglected, then for supernova matter electron and muon lepton numbers are typically $Y_{L,e}\, \equiv Y_e+Y_{\nu_e} = 0.4$ and $Y_{L,\mu}\,\equiv Y_\mu+Y_{\nu_\mu} = 0$, with the lepton partial densities normalized by the baryon density $Y_{e,\mu} = (\rho_{e,\mu}-\rho_{e^+,\mu^+})/\rho_b$, where $e^+$ refers to the positron and $\mu^+$ to the anti-muon.  In the supernova context muonization in the matter can lead to a small (on the order of $10^{-3}$) fraction of muons~\cite{Bollig_2017,Guo:2020tgx} which will be neglected. BNS mergers undergo a different evolutionary path, since, initially, there are two cold neutron stars whose lepton fractions are fixed by $\beta$-equilibrium in the cold baryonic matter. In this case, the typical values of the lepton fractions are $Y_{L,e} = Y_{L,\mu}= 0.1$, which are consistent with the compositions of pre-merger cold neutron stars.

The weak equilibrium conditions for baryons are given by Eqs. \eqref{eq:c1}-\eqref{eq:c4}. The charge neutrality conditions are given by Eq.~\eqref{eq:charge_neutraility}, which can be written in terms of 
partial charge densities normalized by the baryonic
density $Y_Q = \rho_Q/\rho_B$ as
\begin{equation}
\label{eq:charge_neutraility_Y}
Y_Q = Y_e + Y_{\mu}.
\end{equation}
The electron and muon chemical potentials are given by 
\begin{subequations}
\begin{eqnarray}
\label{eq:mu_condition}
  \mu_e &=& \mu_\mu =-\mu_Q = \mu_n-\mu_p, \quad \textrm{(free streaming)} , \\
  \mu_e& = & \mu_{L,e} - \mu_Q, \quad \mu_{\mu} =  \mu_{L,\mu} -
  \mu_Q,
  \quad \textrm{(trapped)} ,
\end{eqnarray}
\end{subequations}
where $\mu_{L,e/\mu}$ are the lepton chemical potentials associated with the lepton numbers $Y_{L,e} = Y_{e} + Y_{\nu_e}$ and $Y_{L,\mu} = Y_{\mu} + Y_{\nu_\mu}$. These are conserved separately.  Combining the weak-equilibrium and charge neutrality conditions we are now in the position to compute the EoS of stellar matter for both the trapped and free-streaming neutrino regimes. Note that in general, the $\beta$-equilibrium conditions are not strictly satisfied for various reasons. For example, at finite temperatures, the rate of the processes $n\to p + e+\bar\nu$ and $ p + e+\nu\to n$ may not be balanced as the neutrino and anti-neutrino distributions may deviate from thermal equilibrium at the same temperature. In this case an additional ``isospin chemical potential'' arises~\cite{Alford2019a,Alford2021b}. Note also that the $\beta$-equilibrium condition
occurs at finite temperature, i.e.,  the particles are not constrained to their Fermi surfaces.

\begin{figure}[bt] 
\includegraphics[width=0.91\linewidth]{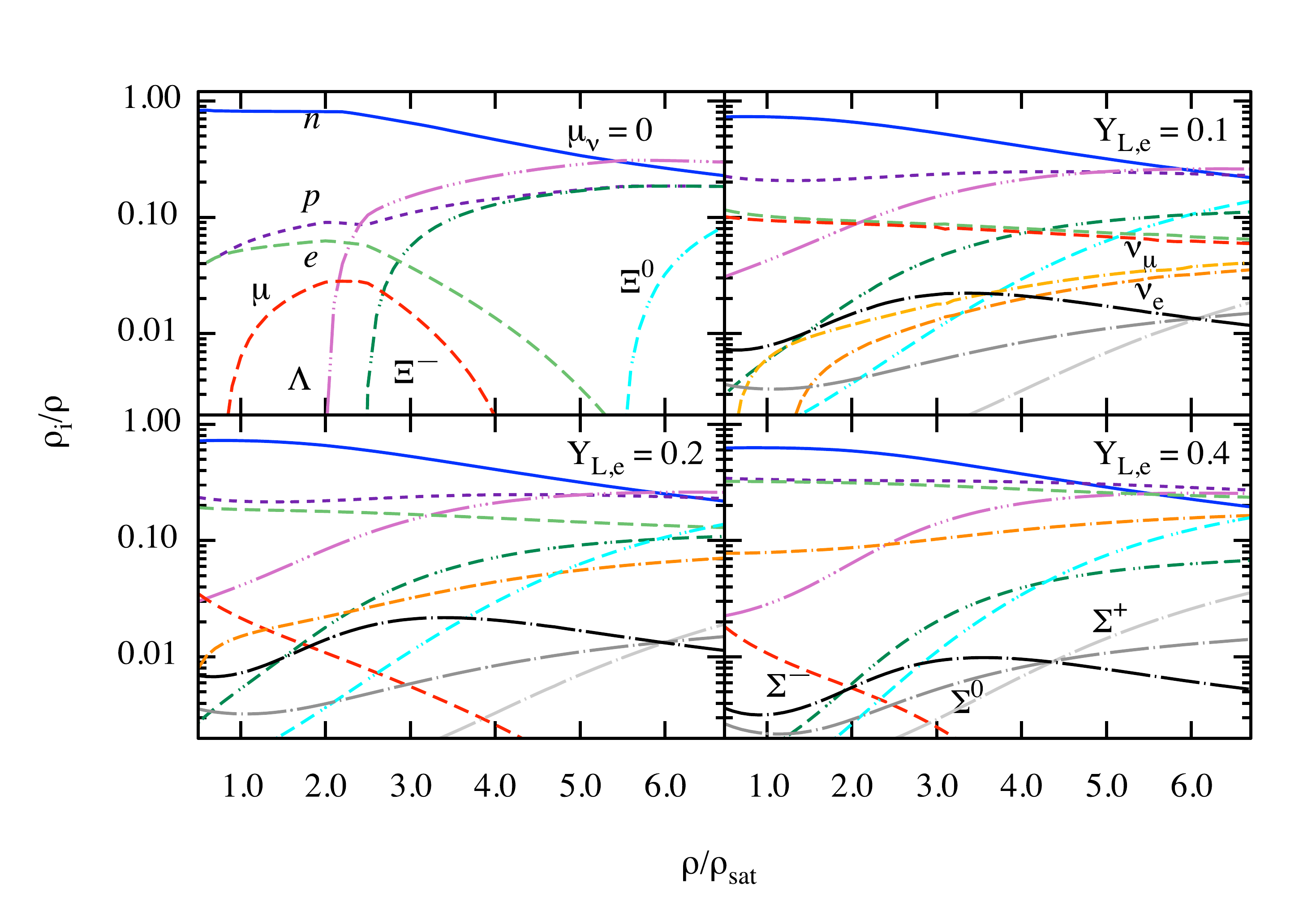}
\caption{Particle fractions $\rho_i/\rho$ as a function of baryon
  density $\rho$ normalized to nuclear saturation
  density $\rho_{\rm sat}$ \cite{Sedrakian2021Univ}. The parameter choices in the 
  different panels are the same as in Fig.~\ref{fig:EoS_Uni}, with the exception
  that only one temperature $T=50$~MeV is shown in the upper right and lower panels.
  At low temperatures (upper left panel)
  one observes a sharp increase in the hyperonic fractions at their density thresholds. 
  In the remaining panels the density thresholds are significantly
  lowered and the appearance of all three states of the $\Sigma$ hyperon is favored for
  $T=50$~MeV.}
\label{fig:abundances-hyperons_Uni} 
\end{figure}
To illustrate the new features of finite-temperature hypernuclear matter, we turn to the numerical solutions of the self-consistent equations for the meson fields and scalar and baryon densities for fixed values of temperature, density, and lepton numbers $Y_{L,e}$ and $Y_{L,\mu}$.
The $\Delta$-meson couplings are fixed as $R_{\Delta\omega} = R_{\Delta\rho} = R_{\Delta\sigma} = 1$, which correspond (roughly) to the case $x_{\Delta N}=1$.
The EoS of nucleonic and hyperonic matter are compared in Fig.~\ref{fig:EoS_Uni}, where we observe the well-known softening (shift of pressure toward lower values) of the EoS with the onset of hyperons. 

Figure~\ref{fig:abundances-hyperons_Uni} shows the relative particle number densities $\rho_i/\rho$ in matter containing the full baryon octet as a
function of $\rho/\rho_{\rm sat}$ ($\rho_{\rm sat} = 0.152$~fm$^{-3}$).  For the parameterization used in Ref.~\cite{Sedrakian2021Univ}, $\Lambda$, $\Xi^-$, and $\Xi^0$ hyperons appear at densities above the saturation density in the low-temperature matter. Early calculations based on non-interacting models~\cite{Ambartsumyan1960SvA} or models based on weakly repulsive hyperonic potentials~\cite{Colucci2013} have found that the $\Sigma^-$ is the first hyperon present, although it is heavier than the $\Lambda$. More recent studies have shown that the onset of the $\Sigma^-$ is shifted to high densities due to their probable repulsive potential in nuclear matter~\cite{BartPhysRevLett,DOVER1984171,Maslov:2015wba,LopesPhysRevC2014,Gomes:2014aka,Miyatsu_2015}.  The situation is changing at higher temperatures as the  $\Sigma^{\pm,0}$ triplet is found to be present for $T=50$~MeV, independent of the value of the lepton number.  One may observe that at finite temperatures the hyperon thresholds move to lower densities, which we will discuss in detail within a different model below.  At high densities, the $\Lambda$ hyperon is the most abundant baryon in the matter, with its fraction even exceeding that of neutrons (this occurs in Fig.~\ref{fig:abundances-hyperons_Uni} for $\rho/\rho_{\rm sat} \gtrsim 5.5$).  This is a consequence of the weaker (repulsive) coupling of $\Lambda$s to the $\omega$ meson than for neutrons. Note also that the effective masses are density-dependent, but this is not a key factor here.  Finally, note that the difference between the nearly equal abundances of leptons for $Y_{L,e} = 0.1$ and the remaining cases $Y_{L,e} = 0.2,\, 0.4$ is due to the assigned value of $Y_{L,\mu}$, as already discussed above. This also affects the abundances of electron and muon neutrinos. These were present in almost equal amounts in the first BNS merger case. In supernova matter, the muon--neutrinos are replaced by a smaller amount of muon--antineutrinos.

\begin{figure}[tb]
    \centering
\includegraphics[width=.9\linewidth]{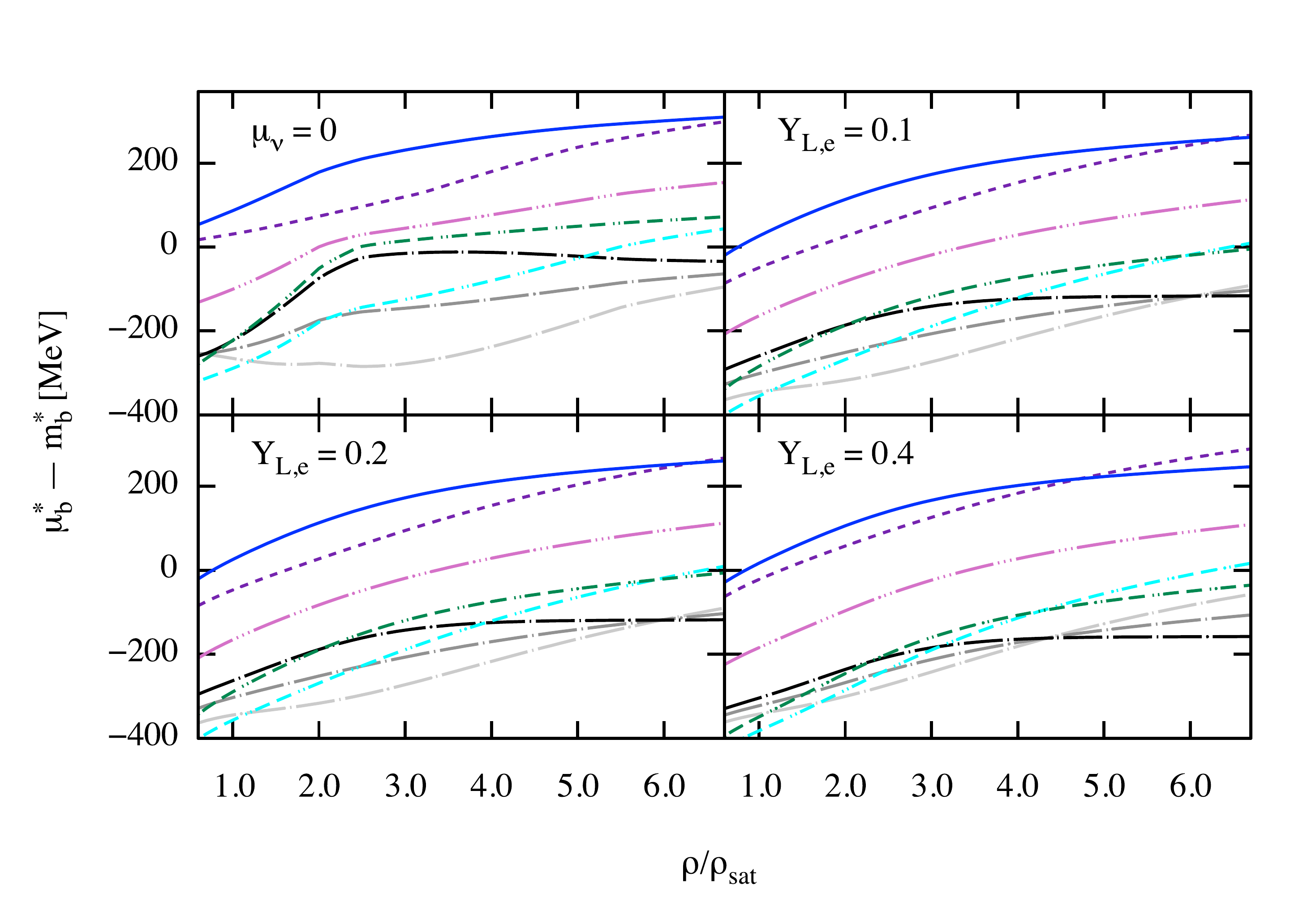}
\caption{Baryon effective chemical potentials (computed
  from their effective masses) as a function of normalized baryon number density~\cite{Sedrakian2021Univ}. The labeling of curves and parameters
   are the same as in Fig.~\ref{fig:abundances-hyperons_Uni}. }
\label{fig:mu-hyperons} 
\end{figure}

As can be seen in Fig.~\ref{fig:abundances-hyperons_Uni}, the abundances of baryons within the same isospin multiplet are intersecting at some special point, which may be called {\it isospin degeneracy point}~\cite{Sedrakian2021Univ}.  The position of this point depends on the choice of $Y_{L,e}$.  The origin of the isospin degeneracy can be traced to the equality of the effective chemical potentials at the isospin degeneracy point. To illustrate this feature, we refer to Fig.~\ref{fig:mu-hyperons} which shows 
the effective baryon chemical potentials minus the corresponding effective baryon masses. It should be noted that within the CDFs that have been discussed in Sec.~\ref{sec:Hyper_DFT}, the effective masses within each isospin multiplet are equal. This is a consequence of the minimal number of non-strange and strange mesons included in the Lagrangian~\eqref{eq:Lagrangian_B}. With a larger set of mesons, this degeneracy can be lifted.  A similar behavior of the baryon abundances was also found in Refs.~\cite{Malfatti_PRC_2019,Raduta2020,Sedrakian2022EPJA}, where the composition of hot stellar matter (including the quartet of $\Delta$-resonances) was shown for a constant entropy-per-baryon ratio. Note that, if the isospin symmetry were exact, the intersection points of the three isospin-multiplets, $n$-$p$, $\Sigma^{0,\pm}$, and $\Xi^{0,-}$, would be located exactly at the same density. 

\begin{figure}[bt]
  \centering
\includegraphics[width=0.7\linewidth]{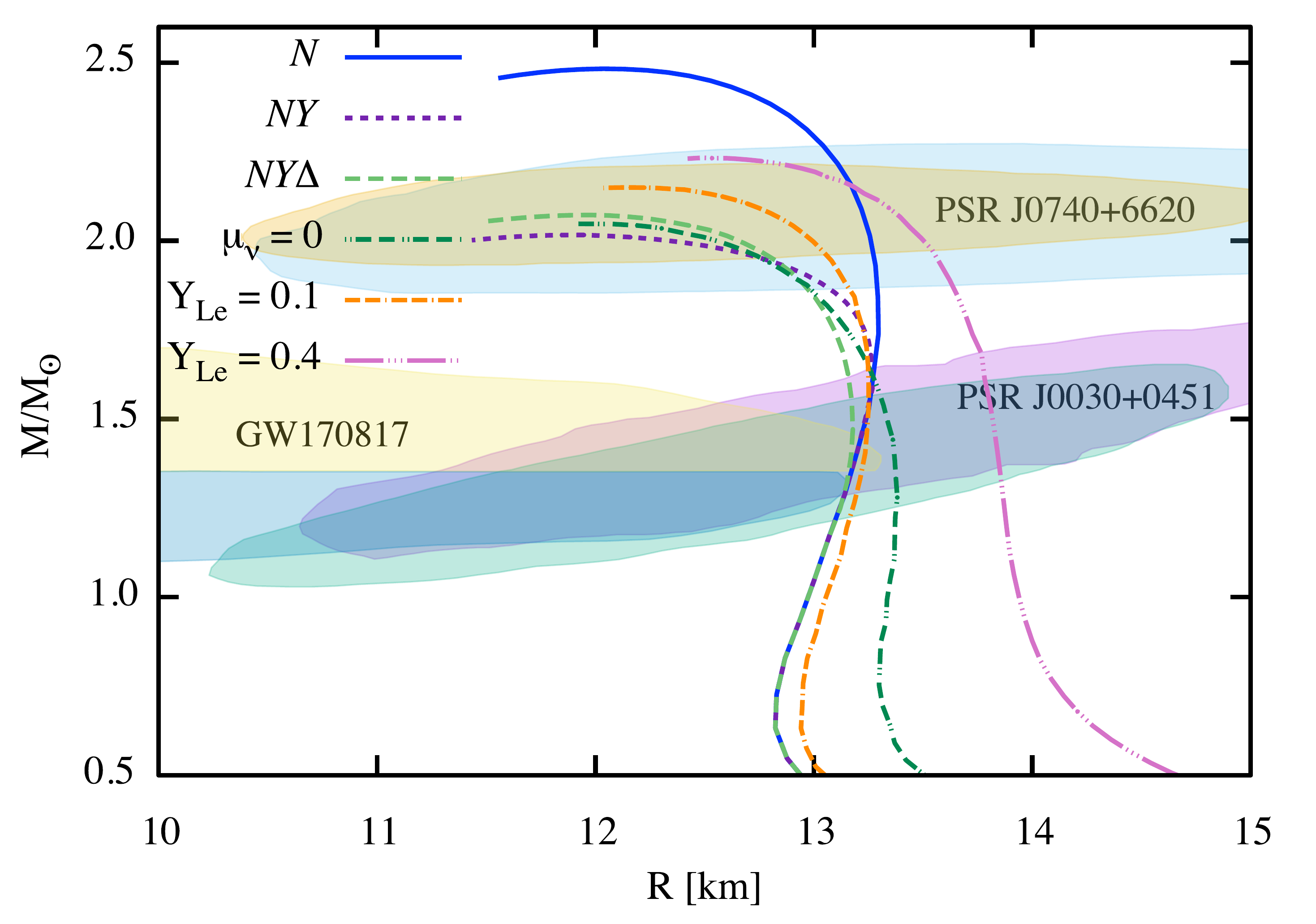}
\caption{ MR relations  for non-rotating spherically-symmetric compact stars. 
Three stellar sequences are shown for $\beta$-equilibrated, neutrino-transparent stars with nucleonic  ($N$), hypernuclear ($NY$), and $\Delta$-admixed hypernuclear
($NY \Delta$) compositions at $T=0.1$~MeV. The underlying nucleonic CDF is based on DDME2 parameterization and  $\Delta$-couplings are
$R_{\rho\Delta}=R_{\omega\Delta}=R_{\sigma\Delta}  =1$; the hyperonic
couplings are specified in~Ref.~\cite{Sedrakian2022EPJA}. In addition, we show
sequences of fixed $S/A=1$ neutrino-trapped, isentropic stars  composed
of $NY \Delta$ matter for constant lepton fractions of $Y_{Le}=Y_{L\mu}=0.1$ and $Y_{Le}=0.4,\,Y_{L\mu}=0$ as well as for neutrino-free matter ($\mu_{\nu}=0$). The
  ellipses show 90\% CI regions for pulsars PSR J0030+0451, PSR J0740+6620 and the gravitational-wave event GW170817 (see text for details). }
\label{fig:isoentropic-stars} 
\end{figure}
The finite-temperature EoS of hypernuclear matter can be used to construct static configurations
of isentropic stars, which are governed by the TOV equation describing 
static solutions of Einstein's field
equation, see Fig.~\ref{fig:isoentropic-stars}. For reference, the configurations
obtained for cold matter made of purely nucleonic ($N$), hyperonic ($YN$),
  and $\Delta$-resonance admixed hypernuclear matter ($YN\Delta$) are shown in this figure. 
  The purely nucleonic model has the highest maximal mass $M_{\rm max}=2.48M_{\odot}$
  of all models, and its radius is $R_{\rm max} = 12.1$ km.
  As already discussed, the softening of the EoS in the cases of 
  $YN$ and $YN\Delta$ compositions of cold matter reduce the
  maximum stellar mass.  Obviously, the
  properties of these three models differ only for stars beyond the bifurcation
  point at about $\sim 1.55 M_{\odot}$. As examples of isentropic stars made of  $YN\Delta$ matter at finite temperature,  
  we show in Fig.~\ref{fig:isoentropic-stars} the MR relationships of 
  $Y_{Le}=0.1$ and $Y_{L\mu}={0.4}$ stars at a
  fixed entropy of $S/A=1$. (We note that in realistic models the entropy per baryon  may vary 
 within  the star in a time-dependent manner. Our approximation allows us to avoid any references to the dynamical modeling of a supernova or a BNS merger.) Three effects are readily observed for the neutrino-trapped cases: (a)
   the maximum masses of the stars are shifted towards larger values; (b)  the radii of the stars can be significantly larger than those of their cold counterparts; (c)  the increase in the radius is larger for larger lepton fractions.

\subsection{Matter at low densities and clustering}
\label{ssec:Clusters}

Below nuclear saturation density, the warm nuclear matter is composed of nucleons and clusters that are characterized by a mass number and a charge. Such matter is conveniently treated within the nuclear statistical equilibrium approach~\cite{Typel_PRC_2010,Botvina2010NuPhA,Raduta2010,Hempel2012ApJ,Hempel2015PhRvC,Gulminelli2012PhRvC,Gulminelli2015PhRvC,Furusawa2016,Avancini2017,Zhang2019PhRvC,Grams2018PhRvC,Raduta2019,Ropke2020,Pais2020}.  Besides light clusters, the pionization of matter at sufficiently high temperatures~\cite{Peres2013,Ishizuka2008,Colucci2014PhLB,Fore2020} and the formations of condensates of deuterons~\cite{Lombardo2001PhRvC,Sedrakian2006PhRvC,Sedrakian:2018ydt} and alpha-particles at low temperatures \cite{Wu2017JLTP,Zhang2017,Zhang2019PhRvC,Satarov2019PhRvC,Satarov2021PhRvC,Furusawa2020} have been studied. 

In the present context, one may ask about the possible role of hyperons and resonances in low-density matter containing clusters and condensates. As is well known, nuclear systems show rich nucleon-$\Delta$-pion dynamics, which were extensively explored in the context of heavy-ion physics. Similarly, the role of the quartet of $\Delta$-resonances and the isotriplet of pions, $\pi^{\pm,0}$, may be important in the treatment of the nucleons and clusters~\cite{Sedrakian:2020cjt} in warm dilute nuclear matter. In a first approximation, the full nuclear statistical ensemble can be approximated by the light clusters with the mass number $A\le 4$ and a heavy nucleus such as $^{56}$Fe. Since the hyperonic interactions are less important at low densities, the lightest hyperon (i.e., $\Lambda$) is expected to contribute predominantly to the strangeness content of matter. The low-density matter with hyperons
in the high-temperature regime was  studied in the supernova and BNS matter~\cite{Sedrakian:2020cjt} and low-temperature crusts of neutron stars~\cite{Menezes:2017svf,Fortin_PASA_2018}. Pions at high temperatures were also studied in the supernova context~\cite{Peres2013,Ishizuka2008,Colucci2014PhLB,Fore2020}. Furthermore, the possibility of the formation of light hypernuclei along with non-strange clusters was studied in Refs.~\cite{Cust2021PhRvC,Cust2022PhRvC}.
\begin{figure}[t]
\centering
\includegraphics[width=16cm]{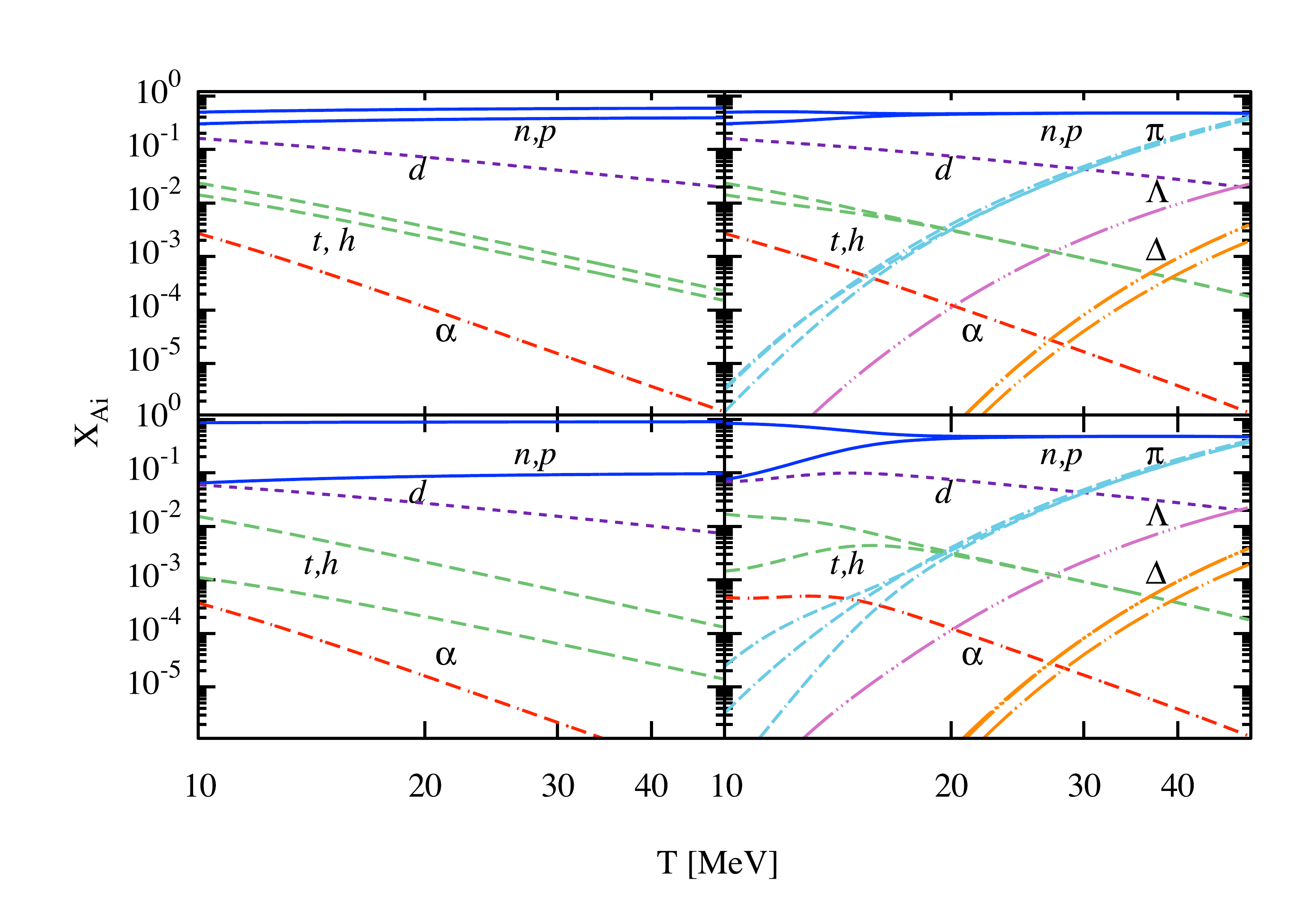}
\caption{Mass fractions of constituents as a function of temperature at 
 fixed density  $10^{-2}\rho_{\rm sat}$ for charge fractions
 $Y_Q = 0.4$ (top panels) and 0.1 (bottom panels)~\cite{Sedrakian:2020cjt}. The two
 left panels include only nucleons and light clusters, whereas the
  two right panels account for the full composition. The full composition includes neutrons $n$, protons $p$, deuterons $d$, triton $t$, helium $h$, $\alpha$-particles, $\Delta$-resonances, $\Lambda$-hyperon, and isotriplet of pions $\pi^{0,\pm}$. The increase of the mass fraction of heavy baryons with temperature is associated with their excitation by temperature, which helps to bridge the difference between their and nucleon rest masses which determines their relative statistical importance.}
\label{fig:1.15}
\end{figure}
\begin{figure}[t]
  \centering
  \includegraphics[width=16cm]{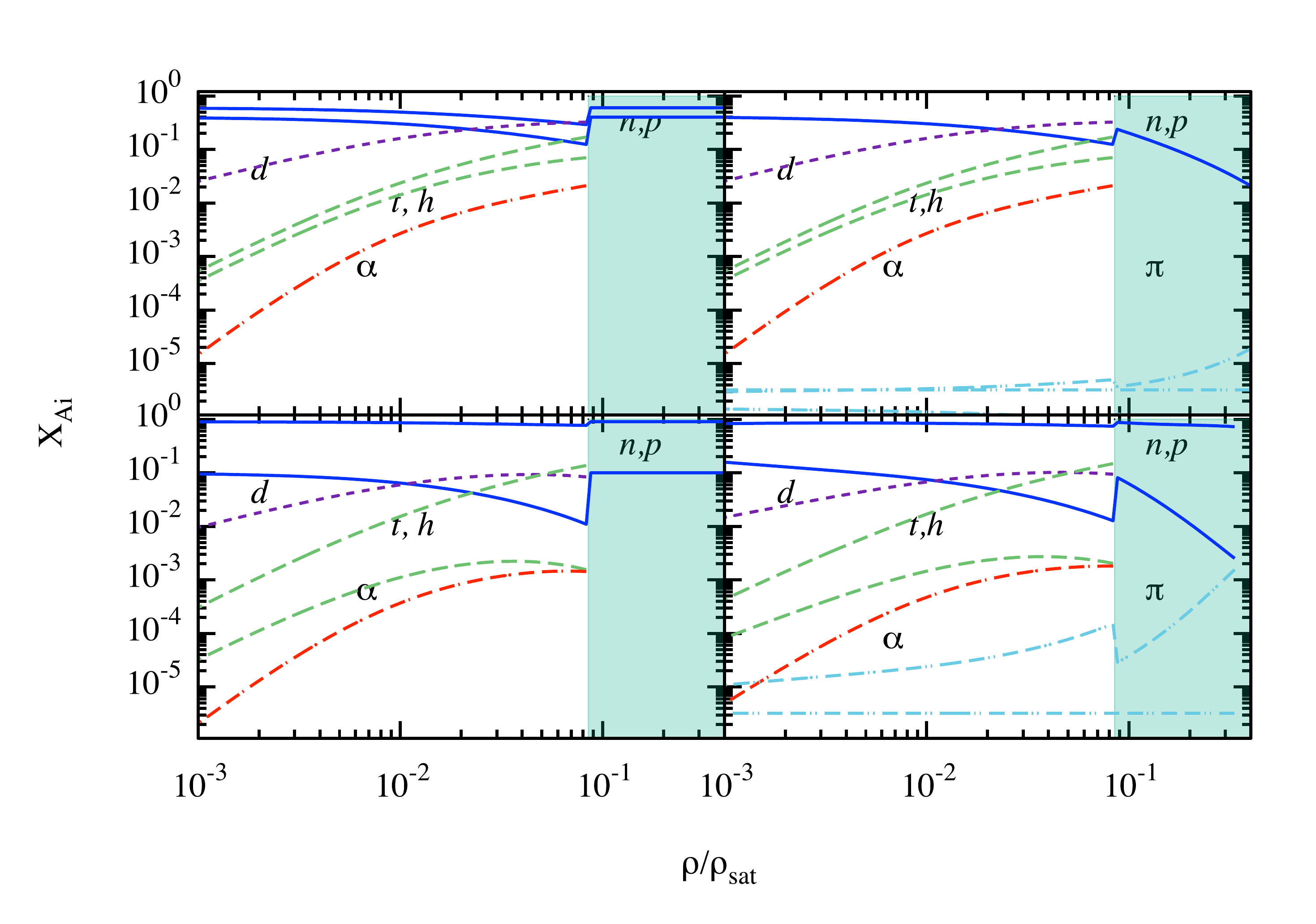}
  \caption{Dependence of the mass fractions of the particles in dilute
   nuclear matter on density, for $T=10$~MeV~\cite{Sedrakian:2020cjt}. The top and lower panels
   correspond to charge fractions $Y_Q = 0.4$ and 0.1. The left and
   right panels correspond to the cases containing only nucleons and
   light clusters and the full composition, respectively. The
   composition includes neutrons and protons (solid lines), deuterons
   (short-dashed), triton and helium (long-dashed), $\alpha$-particles
   (dash-dotted), $\Delta$-resonances (dash-double-dot),
   $\Lambda$-hyperon (dash-triple-dot), and pions
   (double-dash-dot). In the figure on the right-hand-side, the clusters disappear for
   $\rho/\rho_{\rm sat} \ge 9\times 10^{-2}$ (shaded area) due to the
   Pauli-blocking of the phase space. The mass fraction of
   $\isotope[56]{Fe}$ is not visible on the figure's scale.
}
\label{fig:cluster_abundances}
\end{figure}
Let us consider how the thermodynamics of a mixture of clusters is included in the description of  low-density nuclear matter. 
If a nucleus (cluster) is characterized by a mass number $A$ and a charge $Z$, its
chemical potential is expressed as
\be\label{eq:cheq_nuclei} \mu_{A,Z} = (A-Z)\mu_n + Z \mu_p.
\ee
For the chemical potentials of heavy baryons, the relations \eqref{eq:c1}-\eqref{eq:c3} still hold.  The chemical potentials of the pions obey the following relations
\bea \label{eq:cheq_pi}
\mu_{\pi^0} = 0,\qquad
\mu_{\pi^+} = \mu_p-\mu_n,\qquad
\mu_{\pi^-} = \mu_n-\mu_p.
\eea
The baryon number density $\rho$ (the subscript $B$ is dropped for simplicity) and the charge neutrality conditions are changed to the following form,
\begin{subequations}
\bea
\label{eq:n_b}
\rho&=&\rho_n+\rho_p+
\sum_{c}A_c\rho_c +\rho_{\Delta^{++}}+\rho_{\Delta^{+}}
+\rho_{\Delta^-}+\rho_{\Delta^0}+\rho_{\Lambda}, \\
\label{eq:Y_Q}
\rho Y_Q  &=& \rho_p + \sum_{c}Z_c\rho_c+2\rho_{\Delta^{++}}+\rho_{\Delta^{+}}
-\rho_{\Delta^-}+\rho_{\pi^{+}}-\rho_{\pi^-},
\eea
\end{subequations}
where the summations sum the densities of the deuteron ($d$), triton ($t$), $\isotope[3]{He}$ ($h$), $\alpha$-particle and $\isotope[56]{Fe}$ nucleus. Equations~\eqref{eq:n_b} and \eqref{eq:Y_Q} determine the two unknown chemical potentials $\mu_{n}$ and $\mu_{p}$ at a given temperature $T$ for fixed values of $\rho$ and $Y_Q$.

As an illustrative example, we show in Fig.~\ref{fig:1.15} the composition of dilute nuclear matter at
temperatures in the range of $10\le T\le 30$~MeV and at fixed density of $10^{-2}\rho_{\rm sat}$ for charge fractions $Y_Q = 0.4$ (characteristic for supernova matter) and $Y_Q = 0.1$ (characteristic for BNS mergers)~\cite{Sedrakian:2020cjt}.  The clusters' abundances are characterized by a mass fraction $X_i = A_i \rho_i/\rho$, where $A_i$ is the mass number of a constituent and $\rho_i$ is its number density. At the temperatures considered here, the mass fraction of $^{56}$Fe is not visible on the figure's scale; it becomes significant only at low temperatures of the order of a few MeV~\cite{Wu2017JLTP}.  The main observation is that there is a transition from light baryons and clusters to heavy-baryon-containing matter at temperatures of about $T_{\rm tr}\sim 30$ MeV. Once heavy baryons and pions are included the isospin asymmetry in the neutron and proton components is reduced. This affects the abundance of the helion and triton, which are then closer together. The transition temperature itself is dependent on the treatment of the interactions and assumed composition, but this appears to be a generic feature. The emergence of heavy baryons at high temperatures is due to their excitation by temperature which increases their statistical weight compared to nucleons.

The density dependence of the particle abundances at fixed $T=10$~MeV is shown in Fig.~\ref{fig:cluster_abundances}. At a density of $\rho/\rho_{\rm sat} \simeq 0.1$ the clusters abruptly disappear as a consequence of the vanishing of their binding energy with increasing density. This manifests itself in a jump in the density of nucleons and is also reflected in the pion mass fraction.  Heavy baryons do not appear at these low temperatures. The abruptness of the transition due to the vanishing of the cluster binding energies depends on the details of how the Pauli-blocking effects are modeled, in particular on the dependence of the cluster binding energies on the center-of-mass momentum of a cluster, which appear to be smooth in some models~\cite{Typel_PRC_2010}.

In closing, let us mention some of the model-independent aspects of the studies: (a) if heavy nuclei are taken into account, they act to suppress the abundances of light clusters at low temperatures,
but are negligible at high temperatures; (b) the phase-space occupation effectively suppresses the abundances of light clusters at sufficiently low temperatures and for densities roughly $\ge 0.1\rho_{\rm sat}$ due to the Pauli blocking~\cite{Ropke2020,Hempel2015PhRvC,Pais2020}; for experimental studies of signatures of Pauli blocking in heavy-ion collisions see, e.g., Ref.~\cite{Qin2012PhysRevLett}; (c) at low enough temperatures, deuterons can cross-over from a Bose-Einstein to Bardeen--Cooper--Schrieffer pair condensate~\cite{Lombardo2001PhRvC,Sedrakian2006PhRvC,Sedrakian:2018ydt} as the density increases.

The discussion above suggests that the composition of matter in supernovae and BNS mergers can be much richer than studied so far. Additional degrees of freedom, such as heavy baryons or thermally excited mesons affect the neutrino interactions with the matter by providing additional scattering centers and channels of neutrino emission. The presence of heavy baryons may also affect the transport coefficients, such as the bulk viscosity, which is relevant for the damping of the post-merger oscillations~\cite{Alford2021b,Arus2022Particles}.

\section{Pairing in Hypernuclear Matter}
\label{sec:Pairing} 

\subsection{Pairing patterns in compact stars}
Bardeen--Cooper--Schrieffer (BCS) type pairing in compact stars is a complex problem because
of the lack of information about the nuclear interactions at relevant densities and intricate
many-body effects, such as the screening of the nuclear interaction. For the nucleonic components,
in a first approximation, the bare nucleon-nucleon (NN) interaction can be used as the pairing
force in the dominant attractive partial wave channel. Assume a single-component, isotropic, and
homogeneous Fermi system with attractive interaction $v_0(p,p')$ in the $S$-wave, where 
$p$ and $p'$ are the magnitudes of the relative incoming and outgoing momenta of the particles.
According to the BCS theory, the pairing gap $\Delta_{p}$ in the quasiparticle spectrum is given, at the mean-field level, by the  gap equation
\begin{equation}\label{eq:gap_swave}
\Delta_{p} = -
\frac{1}{V}\sum_{\vecp'} v_0(p,p')
\frac{\Delta_{p'}}{2E_{p'}} (1-2f_{p'}),
\end{equation}
where $V$ is the volume and $f_{p} = (e^{E_p/T}+1)^{-1}$ is the Fermi--Dirac
distribution function at temperature $T$.
The quasiparticle energy in the paired states is given  by
 $E_p = \sqrt{\varepsilon_p^2+\Delta_p^2}$, where $\varepsilon_p$ is the quasiparticle
 spectrum in the normal state. For nucleonic pairing at low densities, the 
 BCS calculations can be performed with NN-interaction models that
 fit the scattering data with high precision, i.e., by the so-called ``realistic'' NN interactions
 or chiral interactions to sufficiently high order. Going beyond the bare driving interaction
 entails including such effects as screening of the pairing interaction, which sums up the
 infinite series of loops that describe the screening effects in dense matter. The self-energy
 effects (renormalization of the mass of the fermion participating in the pairing) are also included
 in the approximation of the quasiparticle spectrum $\varepsilon_p$. It is important to note that both
 the pairing interaction and the (renormalized) quasiparticle spectrum are strongly dependent
 on the momentum of the particle. Therefore, solutions based on contact-type interactions
 cannot be applied. The dominant pairing channel in the case of nucleons can be identified from the
 inspection of the phase shifts derived from the analysis of different partial-wave channels $^{2S+1}L_J$
 of the two-nucleon scattering problem. Here $S$, $L$, and $J$ refer to the total spin, orbital angular momentum, and
 total angular momentum of the two-nucleon system. Because of the large disparity between the Fermi energies
 of neutrons and protons, the $T=1$ isospin pairing dominates in neutron-rich matter. The $T=0$ pairing, which competes with $T=1$ pairing in more isospin symmetric systems (nuclei,
 low-density nuclear matter away from beta-equilibrium), is not relevant to our discussion.

 To summarize, in compact stars at low densities, the $T=1$ pairing is
 due to the attraction in the $^1S_0$ partial-wave channel and may
 lead to the formation of neutron--neutron and proton-proton Cooper
 pairs. At higher densities, such as the nuclear saturation density,
 the $^3P_2$--$^3F_2$ coupled partial wave is the dominant attractive
 channel for neutron--neutron pairing as the $^1S_0$-wave interaction
 becomes repulsive.  The peculiarity of the pairing in this state is
 that Cooper pairs have total spin $S=1$ and that there is a multitude
 of ground states which differ by the values of projection of the
 orbital angular momentum.  If at higher densities compact star cores
 develop equal populations of neutrons and protons, $S=1$ and $T=0$
 pairs may be formed in the $^3D_2$ channel, which applies exclusively
 to neutron--proton scattering. For a more detailed discussion of the
 pairing patterns in neutron stars see Ref.~\cite{Sedrakian2019EPJA}.

 \subsection{Hyperonic pairing in compact stars}
\label{ssec:H_papring}

 By analogy to the nucleonic component in compact stars, we anticipate that
 the attractive component of the nuclear force between hyperons at low energies
 can lead to the formation of Cooper pairs made of hyperons. It is immediately clear that
 hyperon-nucleon pairing will be suppressed by the large difference in their masses unless, under some special conditions, their Fermi energies match due to the disparity in their populations.
 We first concentrate on a low-density hyperonic component where the attraction comes
 from the interaction in the $^1S_0$ partial-wave channel. In this case, in complete analogy to
 neutrons and protons, hyperons will form spin-singlet $S=0$ Cooper pairs.
Since the problem of determining the pairing force (i.e., the model for $v_0(p, p')$) can be  decoupled in a first approximation 
 from the problem of determining the quasiparticle spectrum, it has been argued
 that the gap can be determined from a non-relativistic BCS equation with a chosen two-nucleon potential, where the single-particle energies and particle composition are computed from CDFs.
 In the case of nucleonic matter, such an approach was applied first in Ref.~\cite{Kucharek_1991}
 and has been validated by the computations of finite nuclei within the relativistic Hartree--Fock--Bogoliubov theory~\cite{Long2010a,Lijj:2015}. This mismatch of the ingredients
 parallels the many-body computations in non-relativistic models where a similar
 decoupling between the pairing interaction and self-energies have been employed.
 
 The pairing among $\Lambda$ hyperons was studied early on in Ref.~\cite{Balberg1998}.
 More recent work considered particle backgrounds that were derived from CDFs~\cite{Wang_PRC2010}.
 The pairing among $\Xi$ hyperons has been mentioned in Ref.~\cite{Takatsuka_XiPairing},
 but its implications on physical observables remained unexplored.  The $S$-wave pairing between types of hyperons was studied within such an approach in Ref.~\cite{Raduta_2018}, i.e., the composition was taken from a CDF based on the Hartree approximation with density-dependent couplings together with $\Lambda\Lambda$ pairing interaction taken to be the configuration space parameterization of ESC00 potentials~\cite{Rijken_2001}, as presented by Ref.~\cite{Filikhin_NPA2002}. In the case of the pairing among $\Xi^{-}\Xi^{-}$ and $\Xi^{0}\Xi^{0}$ hyperons, the interaction was taken from Ref.~\cite{2016PhRvC..94b4002G}. It is based on the Nijmegen Extended Soft Core ESC08c potential~\cite{Rijken2013}.  Note that the potentials were chosen in Ref.~\cite{Raduta_2018} to maximize the attraction between $\Lambda\Lambda$ and $\Xi\Xi$, therefore the obtained gaps are likely to be upper bounds. Pairing in the $\Sigma \Sigma$ channel can, in a first approximation, be neglected because there is no solid evidence for attraction in this channel at low energies. This is also reflected in the fact that the ESC08c potential for the $\Sigma \Sigma$ channel is repulsive.
\begin{figure}[tb]
\begin{center}
\includegraphics[width=0.9\columnwidth]{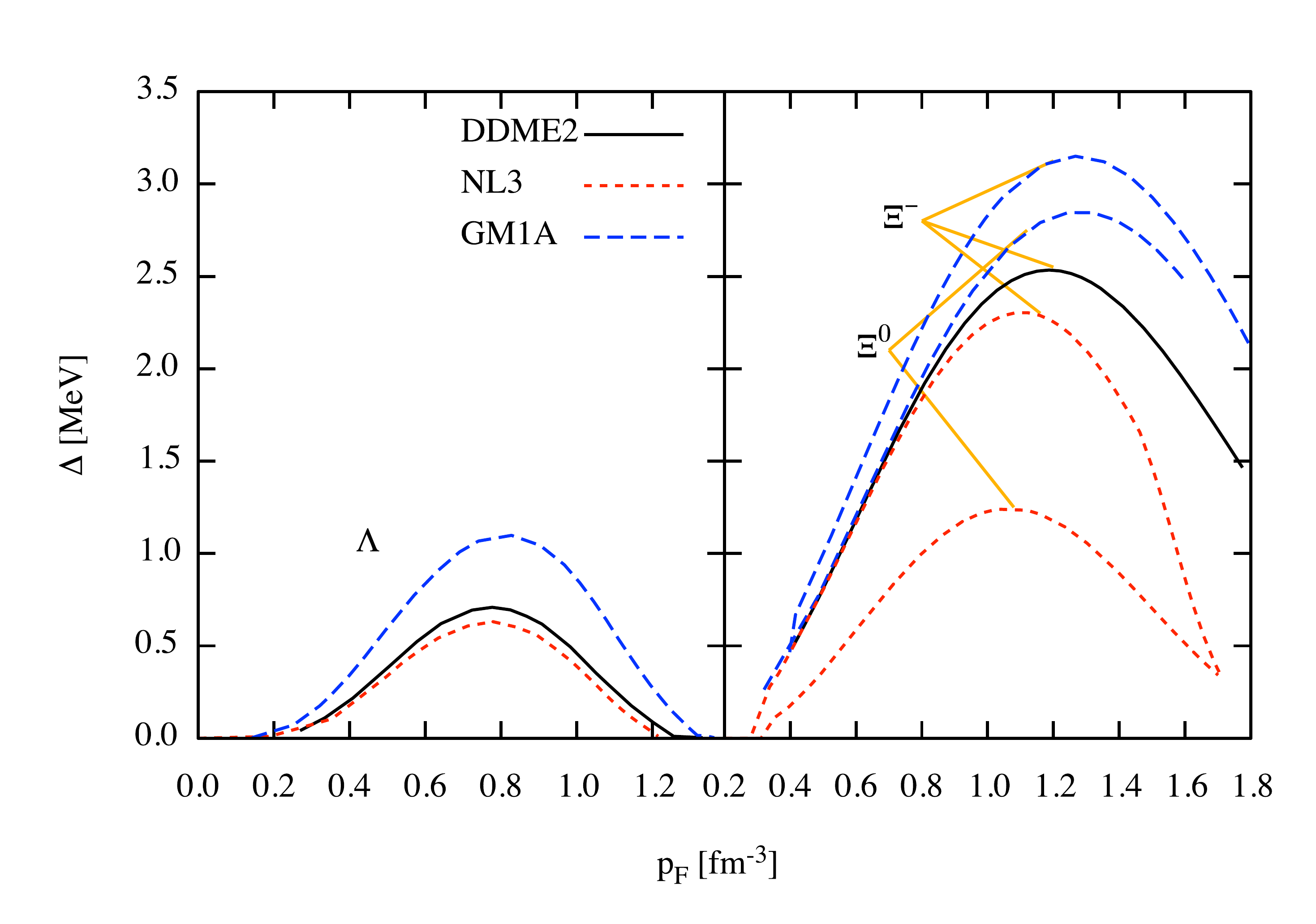}
\end{center}
\vspace{-1.2cm}
\caption{Dependence of $^1S_0$ hyperon pairing gaps on Fermi momentum, according to Ref.~\cite{Raduta_2018}. Three 
compositions are considered based on the DDME2 (solid), NL3 (dashed), and GM1A (long-dashed) CDFs. The left panel shows 
the $\Lambda$ hyperons gaps, the right one --- $\Xi^{0,-}$ hyperon gaps. In the case of the DDME2 CDF, the $\Xi^{0}$ hyperon is absent in matter.}
\label{fig:Gaps_pF}
\end{figure}

 For isotropic $S$-wave pairing, the gap equation for a specific hyperon ($Y$) channel  reads
\begin{equation}
\label{eq:gap_hyp}
\Delta_{Y} (p)=-\frac{1}{4 \pi^2} \int dp' p'^2 \frac{V_{YY}(p,p') 
\Delta_{Y}(p')}{\sqrt{\left[E^Y(p')-\mu_Y \right]^2+\Delta_{Y}^2(p')}},
\end{equation}
where $E(p)$ denotes the single-particle energy of hyperon $Y$. For a Hartree CDF, this
energy is given by
\begin{equation}
E^Y(p)=\sqrt{ p^2+m_Y^{*2}}+g_{\omega Y} 
\omega+g_{\phi Y} \phi+g_{\rho Y} \tau_{3Y} \rho+\Sigma_R .
\label{eq:E_p}
\end{equation}
The quantity $\mu_Y$ is the chemical potential and $m_Y^*$ is the Dirac
effective mass of hyperon $Y$.  In the case of $^1S_0$ pairing, the interaction
in momentum space is related to that in configuration space
$V_{YY}(r)$  via the relation
\begin{equation}
  V_{YY}(p,p')=\langle p | V_{YY} | p'\rangle= 4 \pi \int dr r^2 j_0(p r) V_{YY}(r) j_0(p' r),
\label{eq:V_hyp}
\end{equation}
where $j_0(p r)=\sin(pr)/(pr)$ is the spherical Bessel function of order zero.  Solution of the gap equation \eqref{eq:gap_hyp} are shown in Fig.~\ref{fig:Gaps_pF} for several different choices of background CDFs --- the DDME2 model discussed above, as well as the non-linear models NL3 of Ref.~\cite{Miyatsu_PRC2013} and GM1A of Ref.~\cite{Gusakov_MNRAS2014}. The use of different background models allows us to assess the effects of different compositions of matter on the pairing gaps, which are derived from the same pairing interaction.  The shapes of the pairing gaps resemble those of the nucleonic gaps: the increase of the gaps at low densities is caused by the increase in the density of states; the drop-off at high densities is caused by the decrease of the pairing force at high densities in the $S$-waves. The reduction of the hyperon masses at high densities also contributes to the reduction of the gaps as the density increases. The stronger reduction of the $\Lambda\Lambda$ pairing gap compared to the $\Xi^-\Xi^-$ and $\Xi^0\Xi^0$ gaps can be attributed to the stronger reduction of the effective mass of $\Lambda$s.  The $\Lambda$ $S$-wave pairing gap vanishes at densities that are relevant phenomenologically. At higher densities, $P$-wave $\Lambda$-pairing is a possibility which will affect the cooling behavior of hypernuclear stars~\cite{Raduta_2019}. The pairing of $\Lambda$s, as well as protons in the high-density regions of neutron stars, remains however an open issue.

\begin{figure}[tb]
\begin{center}
\includegraphics[width=0.75\columnwidth]{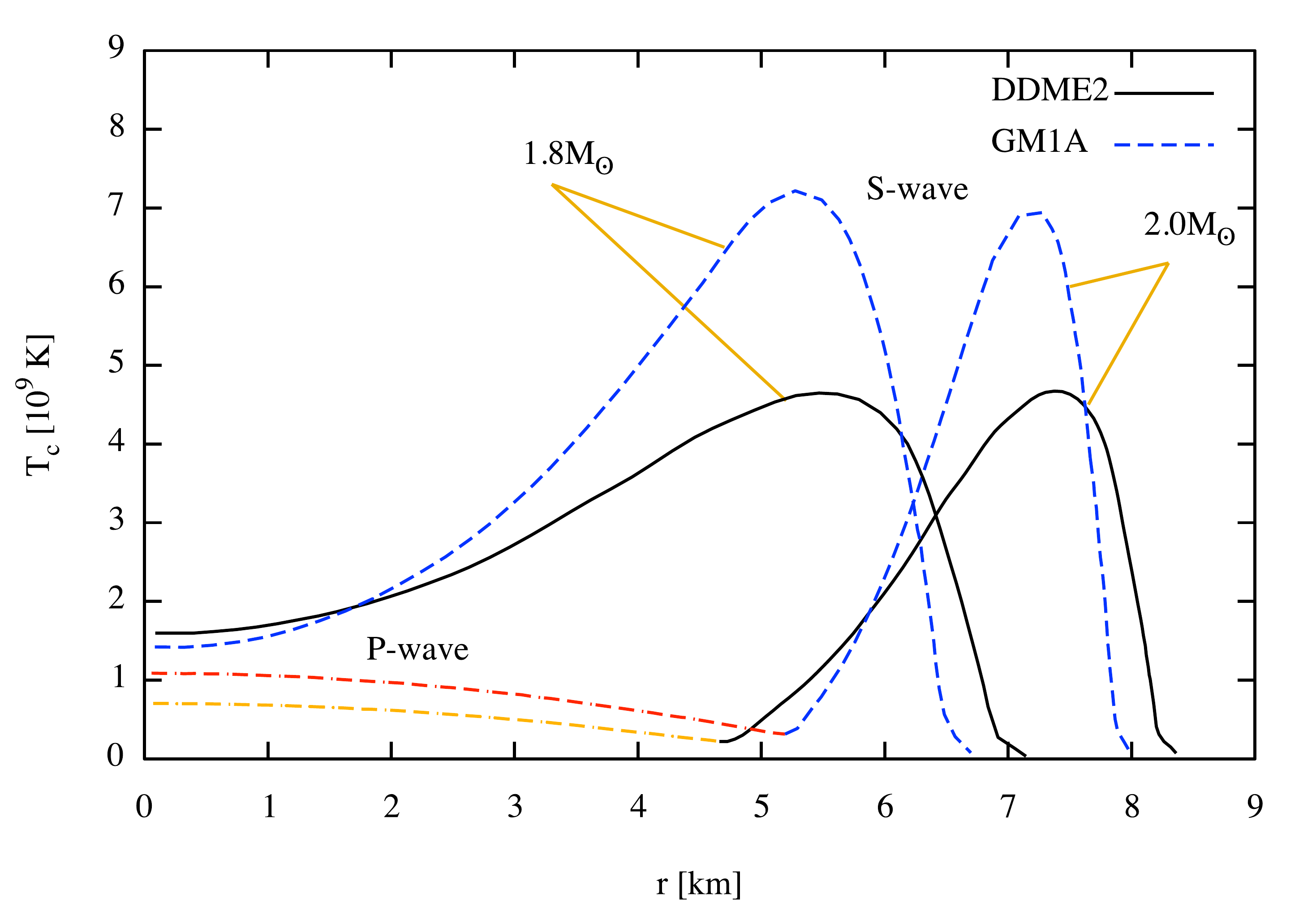}
\end{center}
\caption{Critical temperature of $\Lambda$ superfluidity
as a function of radial distance inside of $1.8M_{\odot}$ and $2\,M_{\odot}$ mass stars for two CDFs~\cite{Raduta_2019}. $P$-wave pairing appears in the $2\,M_{\odot}$ mass star, as shown by the dash-dotted lines. }
\label{fig:TcL}
\end{figure}

Qualitative estimates of the gaps of baryons in higher partial waves can be
made using the knowledge of the $P$-wave phase-shifts and gaps in neutron
matter~\cite{Raduta_2018,Raduta_2019}. In the case of protons, the isospin invariance of nuclear
forces suggest that $^3P_2$-$^3F_2$ pairing should emerge for a sufficiently
dense proton component.  A simple rescaling of the density of states allows
one to obtain the proton $^3P_2$-$^3F_2$ pairing gap, which amounts to 
replacing the neutron Landau mass $m_n^*$ by the proton Landau mass
$m_p^*$, which results in a rescaling of the  dimensionless pairing
interaction by the factor $\alpha_p^{-1} = m^*_p/m^*_n$. As a result, 
the weak-coupling estimate for the $P$-wave pairing gap for protons
can be written as 
\begin{eqnarray}
\label{eq:gap_ratio_p}
\Delta_{p} = E_{{F_p}}
 \left(\frac{\Delta_n}{E_{{F_n}}}\right)^{\alpha_p},
\end{eqnarray}
where $E_{{F_i}}$ is the Fermi energy of baryon $i$. This procedure provides an elementary estimate of the $P$-wave pairing gap for protons under the assumption that the BCS weak-coupling formula 
$
\Delta_{i} = \epsilon_{Fi} \exp[-1/(\nu_i V_i)]
$
can be applied. Here $\nu_i$ and $V_i$ denote the density of states and the
pairing matrix element of a baryon of type $i$.

Similar estimates can be made for $P$-wave pairing among $\Lambda$'s.  Applying the same argument as in the case of protons and additionally rescaling the pairing interaction among neutrons by a factor of $2/3$ following the spirit of SU$(6)$ spin-flavor quark model.  Combining all the factors one finds that the dimensionless pairing interaction is rescaled by the factor $\alpha_{\Lambda}^{-1} = 2m_\Lambda^*/3m_n^*$ and, consequently,
\begin{eqnarray}
\label{eq:gap_ratio_Lambda}
\Delta_{\Lambda} = E_{{F_\Lambda}}
\left(\frac{\Delta_n}{E_{F_n}}\right)^{\alpha_{\Lambda}}.
\end{eqnarray}
The inclusion of the higher partial-wave pairing may have an important influence on the cooling of compact stars, as will be discussed later in Sec.~\ref{sec:Cooling}. However, we point out here already that for the CDF parameterizations
DDME2, NL3 and GM1A studied in Ref.~\cite{Raduta_2019}, the $^3P_2$-$^3F_2$ proton superfluidity has only a marginal
effect on the cooling, since in the innermost regions of the stars the critical temperature is below the characteristic
stellar temperature.
In the case of $\Lambda$s, their concentration could be large enough so that one finds
a $P$-wave $\Lambda$ superfluid in the inner core of a compact
star. Figure~\ref{fig:TcL} illustrates the critical temperature of $\Lambda$ pairing
inside of a $2M_{\odot}$ star for two models for the EoS. 

Better models for the $P$-wave pairing of protons and $\Lambda$s
are needed to draw more definitive conclusions about the role of
superfluidity in the deep interiors of baryonic compact stars. 

\section{Cooling of Hypernuclear Stars}
\label{sec:Cooling}

\subsection{Neutrino radiation reactions}
\label{ssec:Nu_radiation}

The long-term cooling of compact stars serves as a sensitive indicator of their
  interior composition~\cite{Shapiro:1983du,SCHAAB1996531,PAGE2006497,Page2009ASS,Sedrakian2007PrPNP,Potekhin2010PhyU}.
It is characterized by neutrino emission from the bulk of the stellar interior during the first $\sim 10^5$ years after the star's birth, which is followed by late-time photon cooling from the surface, assuming that there is no internal heating mechanism operating at any stage of evolution. With the advent of CDFs, which were tuned to reproduce the available astrophysical and laboratory data, it became possible to perform simulations of cooling of compact stars with hyperonization in a more constrained way than was ever possible before~\cite{Raduta_2018,Raduta_2019,Grigorian:2018bvg,Negreiros:2018cho,Fortin:2021umb}.

Hyperonic matter cools via the direct Urca (dUrca) processes~\cite{Prakash:1992zng},
\begin{subequations}
\begin{eqnarray}
\label{eq:UrcaLambda} 
\Lambda &\to& p + l  + \bar\nu_l,\\
\label{eq:UrcaSigmaminus}
\Sigma^- &\to& \left(\begin{array}{c} n  \\
                       \Lambda \\
\Sigma^0
\end{array} \right) + l + \bar\nu_l,\\
\label{eq:UrcaXiminus}
\Xi^- &\to& \left(\begin{array}{c} \Lambda  \\ 
\Xi^0
                        \\
\label{eq:UrcaSigmazero}
\Sigma^0 \end{array} \right) + l + \bar\nu_l,\\
\label{eq:Xizero}
\Xi^0 &\to& \Sigma^+ + l  + \bar\nu_l,
\end{eqnarray}
\end{subequations}
where $l$ stands for a lepton, either electron or muon, and
$\bar\nu_l$ is the associated anti-neutrino.  

There exist density thresholds for these processes to operate, which are dictated by the kinematics involved in these reactions. But these densities are very low for hyperons, i.e., the reactions start operating at a density that is slightly above the onset density for a hyperon participating in a reaction. Baryon pairing is known to suppress the rates of the Urca (and other baryonic) processes. Therefore, another unknown in the cooling simulations is the magnitude of the gaps in the spectra of various hyperons~\cite{Raduta_2018,Raduta_2019}.

In addition to the dUrca processes, the neutrino emission via the Cooper pair-breaking and formation (PBF) mechanism 
\begin{eqnarray}
\label{eq:Y_PBF}
\{YY\} \to Y+Y + \nu + \bar \nu, \quad 
Y+Y\to \{YY\} + \nu + \bar \nu, 
\end{eqnarray}
contributes to neutrino emission, where $\{YY\}$ stands for a hyperonic Cooper pair. Note that this charge neutral reaction produces neutrino--anti-neutrino pairs of all flavors. One needs to include the  PBF processes also for the neutron $S$- and $P$-wave superfluids and proton $S$-wave superconductor. The expression for the $S$-wave $\Lambda$ condensate emission can be easily obtained from the analog expression for the neutrons with appropriate changes in the condensate parameters and the weak charges, see Ref.~\cite{Raduta_2018}. The emission from the $P$-wave hyperonic condensate can be obtained, in a similar fashion, from the emission rate derived for neutrons in Ref.~\cite{Leinson2017}.

\subsection{Cooling tracks}
\label{ssec:CTracks}

Despite the uncertainties in the pairing gaps for baryons, especially in the high-density patterns of pairing, it is possible to obtain some generic insights into the cooling of compact stars~\cite{Raduta_2018,Raduta_2019}. To quantify the cooling evolution models that are based on an alternative parameterization of the DDME2 CDF were used
with the $\sigma$ meson couplings
to hyperons defined by $R_{\sigma\Lambda} = 0.6154$, $R_{\sigma\Sigma} = 0.474$, and $R_{\sigma\Xi} = 0.3259$. This, alternative, choice leads to larger populations of $\Sigma$-hyperons~\cite{Raduta_2018}.
As a consequence, $\Sigma^-$ appears in larger amounts and at lower densities than in the models discussed in Sec.~\ref{sec:Hyper_DFT}.

There is a mass hierarchy with respect to cooling behavior. The lightest stars that contain hyperons with $M/M_{\odot} \gtrsim 1.5 $ cool via the Urca process $\Lambda\to p + l +\bar\nu_l$.  The appearance of the $\Xi^-$ in slightly more massive stars opens a competing cooling channel via $\Xi^-\to \Lambda + l +\bar\nu_l$, the degree of its efficacy depending on the pairing gaps of the $\Xi^-$.  For very massive stars with $M/M_{\odot}\sim 2 $, the $S$-wave gaps of protons and $\Lambda$'s will vanish at high densities [because the interactions will become repulsive at high (Fermi) energies]. There are three alternatives in this case: (a) both components are non-superfluid and therefore the Urca process involving $\Lambda$s and protons \eqref{eq:UrcaLambda} will be the dominant one; (b) both components are superfluid, and there is a competition between processes on $\Lambda$s and $\Xi$s; (c) the third possibility, which is the more likely case, would be that the proton $P$-wave gap is too small to be relevant so the suppression of neutrino emission is solely due to $\Lambda$ pairing and the process \eqref{eq:UrcaLambda}  dominates as in the case (a).  Note that here and in all models that will be discussed below the $(\Sigma^-, n)$ channel \eqref{eq:UrcaSigmaminus}  is energetically forbidden.  This is not surprising given the large difference in their abundances.

\begin{figure}[tb]
\begin{center}
\includegraphics[angle=0,width=0.9\columnwidth]{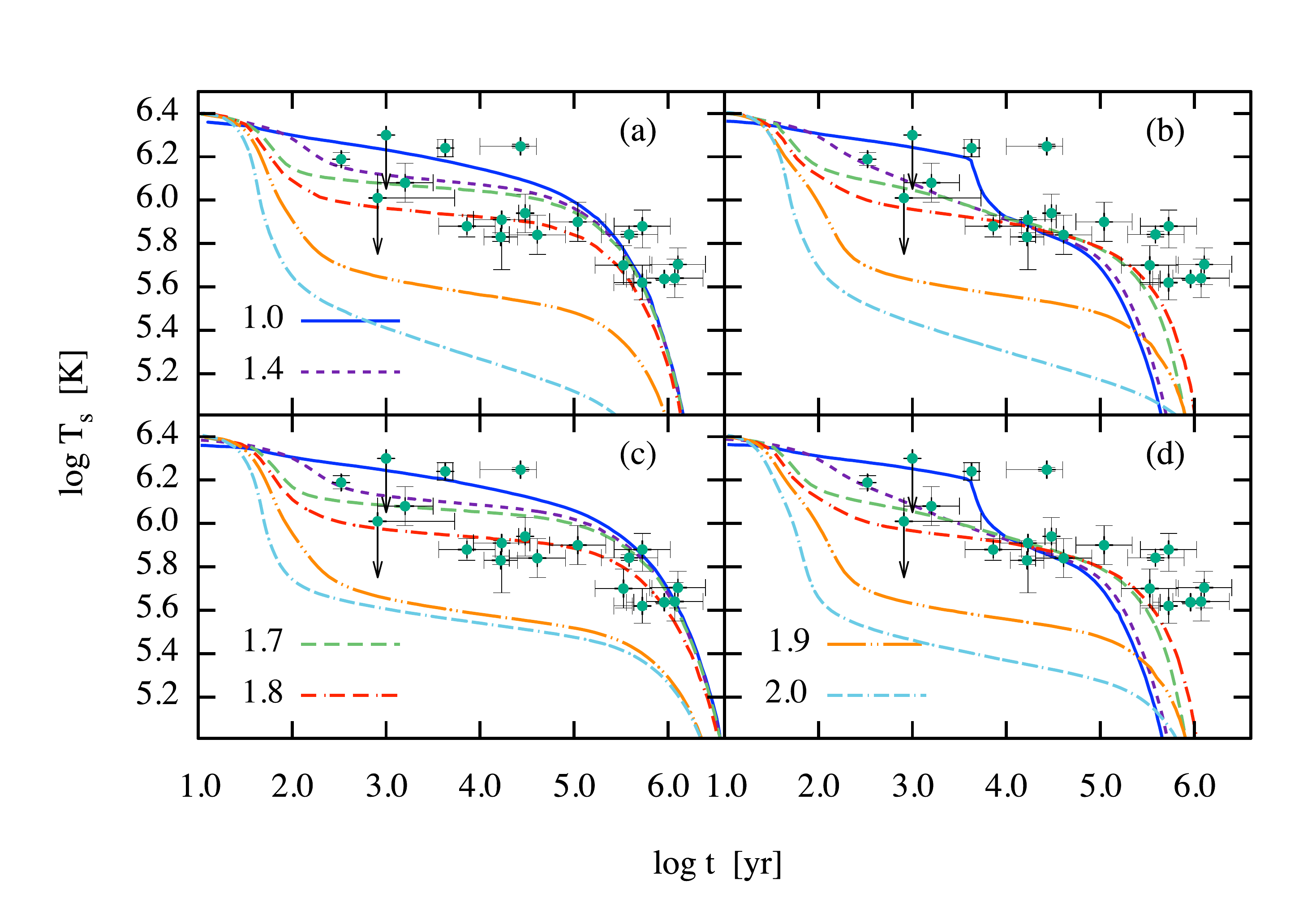}
\end{center}
\vspace{-1.2cm}
\caption{Dependence of surface temperature of compact stars on their
  age for models that are based on the DDME2 CDF EoS with masses in
  the range $1\le M/M_{\odot} \le 2$~\cite{Raduta_2019}, as indicated
  in the plot. The left panels (a) and (c) use the neutron $^1S_0$
  gap from Ref.\cite{SFB_2003}, proton $^1S_0$ gap from
  Ref.\cite{Chen_NPA1993}, and we assume that the neutron $P$-wave
  pairing gap is zero. The right panels use the same proton gap, but
  the $S$- and $P$-wave neutron pairing gaps are from
  Ref.~\cite{Ding_PRC_2016}.  The $\Lambda$ $^1S_0$-pairing gaps are
  included in all panels and are based on the BCS computations of
  Ref.~\cite{Raduta_2018}. The $\Lambda$ $^3P_2$-gaps are included in
  panels (c) and (d) and are obtained from a scaling procedure
  described in Ref.~\cite{Raduta_2019}, which uses the neutron
  {$P$}-wave results of Refs.~\cite{Baldo_PRC58} in panel (c)
  and those of Ref.~\cite{Ding_PRC_2016} in panel (d). For the
  observational data and references, see Fig.~1 of
  Ref.~\cite{Beznogov2015}. The arrows indicate upper
  limits. }
\label{fig:Teff_DDME2}
\end{figure}

Cooling simulations for DDME2 CDF are shown in Fig.~\ref{fig:Teff_DDME2} to illustrate some generic features that are observed also for other CDFs.
For this CDF model, the allowed neutrino emission channels 
  as a function of density are as follows:
\begin{itemize}
\item the $(\Lambda, p)$ process \eqref{eq:UrcaLambda} is
  active in the density range $0.34 \leq \rho_b \leq 1.02$ fm$^{-3}$
  for stars with $M/M_{\odot} \geq 1.4$,
\item the $(\Xi^-, \Lambda)$ channel from Eq.~\eqref{eq:UrcaXiminus} is active
  in the range $0.37 \leq \rho_b \leq 0.98$ fm$^{-3}$ for stars with 
  $ M/M_{\odot} \geq 1.54$,
\item the $(\Sigma^-, \Lambda)$ channel from Eq.~\eqref{eq:UrcaSigmaminus} is
  active in the range $0.39 \leq \rho_b \leq 0.60$ fm$^{-3}$ for stars with masses
  in the range $1.6 \leq M/M_{\odot} \leq 2$.
\end{itemize}
The interior-to-surface temperature relation adopted in these
simulations assumes that the surface is made of iron, which predicts a
lower surface temperature for a given interior temperature than is the
case for a light--element atmosphere.  However, it is expected that
young compact objects may have some light components in their
atmospheres such as hydrogen, helium, or carbon. For example, the
compact central object (CCO) in Cassiopeia A with an estimated age of $t
\sim 330 $~yr, or XMMU J173203.3-344518 with $t \sim 2.7 \times
10^3$~yr, are expected to be such stellar objects. Therefore, these
simulations are likely to underestimate the actual temperatures of
young stars.

  The cooling tracks corresponding to the constant mass models within the mass range $1\le M/M_{\odot}\le 1.85$ cover the observed range of temperatures well~\cite{Raduta_2018,Raduta_2019}, thus showing a clear mass hierarchy of the cooling behavior of compact stars in the neutrino cooling era, $ t\le 10^5$ yr. Interestingly, this hierarchy is inverted at the very early ($t\le 10$ yr) stages of thermal evolution, which is however observationally insignificant.

Let is consider some more specific features of cooling tracks. An inspection of the simulation curves shows that switching on and
off neutron $P$-wave  pairing has an important effect on the cooling of
  low-to-intermediate mass stars, as this pairing significantly reduces
  the heat capacity of the core and induces one of the dominant
  cooling processes via PBF.  For the late cooling era of $t\ge 10^5$~yr,
  one finds that the cooling tracks computed for zero neutron $P$-wave
  pairing gaps are in better agreement with the data than the cooling
  tracks computed for non-zero gaps. 
Let us turn to the role of the hyperonic component. As discussed above, the hyperonic dUrca processes start to operate in stars with masses above the canonical mass of $M/M_{\odot} \geq 1.4$.  Thus, stars in the mass range $1.4 \leq M/M_{\odot} \leq 1.6$ would cool down via the $(\Lambda, p)$ channel, 
which is the only active dUrca process in their interiors.  For more massive stars, $M/M_{\odot} \geq 1.7$, several reactions are triggered of which the 
$(\Sigma^-,\Lambda)$ channel is the dominant one.  Nevertheless, the
  cooling of stars with $M/M_{\odot} \leq 1.85$ is not that fast because of the presence of a  proton and/or $\Lambda$ superfluidity.

For stars with a larger mass, the central density exceeds the density at which the $\Lambda$ $S$-wave gap disappears.  In this case, the stars cool fast via the $(\Sigma^-,\Lambda)$ dUrca processes, because at high densities $\Lambda$s cannot pair in the $S$-wave state.  However, if one allows for a higher $P$-wave pairing of $\Lambda$s, this slows down the
  cooling of the $M/M_{\odot}=2$ model, as seen in
  Fig.~\ref{fig:Teff_DDME2}. The $\Lambda$ $P$-wave pairing scaled
  from the neutron $P$-wave pairing results of Ref.~\cite{Ding_PRC_2016} (right
  panel) is less effective than the one obtained from that of 
  Ref.~\cite{Baldo_PRC58} (left panel).  The reason is that the latter
  spans a wider range of densities and its maximum is located at a
  larger value.

Given the mass hierarchy, the observations of the surface temperatures of neutron stars can be explained by the variation of their masses within their population (the light objects are hot, and the heavier ones are cold). Note that the massive models may not develop normal cores of hyperons due to the possible pairing in the $P$-wave channel in high-density matter, in which case fast cooling will not take place~\cite{Raduta_2019}. The pairing in the hyperonic sector remains the main unknown for cooling simulations of hypernuclear stars. Given this uncertainty, some studies neglect the hyperonic pairing altogether~\cite{Grigorian:2018bvg,Negreiros:2018cho}.

There are uncertainties in the studies of the cooling of neutron stars that are unrelated to the hyperonic component, which we list here for completeness. These include the composition of the atmosphere~\cite{Potekhin:2020ttj}, which substantially affects the surface temperature of a star, and the pairing gaps of neutrons and protons in the domains where interactions are attractive~\cite{Sedrakian:2018ydt}. Large magnetic fields are a contributing factor, too, as they dissipate sufficient energy to heat up a star~\cite{Vigano:2021olr}, which can counterbalance the fast cooling by Urca processes~\cite{Amzuini_2022a_MNRAS,Amzuini_2022b_MNRAS}. Additional heating may arise due to the frictional effects between the superfluid and normal components inside of a neutron star~\cite{Schaab1999}. 


\section{{Universal Relations}}
\label{sec:Universality}

\subsection{Relations for static compact stars}
\label{ssec:Uni_static}

The integral quantities of a compact star such as the mass, radius, moment of inertia, quadrupole moment, etc. sensitively depend on the EoS. However, {\it universal relations} among some integral quantities have been established ~\cite{HaenselZdunik:1989,Friedman:1989,Shapiro:1989,Haensel:1995,Haensel:1996,Haensel:2009} and have been intensively studied in recent years.  
The universality of the relations refers to the fact that they are highly insensitive to the input EoS. 
The pair-wise universal relations between the (dimensionless) moment of inertia, quadrupole moment, and tidal deformability were established in Ref.~\cite{Yagi_PRD_2013}, which raised interest in universal relations
among and beyond these quantities under various conditions such as rapid rotation, strong magnetic fields, etc. Specifically, 
universal relations have been derived for  cold non-rotating stars~\cite{Yagi2013Sci,Sotani2013MNRAS, Majumder2015PhRvD,Steiner2016EPJA,Lenka2017,Wei2019JPhG,Kumar2019PhRvD,Raduta2020,Suleiman2021PhRvC}, slowly and rapidly rotating
stars~\cite{Silva2016MNRAS,Breu_MNRAS_2016,Paschalidis:2017qmb,Riahi2019,Bozzola:2019tit,Khadkikar2021,Koliogiannis:2020}, magnetized stars~\cite{Haskell2014MNRAS}, finite temperature stars~\cite{Raduta2020,Khadkikar2021}, as well as compact stars in a binary~\cite{Manoharan2021}. The universal relations have also been studied in alternative theories of gravity~\cite{Doneva2014PhRvD,Pappas2019,Popchev2019EPJC,Yagi2021PhRvD}. For a review, see Ref.~\cite{Yagi2017x}.
The reason for this universal behavior is not yet fully understood, but its practical use is undisputed, as it helps us constrain quantities that are difficult to access observationally and eliminate the uncertainties associated with the EoS when analyzing data. For example,  in the context of gravitational-wave analysis, the universalities allow one to break degeneracies between integral quantities, e.g., the quadrupole moment and the neutron-star spins in binary inspiral waveforms.

\begin{figure}
\begin{center} 
\includegraphics[width=0.95\columnwidth]{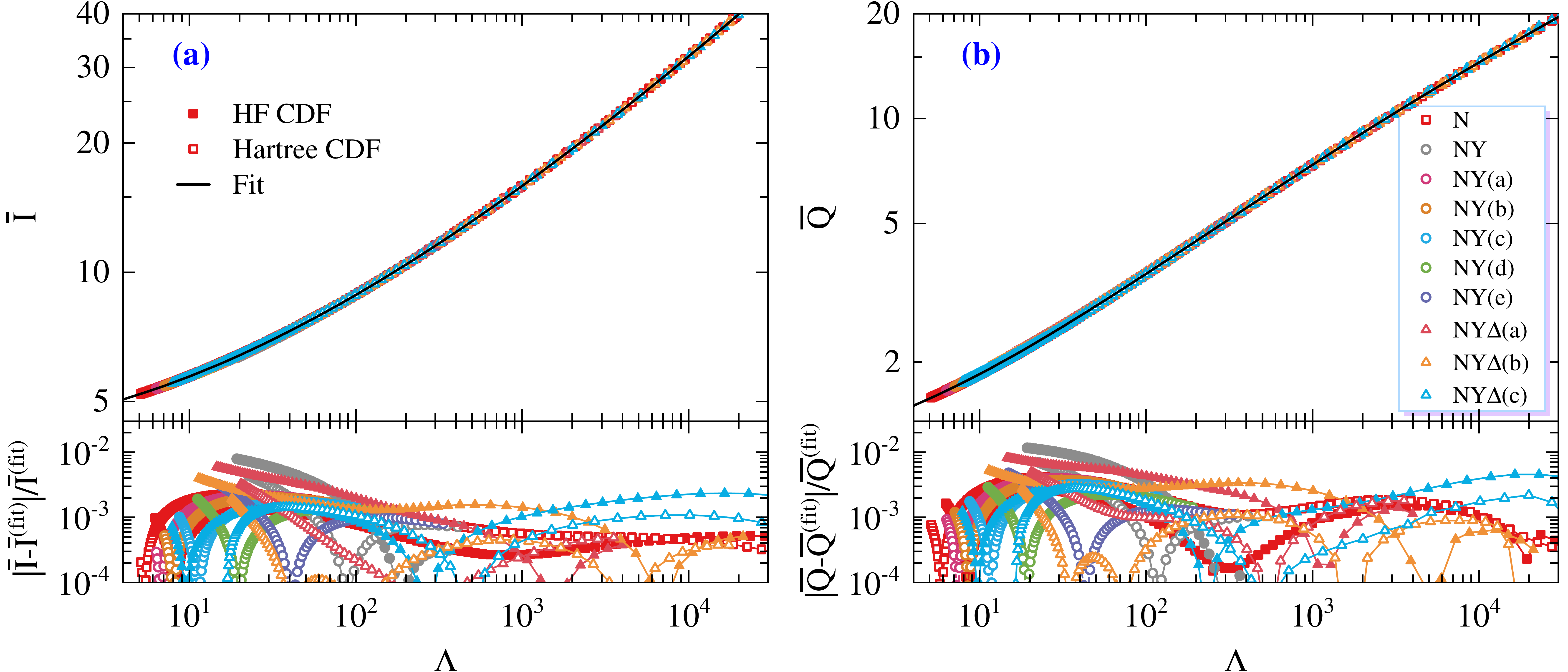} 
\end{center}
\caption{The $I$-$\Lambda$ and $Q$-$\Lambda$ relations for
  non-rotating cold compact stars. The top panels show universal relations for
  various EoS models and matter compositions, together with fitting
  curves; bottom panels show fractional errors between the fitting
  curve and numerical results. The results are derived
  from a collection of EoS models with  nucleonic ($N$),
  hypernuclear ($NY$), and hypernuclear-$\Delta$-admixed
  ($NY\Delta$) compositions. The Hartree models are based on the DDME2
  parameterization (open symbols), the Hartree-Fock models are based on the
  PKO3 parameterization (filled symbols).  The bottom panels show fractional errors between the fitting curve and numerical results. }
\label{fig:I_Love_Q}
\end{figure}

The universalities are proven empirically by testing a large collection of EoS that are already constrained by astrophysical and nuclear laboratory data.  
The so-called $I$-Love-$Q$ relations for non-rotating compact stars can be numerically fitted with a polynomial on a log scale~\cite{Yagi_PRD_2013}:
\begin{align}\label{Eq:I-love-Q}
\ln y = a_0+a_1\ln x+a_2\,(\ln x)^2 + a_3\,(\ln x)^3+a_4\,(\ln x)^4,
\end{align}
where pairs $(x, y)$ represent $(\Lambda, \bar{I})$, $(\Lambda, \bar{Q},)$ and $(\bar{Q}, \bar{I})$. The normalized quantities are defined as $\bar I=I/M_G^3$, $\bar Q=Q M_G/J^2$, where $M_G$ is the gravitational mass and $J$ stands for the angular momentum.
In Fig.~\ref{fig:I_Love_Q}, we plot the results
for $\bar I(\Lambda)$ and $\bar Q(\Lambda)$ for compact
stars constructed using EoS with different matter 
compositions -- nucleonic (labeled $N$), hypernuclear ($NY$) and hypernuclear--$\Delta$-resonance-admixed ($NY\Delta$). 
The EoS collection includes Hartree (based on DDME2 parameterization)
and Hartree-Fock (based on PKO3 parameterization) CDF modeling of dense stellar matter, with the radius and the
maximum mass, respectively, covering $12 \lesssim R_{1.4} \lesssim 14$\,km and $2.0 \lesssim M_{\rm max} \lesssim 2.5\,M_{\odot}$, where $R_{1.4}$ is the radius of the $M=1.4M_{\odot}$ star.
It is seen in Fig.~\ref{fig:I_Love_Q} that the maximum absolute fractional difference is about $1\%$.
Clearly, the universality holds for the third pair $\bar I-\bar Q$ as well.

The search for universal relations among global parameters of compact
stars stretches back to the work of Refs.~\cite{Ravenhall_ApJ_1994, Lattimer_ApJ_2005}, who established the universality for the combination
involving the moment of inertia $\tilde I=I/\left(M_GR^2 \right)$, specifically
\begin{equation}
\tilde I \approx 0.21 (1-2C)^{-1}, \qquad
\tilde I=c_0+c_1 C+ c_2 C^2 + c_3 C^3 + c_4 C^4,
\label{eq:ItildeC}
\end{equation}
with the compactness of a star $C=M_G/R$.
For the moment of inertia $\bar I=I/M_G^3$
(introduced above in the context of $I$-Love-$Q$ relations) universality was also established according to the relation \cite{Breu_MNRAS_2016}
\begin{equation}
\bar I= a_1 C^{-1}+a_2 C^{-2}+ a_3 C^{-3}+ a_4 C^{-4}.
\label{eq:Ibar}
\end{equation}
Reference~\cite{Maselli_PRD_2013} showed the universality of compactness as a function of normalized tidal deformability according to
\begin{equation}
C=b_1 +b_2\ln{\Lambda} + b_3 \left(\ln{\Lambda} \right)^2.
\label{eq:CLove}
\end{equation}
Polynomial fits for $\bar Q$ as a function of inverse compactness
$C^{-1}$ were found in Ref.~\cite{Raduta2020}, 
\begin{equation}
\bar Q=e_0+e_1 C^{-1}+e_2 C^{-2}+e_3 C^{-3} ,
\label{eq:QC}
\end{equation}
which is an analog of Eq.~\eqref{eq:Ibar} for $\bar I$.

Another interesting quantity is the compact star's binding energy, which is defined as the difference between the baryonic and gravitational masses, $E_B=M_B-M_G$. Its insensitivity 
towards the input EoS model, if normalized by the gravitational mass, was shown in Refs.~\cite{Lattimer_ApJ_2001,Breu_MNRAS_2016}.
A possible fit is~\cite{Lattimer_ApJ_2001}
\begin{equation}
\label{eq:BE}
\frac{E_B}{M_G}=\frac{d_1  C}{1-d_2 C}.
\end{equation}
The values of the fitting parameters $a_i$, $b_i$, $c_i$, $d_i$, $e_i$ 
entering Eqs.~\eqref{eq:ItildeC}-\eqref{eq:BE} can be found in tabulated 
from in Ref.~\cite{Raduta2020}. 
\begin{figure}[tp]
\begin{center} 
\includegraphics[width=0.95\columnwidth]{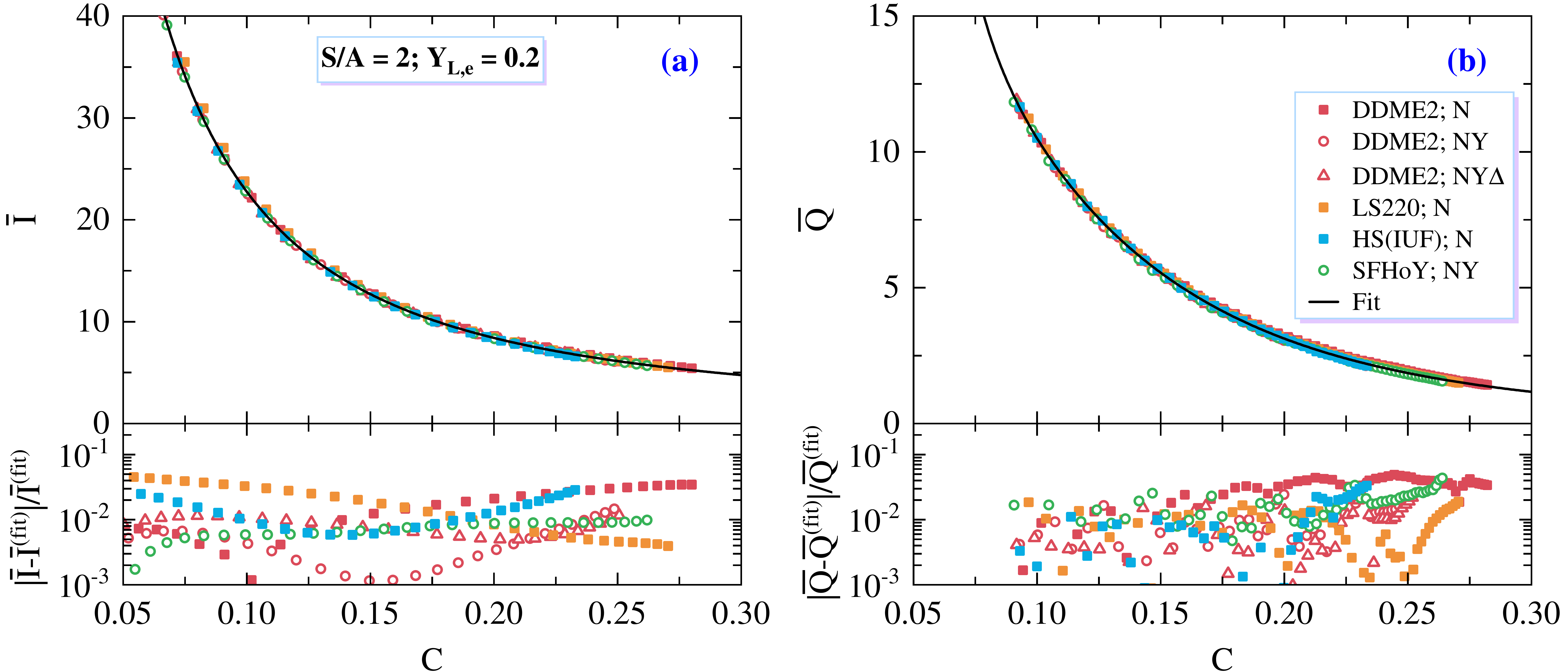} 
\end{center}
\caption{Dependence of $\bar Q$ and $\bar I$ on the compactness $C$ at
  constant $S/A=2$ and $Y_{L,e}=0.2$ for compact stars based on
  different finite-temperature EoS with nucleonic only ($N$),
  hyperonic ($NY$), and $\Delta$-admixed hyperonic ($NY\Delta$)
  compositions according to Ref.~\cite{Raduta2020}. The bottom
  panels show fractional errors between the fitting curve and
  numerical results. }
\label{fig:Qbar_C}
\end{figure}

Let us now turn to the discussion of universality at nonzero temperature~\cite{Martinon_PRD_2014,Marques_PRC_2017,Lenka_JPG_2019, Raduta2020}. Reference~\cite{Martinon_PRD_2014} demonstrated the violation of the universality of $I$-Love-$Q$ relations in the presence of entropy gradients that exist during the early evolution of proto-neutron stars. Similar conclusions were reached in Ref.~\cite{Marques_PRC_2017},  which suggested that these universalities are broken when thermal effects become important. It is now firmly established that if  proto-neutron stars can be approximated by constant entropy and electron fractions, the universalities are restored, as shown for the moment of inertia in Ref.~\cite{Lenka_JPG_2019}, and for all the quantities listed above in Ref.~\cite{Raduta2020}. These universalities have been established using EoS that include hyperonic components~\cite{Marques_PRC_2017,Lenka_JPG_2019,Raduta2020}. The last of these references includes also $\Delta$-resonance featuring EoS. In Fig.~\ref{fig:Qbar_C}
we show the universality of $\bar I(C)$ and $\bar Q(C)$
for a collection of EoS, for $N$, $NY$, or $NY\Delta$ compositions 
based on DDME2 parameterization. Also shown in this figure are the results obtained with the purely nucleonic EoS models LS220~\cite{LS_NPA_1991} and HS(IUF)~\cite{Fischer2014} as well as the hyperonic SFHoY EoS \cite{Fortin_PASA_2018}. The universality visible in Fig.~\ref{fig:Qbar_C} holds to high accuracy with the deviations being at the level of a few percent. The general conclusion is that the universality holds well for all quoted relations also at finite temperature if the same thermodynamic conditions are considered and the precision is similar to that of cold $\beta$-equilibrated matter. There are only a few exceptions for the $E_B/M_G(C)$ relation \eqref{eq:BE}, where the fits can deviate by up to $20\%$. Clearly, the constant entropy approximation will break down to some degree under realistic conditions. While we can be certain that there are no physical obstacles in maintaining the universality in the finite-temperature domain, a formulation that accounts for entropy gradients {\it and} maintains the same thermodynamic conditions (if enforceable at all) for the finite-temperature EoS is necessary to extend the validity of universal relations to dynamical (time-dependent) settings of supernova and BNS mergers.

\subsection{Relations for rapidly rotating compact stars}
\label{ssec:Uni_Kepler}

\begin{figure} \begin{center} \includegraphics[width=0.95\columnwidth]{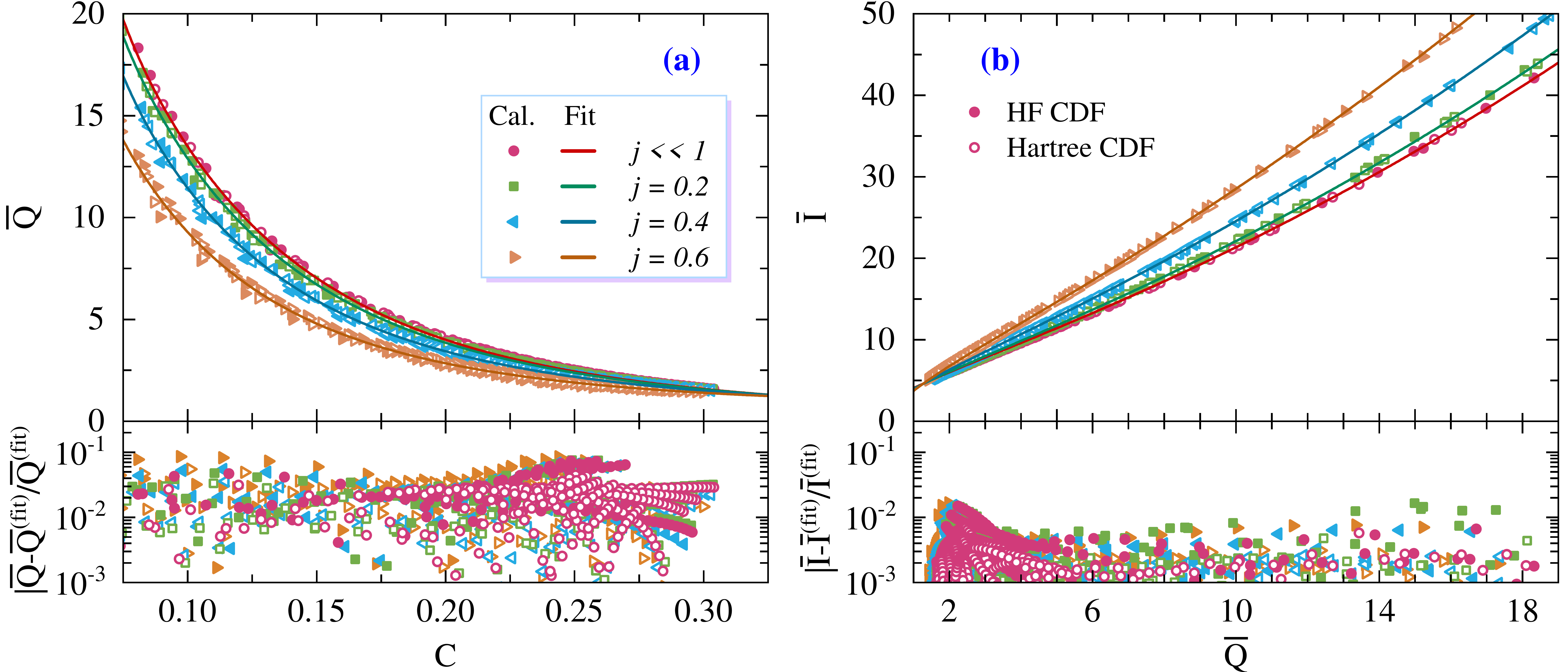} \end{center} \caption{$\bar Q$-$C$ and $\bar I$-$\bar Q$ relations for rotating cold compact stars with different values of the normalized angular momentum.  The top panels show universal relations for various EoS models and matter compositions, together with fitting curves.  The bottom panels show fractional errors between the fitting curve and numerical results. The EoS collection is the same as in Fig.~\ref{fig:I_Love_Q}.}
\label{fig:I_Com_Q}
\end{figure}

The zero- and finite-temperature universalities for hyperonic models of compact stars discussed above can be extended further to rapidly rotating stars without loss of accuracy~\cite{Khadkikar2021}.  When considering rapidly rotating compact stars, the quasi-universal relations between dimensionless compactness $C$, the moment of inertia $\bar{I}$ and quadrupole moment $\bar{Q}$ can be constructed from a stellar sequence with fixed spin parameter $j=J/M_G^2$~\cite{Breu_MNRAS_2016}.  In Fig.~\ref{fig:I_Com_Q}, we show the $\bar{Q}$-$C$ and $\bar{I}$-$\bar{Q}$ relations for four values of $j$. The adopted EoS models are the same as in Fig.~\ref{fig:I_Love_Q}. The fitting curves correspond to Eqs.~\eqref{eq:QC} and \eqref{Eq:I-love-Q}, respectively.

The universalities are of practical importance in determining the maximum masses of compact stars from the analysis of gravitational-wave events, such as GW170817~\cite{Abbott2017a}. The fact that the merger remnant apparently collapsed to a black hole after producing the electromagnetic emission suggests a scenario in which initially a hypermassive compact star evolves into a supramassive, maximally fast rotating star which then collapses into a black hole. Then, universal relations can be used to find the maximum mass of a static star from the deduced mass of the supramassive star, see for more details Refs.~\cite{Margalit:2017dij,Rezzolla:2017aly,Shibata:2019ctb,Khadkikar2021}. Thus, in this context, it is important to establish the relation between the maximum masses of maximally rapidly (Keplerian) rotating compact stars and their static counterparts.  If both stars are taken at finite temperature and at the same thermodynamic state (for example, constant entropy and electron faction) then the relation between these masses is universal.  However, if the static counterpart is cold, then this relation breaks down. Now, because the universality is broken between the cold and hot masses on the Keplerian and static branches, the theory does not work all the way for deducing the static and cold maximum masses. Rather, one is only able to deduce  the mass of the hot counterpart of the static maximum-mass configuration~\cite{Khadkikar2021}.
       
\section{Summary and Outlook}
\label{sec:Conclusions}

In this review article, we presented the CDF approaches to nuclear and hypernuclear matter and their application to the astrophysics of compact stars. As is evident from the discussion, the density functional approach allows one to develop a flexible theoretical framework that can be adjusted to experimental data from laboratory and astrophysical observations. Additional degrees of freedom, such as $\Delta$-resonances and a kaon condensate, can be included in the CDF approach. This framework provides an avenue for further studies towards a CDF  that is highly constrained by the available data. So far, such an approach has provided one of the few avenues to study hypernuclear stars with significant depth, including, for example, their structure (Secs.~\ref{ssec:TOV}, \ref{ssec:TD}),   rapid rotation~(Sec.~\ref{sec:Rapid_rotation}), magnetism (Sec.~\ref{ssec:Bfields}), and thermal evolution~(Sec.~\ref{ssec:CTracks}).

The central motivation to study heavy baryonic degrees of freedom comes from the observation of massive compact stars that are in tension with the low values of maximum masses of compact stars with hyperons in a multitude of models. This hyperon puzzle drew attention to CDF approaches as universal tools for fitting the input  directly to the astrophysical data, rather than taking the circuitous route of fitting to the microscopic quantities, such as phase shifts, on which to build a many-body theory. The uncertainties in the CDF are easily quantified in terms of observable low-order characteristics of nuclear matter, see Sec.~\ref{ssec:Characteristics}. 

The static properties of hypernuclear stars can be constructed from several extensions of nucleonic CDFs in a manner that allows for massive static stars that are consistent with the currently observed heavy millisecond pulsars.  The CDFs that involve couplings consistent with the SU(6) quark model (in the vector meson sector) predict compact stars containing a significant fraction of hyperons in their cores, of the order of $20\%$,  with $\Lambda$s becoming the most abundant species at asymptotically high densities, see Secs.~\ref{ssec:TOV} and \ref{ssec:TD}. However, to interpret the 2.5-2.8 $M_{\odot}$ mass objects identified in gravitational-wave mergers, where the primary is a massive black hole, one needs models which assume SU(3) symmetry with extreme values of parameters, leading to very stiff hyperonic EoS. However, these extreme SU(3) models predict a very low fraction (on the order of a few percent) of hyperons. Therefore, one is essentially dealing with nucleonic stars. Thus, one may conclude that the hypernuclear stars with significant (above a few percent) hyperons must have masses close 2.1~$M_{\odot}$. Assuming rapid rotation, close or at the Keplerian limit, increases the stellar masses by 20$\%$, which may also be an option for the origin of highly massive compact stars, see~Sec.~\ref{ssec:SU(3)}.

The hypernuclear CDFs can be extended straightforwardly to finite temperatures and conditions that are relevant for supernovae and BNS mergers by including trapped neutrinos, see Sec.~\ref{sec:FiniteT}. The count rates of nearby type-II supernovae leave open the prospect of observations of such events in the near future. The prospects of observations of BNS mergers, in particular their post-merger phases that is sensitive to the composition of matter, are much better and can potentially provide independent constraints on the CDF parameters. Interestingly, finite temperatures promote the appearance of hyperons at low densities, making them relevant for the description of matter below nuclear saturation density.   

It has been well established that nucleonic pairing has a crucial impact on the cooling of neutron stars by modifying the rates of neutrino emission and the specific heat of nucleonic matter. Similar studies for hyperons in hypernuclear stars with modern CDFs are in the early stages. The key feature of hyperons is the rapid direct Urca cooling, which can be moderated by the pairing gaps in the spectra of hyperons. As discussed in Sec.~\ref{ssec:CTracks}, the combined effect of direct Urca processes involving hyperons  and  hyperon pairing leads to a mass hierarchy in the cooling behavior of hypernuclear stars, such that the heavier the star, the faster it cools. This feature accounts for a wide diversity of temperature distributions of compact stars in their temperature--age diagram. In the future, an improved understanding of hyperonic pairing and cooling will be needed to test CDFs and their applications to the thermal evolution of compact stars. 

Universal relations among the various integral parameters of neutron stars have been established during the last decades. They have become useful for making EoS-independent predictions and inferences. Hypernuclear stars obey these universalities both in the static and rapidly rotating limits with an accuracy that is comparable to their nucleonic counterparts, see Sec.~\ref{sec:Universality}. Therefore, methods and astrophysical scenarios that are based in part or entirely on the assumption of universality can be applied to hypernuclear stars as well. They can also be applied at finite temperatures to isentropic stars with the same compositions. A deeper understanding of the physical origins of universal relations will be needed 
to reveal their insensitivity to the internal composition (e.g., nucleonic vs. hyperonic and $\Delta$-admixed) and identify the astrophysical signatures of hypernuclear stars that distinguish them from nucleonic stars. This will eventually allow one to answer the fundamental question of the existence of hyperons in compact stars. 

To conclude, the hyperonization of dense matter and the astrophysics of compact stars featuring hyperons and $\Delta$-resonances is an enormously vivid and fruitful field of research, with a strong relation to the observational multimessenger astronomy of compact stars. The recent advances on the observational front, which include the observation of gravitational waves from BNS mergers, notably GW170817, simultaneous measurements of masses and radii of neutron stars, and the discovery of very massive pulsars motivate further improvements to existing theoretical models and exploration of their implications for novel astrophysical scenarios. The ongoing and future hypernuclear programs at CERN, BNL, JLab, GSI-FAIR, and J-PARC, and elsewhere will eventually provide us with extremely valuable new insight into hypernuclei. This can be used to narrow down the parameters of CDFs but also constrain other theoretical models. The results achieved so far suggest that the combination of theoretical work and observational progress has the potential to reveal the detailed features of high-density matter in compact stars in the near future.

\section*{Acknowledgments}

It is a great pleasure to thank M.~Alford, G.~Colucci, A.~Harutyunyan,
W.~Long, M.~Oertel, A.~Raduta, M.~Sinha, and V.~Thapa for
collaborations from which in part this review derives.
A.~S. acknowledges support by Deutsche Forschungsgemeinschaft Grant
No. SE 1836/5-2 and the Narodowe Centrum Nauki (NCN) Grant
No. 2020/37/B/ST9/01937 at Wroc\l{}aw University.  J.~L. is supported
by the National Natural Science Foundation of China (Grant
No. 12105232), the Venture \& Innovation Support Program for Chongqing
Overseas Returnees (Grant No. CX2021007), and by the Fundamental
Research Funds for the Central Universities (Grant No. SWU-020021).
F.~W. acknowledges support from the U.S. National Science Foundation
under Grant PHY-2012152.


\end{document}